\documentclass[a4paper,11pt]{article}
\pdfoutput=1
\usepackage[text={17.1cm,24.6cm},centering]{geometry}
\usepackage[utf8]{inputenc}
\usepackage{physics}
\usepackage{amsmath,amssymb,mathtools,amsthm,mathrsfs}
\usepackage{subcaption}
\usepackage{pst-node}
\usepackage{tikz-cd} 
\usepackage{graphicx}
\usepackage{authblk}
\usepackage{cite}
\usepackage[colorlinks=true,linkcolor=blue, citecolor=red, urlcolor=blue, bookmarks]{hyperref}
\usepackage{dsfont}
 
\title{Symmetry resolved relative entropies and distances in conformal field theory}
\author[1]{Luca Capizzi\footnote{email: lcapizzi@sissa.it}}
\author[1,2]{Pasquale Calabrese}
\affil[1]{SISSA and INFN, Via Bonomea 265, 34136 Trieste, Italy}
\affil[2]{International Centre for Theoretical Physics (ICTP), Strada Costiera 11, 34151 Trieste, Italy}
\date{}                     
\def \be {\begin{equation}} 
\def \ee {\end{equation}} 

\def \bea {\begin{eqnarray}} 
\def \eea {\end{eqnarray}} 

\def \l {\left(} 
\def \r {\right)} 
 
\def \la {\left\langle} 
\def \ra {\right\rangle}

\begin{document}
\maketitle

\begin{abstract}

We develop a systematic approach to compute the subsystem trace distances and relative entropies for subsystem reduced density matrices associated 
to excited states in different symmetry sectors of a 1+1 dimensional conformal field theory having an internal U(1) symmetry. 
We provide analytic expressions for the charged moments corresponding to the resolution of both relative entropies and distances 
for general integer $n$. 
For the relative entropies, these formulas are manageable and the analytic continuation to $n=1$ can be worked out in most of the cases.  
Conversely, for the distances the corresponding charged moments become soon untreatable as $n$ increases. 
A remarkable result is that relative entropies and distances are the same for all symmetry sectors, i.e. they satisfy entanglement equipartition, like the entropies. 
Moreover, we exploit the OPE expansion of composite twist fields, to provide very general results when the subsystem is a single interval much smaller than the total system. 
We focus on the massless compact boson and our results are tested against exact numerical calculations in the XX spin chain.

\end{abstract}

\newpage
\tableofcontents

\section{Introduction}
Nowadays entanglement is a central theme in the description of extended quantum systems such as in field theories and many-body condensed matter. 
Different communities, both experimental and theoretical ones, started looking into entanglement for so many different reasons that it is impossible to give 
the right credit to all the ideas and concepts that came to the light in the last two decades or so.
Just to quote few examples, entanglement is an extremely powerful tool to characterise different phases of matter \cite{Amico,Laflorencie:2015eck,ccd-09}, 
in particular reference to topological order \cite{tord1,tord2,tord3,lh-08}.
It is also a fundamental object to understand equilibration and thermalisation of isolated non-equilibrium quantum systems \cite{kaufman2016quantum,c-20}. 
It turned out to have a key role in the black hole information loss paradox \cite{binfo1,binfo2,binfo3,binfo4} and confinement in gauge theories \cite{bp-08,gauge,conf1}.
Furthermore in the AdS-CFT correspondence, the Ryu-Takayanagi formula \cite{rt-06,nrt-09,Hol-Ren} opened the route for a deeper understanding of the emergence of space-time 
from the entanglement itself \cite{v-10}.
Finally, the recent pioneering experiments measuring the many-body entanglement in cold-atom and ion-trap 
settings \cite{kaufman2016quantum,islam2015measuring,elben2018renyi,lukin2018probing,brydges2018probing,ekh-20}
further boosted the field. 
A pivotal contribution to all these developments, central to both high energy and condensed matter,  came from two dimensional conformal field theory (CFT)
that led to a pletora of remarkable universal results for many entanglement related quantities 
\cite{hlw-94,cc-04,cc-09,calab-2int1,calab-2int2,Neg1,Neg2,lashkari2014,lashkari2016,Sierra1,Sierra2,Rel_Paola,Rel-CFT1,Trace_Paola,Zhang2}
(here we mention only some references that will be useful later on), in particular for the celebrated entanglement entropy defined as $S=-\tr \rho_A\log \rho_A$, with 
$\rho_A$ the reduced density matrix (RDM) of the subsystem $A$.

The large majority of these studies concerned the entanglement in a single quantum state.
However, quantum information ideas provide also insightful ideas when considering two different quantum states.
In this respect, the most studied  quantity so far is surely the \emph{relative entropy} \cite{op-04,a-76}
\begin{equation}
S(\rho || \sigma) = {\rm Tr} (\rho \log \rho) -{\rm  Tr }(\rho \log \sigma),
\label{RelDef}
\end{equation}
for two (reduced) density matrices $\rho$ and $\sigma$.
The relative entropy is often interpreted as a measure of distinguishability of quantum states.
The relative entropy attracted a lot of interests from the field theory community, see e.g. 
\cite{casini-2016,clt-16,black-hole-thermodynamics,bekenstein-bound,hol-rel-entropy,lashkari2014,lashkari2016,Rel-CFT1,Rel-CFT2,ugajin2016-2, Rel_Paola,mrc-18,fr-19,nsu-19,h-20,nu-17,a-20,cgd-19,jlm-16}, 
also, but not only, for its relation with the modular Hamiltonian \cite{chm-11,bntu-13} and quantum null energy condition \cite{bfkw-19}. 

However, the relative entropy has a major drawback as a measure of distinguishability. 
Indeed, a proper measure of the difference between states should be a {\it metric} in a mathematical sense, 
meaning it should be nonnegative, symmetric in its inputs, equal to zero if and 
only if its two inputs are the same, and should obey the triangular inequality. 
Clearly, the relative entropy does not match these requirements (it is not even symmetric in its entries). 
An important family of distances, all satisfying the above rules, is given by the Schatten distances
\be \label{DnDef}
D_n(\rho,\sigma) = \frac{1}{2^{1/n}} \| \rho - \sigma \|_n,
\ee
where $n \geq 1$ is a generic real parameter. Here $||\cdot||_n$ stands for the $n$-norm, see below.
It is well known that the \emph{trace distance}
$D(\rho,\sigma)  = \frac12 \| \rho - \sigma \|_1$ (i.e. \eqref{DnDef} for $n=1$) has several properties that makes it special and more effective compared 
to the others values of $n$ and even compared to other distances, see e.g. the examples and discussions 
in Refs. \cite{nc-10,watrous2018theory,fe-13,zcdr-20,Zhang2,bgkk-21}. 
Also the Schatten distances have been studied in field theories \cite{Trace_Paola,Zhang2, Zhang1,zcdr-20, za-20,zr-20,zr-21}, 
but not as much as the relative entropy, most likely because of the 
more difficult replica approach (see below) necessary for their determination.

In very recent times, it has been also understood that many genuine quantum features can be characterised by studying 
the relation between entanglement and symmetries and in particular how entanglement is shared between the various symmetry sectors 
of a theory \cite{lr-14,lukin2018probing,G1,Sierra3}.
To date there are many results concerning the resolution of the entanglement in 
a given state \cite{G1,G2,G3,G4,G5,Ric1,Ric2,Sara2,Sara3,Sara4,Sara5,Sierra3,Purif,ncv-21,S-int,Hol-CS,Zn,Multi-bos,Random1,Capizzi,hcc-21,Ric3,fg-21}, but none 
for the distinguishability of two different states. 
The goal of this work is to start filling this gap by studying both the relative entropies and the distances in the various symmetry sectors of a $U(1)$-symmetric theory. 
The main physical reason why we are interested in this issue is to have a finer description of the similarity between states. 
To be specific, in several different contexts (e.g. for the equilibration after a quench \cite{fe-13} and to test lattice Bisognano Wichmann entanglement Hamiltonians \cite{zcdr-20})  
it is fundamental to study how one reduced density matrix approach another one in the thermodynamic limit (while the full density matrices are still very distant).
Such asymptotic approach is often investigated by studying the scaling of relative entropies and distances with the subsystem size $\ell$. 
However, it is very natural to wonder, whether the distance between the symmetry resolved reduced density matrices can stay finite for large $\ell$ in some sectors 
while the same distance for the entire RDMs tends to zero. 
For this to be possible, the considered sector must have a very little weight (in $\ell$) to ensure that total distance goes to zero; 
this requirement  is not at all odd because, as we shall see, the probability of the various sectors is Gaussian (with a variance proportional only to $\log \ell$ for the states 
of interest here). 
As a consequence, if this would happen, an observable fine tuned on that symmetry sector would have a different value on the two states. 

Studying in general the symmetry resolved distinguishability of reduced density matrices is a very ambitious aim. 
Here, we only consider a much more modest problem and focus on low-lying excited states of conformal field theories and characterise their relative entropies and distances. 
To do so, we have to put together several pieces of a puzzle already present in the literature, namely: 
(I) the construction of the RDM in excited states of CFT \cite{Sierra1,Sierra2}, 
(II) replica trick for relative entropies \cite{lashkari2014,lashkari2016,Rel-CFT1,Rel_Paola} and distances \cite{Trace_Paola,Zhang2}, 
(III) the symmetry resolution of these density matrices via charged moments \cite{G1,Sierra3}. 
This program presents a few technical and conceptual obstacles that will be discussed and tackled in the remaining of the paper

The paper is organised as follows. 
In section \ref{SRdist}, we recap the known tools for the relative entropies and distances and we provide a precise notion their symmetry resolution.  
In section \ref{sec3}, we develop our CFT approach to these symmetry resolved quantities, derive the OPE of  
charged twist fields, and show equipartition of both relative entropies and distances. 
In section \ref{CBos}, we explicitly compute the universal CFT scaling function quantities for the field theory of compact boson (aka Luttinger liquid).
We eventually exploit the knowledge of these functions for the explicit determination of symmetry resolved relative entropies and distances in Sections
\ref{sec5} and \ref{sec6} respectively. 
Finally, section~\ref{sec7} contains our conclusions and a few outlooks. 
Some technical details are relegated to two appendices.

\section{Symmetry resolved relative entropies and distances}\label{SRdist}

In this section, we recap the notion of symmetry resolution of entanglement measures and provide new definitions for the measures of the subsystem distinguishability of two states
within the symmetry sector. Namely we define \textit{symmetry resolved relative Rényi entropies and subsystem Schatten distances}. 

Let us consider a quantum theory which admits the following decomposition of the Hilbert space $\mathscr{H}$
\be
\mathscr{H} = \bigoplus_q \mathscr{H}_q,
\ee
where $q$ is an index which parametrises the sector $\mathscr{H}_q$ (although we will main interested in decompositions into a direct sum of irreducible representations 
of a group associated with the internal symmetries, this is not yet a required assumption; decomposition of non symmetric states is also a useful idea, see e.g. \cite{ncv-21}). 
Let us denote by $\Pi_q$ the linear projector onto the sector $\mathscr{H}_q$ under consideration. 
For any density matrix $\rho$ satisfying $\trace(\Pi_q \rho )\neq 0$, we can define a conditioned density matrix $\rho(q)$ as 
\be
\rho(q) \equiv \frac{\Pi_q \rho \Pi_q}{\trace(\Pi_q \rho \Pi_q)},
\label{condRDM}
\ee
where the denominator ensures the normalisation $\trace(\rho(q))=1$. Whenever $[\rho,\Pi_q]=0$,  it holds $\Pi_q \rho \Pi_q = \rho \Pi_q = \Pi_q \rho$.
Hereafter, we focus on symmetric states, i.e. such $\rho$ commutes with all $\Pi_q$, so that we can decompose the density matrix in a block diagonal form 
\be
\rho = \sum_q p(q) \rho(q), \qquad p(q)= \trace(\rho \Pi_q), 
\label{pq}
\ee
where $p(q)$ is the probability of the $q$ sector.
The symmetry resolved R\'enyi entropies are then 
\begin{equation}
\label{Snqdef}
S_n(q)=\frac{1}{1-n}\log \mbox{Tr}[\rho(q)^n],
\end{equation}
that in the limit $n\to1$ reduce to the von Neumann entropy $S(q)\equiv S_1(q)$. 
The latter satisfies the important sum rule \cite{nc-10,lukin2018probing}
\be
S= \sum_q p(q)S(q) - \sum_q p(q)\log p(q),
\label{2terms}
\ee
The two terms in \eqref{2terms} are usually referred to as \emph{configurational} and \emph{number} entanglement entropy, respectively \cite{lukin2018probing}.
The former represents the (weighted) sum of the entropies in each charge sector and the latter is the entropy due to the fluctuations of the charge between the two 
subsystems. The two terms have their own interest in the literature \cite{Barg1,Barg2,wv-03,kusf-20,kusf-20b,kufs-21,zfgs-21,ctd-19}, but will not be discussed here.

Let us now consider two density matrices $\rho$ and $\sigma$; 
we can use the relative entropy \eqref{RelDef} for each sector $q$, i.e., 
\be
S(\rho \| \sigma)(q) \equiv S(\rho(q) \| \sigma(q))= \trace(\rho(q) \log \rho(q) )- \trace (\rho(q) \log \sigma(q) ),
\ee
as a measure of distinguishability between the states in that sector, on the same lines of what normally done for the total density matrix. 
In terms of the total density matrices and projectors $S(\rho \| \sigma)(q)$ may be written as 
\be
S(\rho \| \sigma)(q) \equiv  -\frac{\trace(\rho \log \sigma\Pi_q )}{\trace(\rho \Pi_q)} + \frac{\trace(\rho \log \rho \Pi_q)}{\trace(\rho \Pi_q)} - \log \frac{\trace(\rho  \Pi_q)}{\trace(\sigma \Pi_q)}.
\ee
The symmetry resolved relative entropies satisfy the sum rule
\be
S(\rho \| \sigma) = \sum_q p^\rho(q)S(\rho \| \sigma)(q) + \sum_q p^\rho(q)\log \frac{p^\rho(q)}{p^\sigma(q)},
\ee
where
\be
p^\rho(q) \equiv \trace(\rho \Pi_q), \qquad p^\sigma(q) \equiv \trace(\sigma \Pi_q).
\ee
Following Refs. \cite{lashkari2014,lashkari2016,Rel_Paola,Rel-CFT1},  the relative entropy can be obtained as the replica limit $n\to 1$ of the
$n$-th Rényi entropy of the sector $q$
\be
S_n(\rho \| \sigma)(q) \equiv S_n(\rho(q) \| \sigma(q)) = \frac{1}{1-n}\log \frac{\trace(\rho(q)\sigma(q)^{n-1})}{\trace(\rho(q)^n)} = 
\frac{1}{1-n}\log \frac{\trace(\rho\sigma^{n-1}\Pi_q)(\trace(	\rho \Pi_q))^{n-1}}{\trace(\rho^n \Pi_q)(\trace(\sigma \Pi_q))^{n-1}}.
\label{Rel_res}
\ee
(Actually also other more physical forms of R\'enyi relative entropies exist, see e.g. \cite{lashkari2014}, but from a replica perspective they just represent an inessential complication.)

On the same line, we can define the symmetry resolved Schatten $n$-distance $D_n(\rho,\sigma)$ as
\be
D_n(\rho,\sigma)(q) \equiv D_n(\rho(q),\sigma(q))= \frac{1}{2^{1/n}}\| \rho(q)-\sigma(q)\|_n,
\label{Dn}
\ee
It is defined in terms of the $n$-norm of an operator $\Lambda$
\be
\| \Lambda \|_n \equiv \l \sum_i \lambda_i^n\r^{1/n},
\ee
with $\lambda_i$ being the eigenvalues of $\sqrt{\Lambda^\dagger \Lambda}$.
We recall that for infinite dimensional Hilbert spaces, not all distances are equivalent, and thus one has in general different notions of indistinguishability of states. 
Moreover one has to be particularly careful on how the states are regularised in the continuum limit, otherwise the distance can diverge or going to zero in an undesired way,
see e.g. Refs. \cite{Trace_Paola,Zhang2} for practical examples.
Unfortunately, the natural definition \eqref{Dn} of distances between sectors is untreatable analytically (and also very difficult numerically). 
For this reason, we introduce also another notion of (still unnormalised) symmetry resolved distance as 
\be
D'_n(\rho,\sigma)(q) \equiv \frac{1}{2^{1/n}} {\|\Pi_q (\rho-\sigma)\|_n} = \frac{1}{2^{1/n}} (\trace\l |\rho-\sigma|^n \Pi_q \r)^{1/n}.
\label{D1np}
\ee
As we shall see, $D'_n$ is analytically treatable and it is related to the total $n$-distance by the following sum rule
\be
\sum_{q}( {D}'_n(\rho,\sigma)(q) )^n= ({D}_n(\rho,\sigma))^n.
\label{Dprime}
\ee

\subsection{Reduced density matrices and charged moments}

Until this point, everything is valid for arbitrary density matrices, independently of their origin. 
Here we are interested in entanglement properties and so to the case when the density matrices correspond to spatial subsystems of a larger system in a pure state $|\Psi\rangle$,
with $\rho=\ket{\Psi}\bra{\Psi}$.
Such spatial bipartition induces the decomposition of the Hilbert space $\mathscr{H} = \mathscr{H}_A \otimes \mathscr{H}_B$ so that the reduced density matrix of 
the subsystem is 
 \be
 \rho_A \equiv \trace_B(\ket{\Psi}\bra{\Psi}).
\ee
Now we consider a system having an internal $U(1)$ symmetry, meaning that the state $\rho$ commutes with a local charge operator $Q$ \cite{G1} $[\rho,Q] = 0$.
Taking the partial trace of the previous relation, one gets
\be
[\rho_A,Q_A] = 0,
\ee
i.e. $\rho_A$ has a block diagonal form with blocks corresponding to the eigenvalues $q$ of $Q_A$. 
An effective way to write  the projectors $\Pi_{q}$, particularly useful for field theory calculations, is through Fourier transform
\be
\Pi_{q} = \int_{-\pi}^{\pi} \frac{d\alpha}{2\pi} e^{i\alpha Q_A}e^{-i\alpha q}.
\ee
The reason why this technique, introduced in Ref. \cite{G1}, is powerful is that it provides a formalism which connects non local objects, as the 
symmetry-resolved entanglement measures, to local quantities, as correlation functions in a replicated theory.
For example, for the entanglement entropy, in field theory it is convenient to start from the computation of the the  \textit{charged moments} \cite{G1,Sierra3}
\begin{equation}
Z_n({\alpha})\equiv\mbox{Tr}[\rho_A^ne^{i\alpha Q_A}],
\label{Zna}
\end{equation}
whose Fourier transform
\begin{equation}
\mathcal{Z}_n(q)= \int_{-\pi}^{\pi}\frac{d \alpha}{2 \pi} e^{-i q \alpha} {Z}_n(\alpha)\equiv \mbox{Tr}[\Pi_q\rho_A^n],
\label{Znqdef}
\end{equation}
gives the symmetry resolved R\'enyi entropies \eqref{Snqdef} as
\begin{equation}
S_n(q)=\frac{1}{1-n}\log \left[\frac{\mathcal{Z}_n(q)}{\mathcal{Z}_1(q)^n} \right] .
\label{SvsZ}
\end{equation}
The probability $p(q)$ in Eq. \eqref{pq} is $p(q)=\mathcal{Z}_1(q)$.
It is worth to mention that charged moments like (or similar to) those in Eq. \eqref{Zna} have been independently analysed in the 
past \cite{bym-13,d-16,d-17,Caputa,Caputa1,Caputa2,srrc-19}. 

In a very similar manner, charged composite moments for the relative entropies and trace distances can be defined. 
Let us start from the former, although it is a special case of the latter.
For the relative entropy between two RDMs $\rho_A$ and $\sigma_A$, we just need to compute the charged moments
\be
\trace(\rho_A\sigma_A^{n-1}e^{i\alpha Q_A})\,,
\label{exp}
\ee
whose Fourier transform
\be
\trace(\rho_A\sigma_A^{n-1}\Pi_q) =\int_{-\pi}^{\pi}\frac{d\alpha}{2\pi}e^{-iq\alpha}\trace(\rho_A\sigma_A^{n-1}e^{i\alpha Q_A})
\ee
readily provides the R\'enyi relative entropies defined as in Eq. \eqref{Rel_res}.

The replica trick for the subsystem Schatten distance is based on the expansion of  $\tr (\rho_A-\sigma_A)^{n}$ as 
\be \label{expansion}
\tr (\rho_A-\sigma_A)^{n} = \sum_{\mathcal{S}} (-)^{|\mathcal{S}|} \tr \left( \rho_{1_{\mathcal{S}}} \cdots \rho_{(n)_{\mathcal{S}}} \right)  ,
\ee
where the summation $\mathcal{S}$ is over all the subsets of $\mathcal{S}_0 = \{ 1, \cdots, n \}$, $| \mathcal{S}|$ is the cardinality of 
$\mathcal{S}$ and $\rho_{j_{\mathcal{S}}} =  \sigma_A $ if $j \in \mathcal{S}$ and $\rho_A$ otherwise. 
This expression coincides with the Schatten distance only for $n$ even. 
All other (real) values of $n$, including the important $n=1$ being the trace distance, are obtained taking the analytic continuation from 
the sequence of even $n=n_e$, as explained in \cite{Trace_Paola,Zhang2} (and using earlier ideas for the evaluation of absolute value by replicas \cite{k-91}).  
Crucially, each term in the sum appearing in the rhs of Eq.~\eqref{expansion} is related to a partition function on an $n$-sheeted Riemann 
surface.
In the presence of a flux, Eq. \eqref{expansion} is trivially generalised as 
\be \label{exp2}
\tr [(\rho_A-\sigma_A)^{n}e^{i\alpha Q_A}] = \sum_{\mathcal{S}} (-)^{|\mathcal{S}|} \tr \left( \rho_{1_{\mathcal{S}}} \cdots \rho_{(n)_{\mathcal{S}}}e^{i\alpha Q_A} \right)  ,
\ee
whose Fourier transform is exactly $D'_n(q)$ in Eq. \eqref{D1np} for even $n$.
It should be now clear why the distance in Eq. \eqref{D1np} is easily computed by replicas while \eqref{Dn} is not.

\section{From replicas and charged twist fields to symmetry resolved relative entropies and distances}
\label{sec3}

In the replica approach, the moments of the RDM, $\mbox{Tr}\rho_A^n$, are evaluated for any $(1+1)$-dimensional quantum field theory as partition functions over the
$n$-sheeted Riemann surface $\mathcal{R}_n$ in which the $n$ sheets (replicas) are cyclically joined along the subsystem $A$ \cite{cc-04,cc-09}. 
Similarly \cite{G1}, the charged moments find a geometrical interpretation by inserting an Aharonov-Bohm flux through such surface,  so that the total phase accumulated by 
the field upon going through the entire surface is $\alpha$. 
Then the partition function on such modified surface is the charged moments ${Z}_n(\alpha)$ in Eq. \eqref{Zna}.

This partition function can be rewritten in terms of the correlator of twist fields implementing twisted boundary conditions. 
Assuming, without loss of generality, that the Aharonov-Bohm flux is inserted between the  $n$-{th} and first replicas, we can write the action of the charged twist fields on a charged $U(1)$ bosonic field
as \cite{cc-04,Ola,G1}
\begin{equation}
\mathcal{T}_{n,\alpha}(x,\tau) \phi_i(x',\tau)= \begin{cases}\phi_{i+1}(x',\tau)e^{i \alpha \delta_{in}}\mathcal{T}_{n,\alpha}(x,\tau), & \mbox{if } x<x', \\ 
\phi_i(x',\tau)\mathcal{T}_{n,\alpha}(x,\tau), & \mbox{otherwise.}  \end{cases}
\end{equation}
In terms of these composite twist fields, the charged moments for a single interval $A=[0,\ell]$ in the ground state (vacuum of the QFT) are
\begin{equation}
Z_n(\alpha)=\langle  \mathcal{T}_{n,\alpha}(\ell, 0) \tilde{\mathcal{T}}_{n,\alpha}(0,0) \rangle,
\end{equation}
where $\tilde{\mathcal{T}}_{n,\alpha}= \mathcal{T}_{n,\alpha}^\dag$ is known as the anti-twist field. 
We will refer to  $\mathcal{T}_{n,\alpha}$ for $\alpha\neq 0$ as the {\it charged (or composite) twist field} while to $\mathcal{T}_{n}\equiv \mathcal{T}_{n,0}$ as the standard twist field.

In the following, we focus on excited states of conformal field theory. 
To this aim, it is useful to work out first the OPE of the twist fields as done in the following subsection. 

\subsection{Operator product expansion of twist fields}\label{TFields}

In this subsection, we first review the construction of the OPE of standard twist fields (following Refs. \cite{calab-2int1,calab-2int2} for the generation of the primaries
see \cite{Zhang7} for the descendants), and then we generalise these results to charged twist fields.

Let us focus on the holomorphic part of a  CFT (for a non-chiral theory the antiholomorphic sector is similarly treated) of central charge $c$.
We write a full set of primaries as
\be
\{\mathcal{O}_a\}_a.
\ee
We refer to $\text{CFT}_n$ as the theory built with $n$ replicas of the original $\text{CFT}$ with central charge is $cn$. 
A full set of operators which are primary w.r.t all $n$ copies of $\text{CFT}_n$ is
\be
\{\mathcal{O}^{1}_{a_1}\otimes \dots \otimes \mathcal{O}^{n}_{a_n}\},
\ee
where the upper index is a replica index. 
This $\text{CFT}_n$ has a permutation symmetry $\mathbb{Z}_n$ which can be promoted to internal symmetry, leading to the construction of the orbifolded theory 
$\text{CFT}_n/\mathbb{Z}_n$. The operator content of the latter is different from the one of $\text{CFT}_n$ and, in particular, the twist fields appears as local operators 
(a clear and complete treatment of the orbifold construction in the context of entanglement can be found in \cite{Minimal-models}).

Roughly, the twist field $\mathcal{T}_n(z)$ is defined such that its insertion in the spacetime of the orbifolded theory corresponds to an opening of a branch-cut in the time 
slice $ [z,\infty]$ which connects the $j$-th replica to the $j+1$-th \cite{cc-09}. 
The dimension of $\mathcal{T}_n$ is read off from three-point function $\langle\mathcal{T}_n(z) \tilde{\mathcal{T}}_n (z') T(w)\rangle$ with 
$\tilde{\mathcal{T}}_n = \mathcal{T}_n^\dagger$ and the total stress-energy tensor
\be
T =\sum_{j=1}^n T^j,
\ee
where $T^j$ is a short notation for $1 \otimes \dots \otimes T^j \otimes \dots 1$ (the stress-energy tensor of the $j$-th replica).
Through unfolding procedure
induced by the transformation $\zeta(z) = z^{1/n}$ one gets \cite{cc-09}
\be
\frac{\langle \mathcal{T}_n(0)\tilde{\mathcal{T}}_n(\infty)T(z)\rangle}{\langle \mathcal{T}_n(0)\tilde{\mathcal{T}}_n(\infty)\rangle} = \la \sum_{j=1}^n  \l \frac{d\zeta}{dz}\r^2 T(\zeta e^{-i\frac{2\pi j}{n}})+ \frac{cn}{12} \{\zeta,z\}\ra = \frac{c}{24 z^2}\l n-\frac{1}{n}\r,
\ee
which is equivalent to say that the scaling dimension of the twist field is $h_{\mathcal{T}_n} = \frac{c}{24}\l n-\frac{1}{n}\r$.
Moreover, since $\tilde{\mathcal{T}}_n = \mathcal{T}_n^\dagger$, the following fusion is present
\be
[\mathcal{T}_n]\times[\tilde{\mathcal{T}}_n] \rightarrow [1],
\ee
and then all the descendants of the conformal tower of the identity are generated in the OPE.

Similarly, we conclude that a primary (nonidentity) operator $\mathcal{O}^j_a$ is not present in the OPE twist fields, because its one-point function 
$\langle\mathcal{O}^j_a(\zeta)\rangle$ on the plane is zero. 
However, if $[\mathcal{O}_{a_j}]\times [\mathcal{O}_{a_k}] \rightarrow [1]$ (implying that $\mathcal{O}_{a_j}$ and $\mathcal{O}_{a_k}$ have the same conformal dimension 
$h_{a_j} = h_{a_k}$), the following fusion is present
\be
[\mathcal{T}_n]\times[\tilde{\mathcal{T}}_n] \rightarrow [\mathcal{O}^j_{a_j}\mathcal{O}^k_{a_k}],
\ee
and the unfolding leads to 
\be
\frac{\langle \mathcal{T}_n(0)\tilde{\mathcal{T}}_n(\infty)(\mathcal{O}^j_{a_j}\mathcal{O}^k_{a_k})(z=1)\rangle}{\langle \mathcal{T}_n(0)\tilde{\mathcal{T}}_n(\infty)\rangle} = \frac{1}{n^{h_{a_j}+h_{a_k}}}\langle \mathcal{O}^j_{a_j}(\zeta = e^{-i\frac{2\pi j}{n}})  \mathcal{O}^k_{a_k}(\zeta = e^{-i\frac{2\pi k}{n}}) \rangle.
\label{2op_std}
\ee
Here products of operators at coinciding points are intended as applied on different sheets on the unfolded theory, see Ref. \cite{Minimal-models} for details.  
Similarly all the other fusions between the twist fields and the other primaries of the replicated theory can be obtained from $m$-point functions of primaries in the unreplicated theory.

For the charged twist fields, the discussion is almost the same with some additional caveats. 
Let us consider a primary operator $\mathcal{V}_\alpha(z)$ which acts as a symmetry generator, i.e. it inserts an additional flux, in the timeslice $\in [z,+ \infty)$. 
The modified twist field $\mathcal{T}_{n,\alpha}$ is constructed by fusing together $\mathcal{T}_n$ and $\mathcal{V}_\alpha$, which means that it is the lightest operator appearing in the OPE $\mathcal{T}_n(z)\mathcal{V}_\alpha(0)$ (see e.g. \cite{Zn,cadl-11,Levi-12,css-89,ks-90,bhs-98}). 
We use the convention that the additional flux is inserted between the $n$-th and the first replica, hence the fusion  is between $\mathcal{T}_n$ and 
$1\otimes \dots \otimes \mathcal{V}_\alpha$, so that the symmetry generator is inserted only in the $n$-th replica. 
However, this is only a technical point and any other choice does not affect the following discussion in any relevant part.
Once one unfolds the theory, the charged twist field generates an additional insertion $\mathcal{V}_\alpha$ (instead of the identity operator for standard twist fields). 
In order to see this, let us compute the dimension of the modified twist field \cite{kw-90,Bouwknegt-96,G1}
\be
\frac{\langle \mathcal{T}_{n,\alpha}(0)\tilde{\mathcal{T}}_{n,\alpha}(\infty)T(z)\rangle}{\langle \mathcal{T}_n(0)\tilde{\mathcal{T}}_n(\infty)\rangle} = \frac{\la \mathcal{V}_\alpha(0)\mathcal{V}_{-\alpha}(\infty)\l \sum_{j=1}^n  \l \frac{d\zeta}{dz}\r^2 T(\zeta e^{-i\frac{2\pi j}{n}})+ \frac{cn}{12} \{\zeta,z\}\r \ra}{\la \mathcal{V}_\alpha(0)\mathcal{V}_{-\alpha}(\infty) \ra} = \frac{\frac{h_{\mathcal{V}_\alpha}}{n} + h_{\mathcal{T}_n}}{z^2},
\ee
so
\be
h_{\mathcal{T}_{n,\alpha}}= \frac{h_{\mathcal{V}_\alpha}}{n} + h_{\mathcal{T}_n}.
\ee
The fusion $\mathcal{T}_{n,\alpha} \times \l\mathcal{T}_{n,\alpha}\r^\dagger$ is obtained from $(m+2)$-point function of 
$m$ primaries ${\cal O}^k_{a_k}$ and the two charges $\mathcal{V}_\alpha(0), \mathcal{V}_\alpha(\infty)$. 
In particular it holds
\be
\frac{\langle \mathcal{T}_{n,\alpha}(0)\tilde{\mathcal{T}}_{n,\alpha}(\infty)(\mathcal{O}^j_{a_j}\mathcal{O}^k_{a_k})(z=1)\rangle}{\langle \mathcal{T}_n(0)\tilde{\mathcal{T}}_n(\infty)\rangle} = \frac{1}{n^{h_{a_j}+h_{a_k}}}\langle \mathcal{V}_\alpha(0) \mathcal{O}^j_{a_j}(\zeta = e^{-i\frac{2\pi j}{n}})  \mathcal{O}^k_{a_k}(\zeta = e^{-i\frac{2\pi k}{n}})\mathcal{V}_{-\alpha}(\infty) \rangle,
\ee
which is the generalization of \eqref{2op_std} in the presence of a nontrivial flux. 
We stress explicitly that, in the fusion $\mathcal{T}_{n,\alpha} \times \l\mathcal{T}_{n,\alpha}\r^\dagger$, as an important difference with the standard twist fields,
the single primary $\mathcal{O}^j_{a_j}$ appears, as long as the three-point function $\la \mathcal{V}_\alpha(0)\mathcal{V}_{-\alpha}(\infty) \mathcal{O}^j_{a_j}(1) \ra$ 
is non-vanishing.

Summing up, the OPE $\mathcal{T}_{n,\alpha} \times \tilde{\mathcal{T}}_{n,\alpha}$ restricted to the conformal tower of the identity is
\be
\mathcal{T}_{n,\alpha}(z) \tilde{\mathcal{T}}_{n,\alpha}(0) = \langle \mathcal{T}_{n,\alpha}(z) \tilde{\mathcal{T}}_{n,\alpha}(0)\rangle \l 1+z^2\frac{2h_{\mathcal{T}_{n,\alpha}}}{nc} 
\sum_{j=1}^n T^j(0)+\dots \r,
\ee
where $\langle \mathcal{T}_{n,\alpha}(z) \tilde{\mathcal{T}}_{n,\alpha}(0)\rangle = \frac{1}{z^{2h_{\mathcal{T}_{n,\alpha}}}}$ is the normalized correlator among twist fields computed in the vacuum. Similarly, the restriction of the OPE in the space of primaries $\{\mathcal{O}^j\}_j$ reads
\be
\mathcal{T}_{n,\alpha}(z) \tilde{\mathcal{T}}_{n,\alpha}(0) = \langle \mathcal{T}_{n,\alpha}(z) \tilde{\mathcal{T}}_{n,\alpha}(0)\rangle \l \frac{z^{h_{\mathcal{O}}}}{n^{h_\mathcal{O}}}C_{\mathcal{O}\mathcal{V}_\alpha\mathcal{V}_{-\alpha}}\sum_{j=1}^n \mathcal{O}^j(0)+\dots \r,
\ee
and $C_{\mathcal{O}\mathcal{V}_\alpha\mathcal{V}_{-\alpha}}$ is the OPE coefficient of the fusion $[\mathcal{V}_\alpha]\times [\mathcal{V}_{-\alpha}] \rightarrow [\mathcal{O}]$ in the unreplicated theory. In general also the product of two or more non-identity primary operators appears, with an OPE coefficient depending on the theory, 
via a correlation function among primaries. 
We mention that all the other fusions in the corresponding conformal tower can be obtained in analogy with standard operators.


\subsection{Excited states generates by primary fields}

In CFT an excited state $\ket{\Upsilon}$ is written as the action of a local operator $\Upsilon(x,\tau)$  at past infinite imaginary time as
\begin{equation}
\ket{\Upsilon} \sim \underset{\tau \rightarrow -\infty}{\lim} \Upsilon(x,\tau)\ket{0},
\label{UPstate}
\end{equation}
where $\ket{0}$ is the vacuum of the CFT. This mapping is known as state-operator correspondence (see, e.g., the textbooks \cite{pagialle,Muss} for details) 
and applies to any state of the Hilbert state of the CFT. 
The corresponding path-integral representation of the density matrix $\rho=|\Upsilon\rangle\langle\Upsilon|$ presents two insertions of $\Upsilon$ at 
$z=x+i\tau =\pm i\infty$.
Hence, assuming periodic boundary conditions, the worldsheet is an infinite cylinder of circumference $L$. 
We focus on the subsystem $A = [0,\ell]$ (embedded in the system  $[0,L]$) and  we introduce the dimensionless ratio
\be
x \equiv \frac{\ell}{L}.
\ee
Hereafter, we omit explicitly the subscript $A$ for notational convenience, denoting by $\rho_\Upsilon \equiv \text{tr}_{B}(\ket{\Upsilon}\bra{\Upsilon})$ the 
reduced density matrix associated with the state $\ket{\Upsilon}$, referring to the subsystem $A$ only when strictly necessary.

\subsubsection{Moments of the RDM}

For an arbitrary operator $\Upsilon$, $\text{tr}(\rho_{\Upsilon}^n)$ is obtained sewing cyclically along $A$, 
$n$ of the cylinders defining the reduced density matrix $\rho_{\Upsilon}$. 
Consequently, we arrive at a $2n$-point function of $\Upsilon$ on a $n$-sheeted Riemann surface $\mathcal{R}_n$.
Following Ref. \cite{Sierra1}, it is convenient to introduce the universal ratio (with $\rho_{\mathds 1}$ being the vacuum, i.e. ground-state, RDM)  
\begin{equation}
F_{\Upsilon}^{(n)}(x) \equiv \frac{\text{tr}(\rho_\Upsilon^n)}{\text{tr}(\rho_{\mathds 1}^n)}.
\label{FUPdef}
\end{equation}
Keeping track of the  correct normalisation of $\rho_{\Upsilon}$, one obtains \cite{Sierra1,Sierra2}
\begin{equation}
F_{\Upsilon}^{(n)}(x) =  \frac{\left\langle \displaystyle \prod_{k=1}^{n} \Upsilon(z_k^-) \Upsilon^\dagger(z_k^+) \right\rangle_{\mathcal{R}_n}    }{\langle \Upsilon(z_1^-) \Upsilon^\dagger(z_1^+) \rangle_{\mathcal{R}_1}^n },
\label{FU}
\end{equation}
where $z_k^{\mp}$ corresponds to the points at past/future infinite respectively of the $k$-th copy of the system ($k=1,...,n$) in $\mathcal{R}_n$ 
(${\cal R}_1$ is just the cylinder). 
The normalisation  factor of the field $\Upsilon$ does not matter because it cancels out in the ratio \eqref{FU}; 
moreover, $F^{(1)}_\Upsilon(x)=1$ as it should be because of the normalisation of the involved density matrices.

Through the conformal mapping \cite{Sierra2}
\begin{equation}
w(z) = -i\log \left(-\frac{\sin\frac{\pi(z-u)}{L}}{ \sin\frac{\pi(z-v)}{L}}\right)^{1/n},
\label{transf}
\end{equation}
where $u$ and $v$ satisfy $x= \frac{v-u}{L}$, the Riemann surface $\mathcal{R}_n$ is transformed into a single cylinder. 
At this point, exploiting the transformation of the field $\Upsilon$ under a conformal mapping, one relates the ratio \eqref{FU} to the correlation functions of $\Upsilon$ on the plane. 
When $\Upsilon$ is primary, this transformation is 
\begin{equation}
\Upsilon(w, \bar{w}) = \left( \frac{dz}{dw}\right)^h  \left( \frac{d\bar{z}}{d\bar{w}}\right)^{\bar{h}} \Upsilon(z,\bar{z}),
\end{equation}
with $(h,\bar{h})$ the conformal weights of $\Upsilon$. Hence, for primary operators, one can easily express $F^{(n)}_{\Upsilon}(x)$ in terms of correlation functions over the cylinder. 
The final result reads \cite{Sierra2}
\begin{equation}
F_\Upsilon^{(n)}(x) =  n^{-2n(h+\bar{h})}  \frac{\langle\prod_k \Upsilon(w^-_k)\Upsilon^\dagger(w^+_k)\rangle_{\rm cyl}}{\langle\Upsilon(w^-_1)\Upsilon^\dagger(w^+_1)\rangle^n_{\rm cyl}},
\label{FU2}
\end{equation}
where $w_k^{\pm}$ are the points corresponding to $z_k^{\pm}$ through the map $w(z)$, i.e. 
\begin{equation}
 w^-_k = \frac{\pi(1+x)+2\pi (k-1)}{n},\qquad
 w^+_k = \frac{\pi(1-x)+2\pi (k-1)}{n}\,, \qquad {\rm with}\;  {k=1,...,n}.
\label{wkpm}
\end{equation}
We mention that in the literature it is possible to find also some generalisations to descendant states \cite{p-14,p-16,mck-21,bb-21} and boundary theories \cite{txas-13,top-16}.


\subsubsection{Charged moments}
The charged moments of the RDM in primary states of CFTs have been worked out in Ref. \cite{Capizzi}. 
Following this reference, it is useful to introduce a family of generating functions associated to $\ket{\Upsilon}$,
\be
p^{\Upsilon}_n(\alpha) \equiv \frac{\trace(\rho_{\Upsilon}^n e^{i\alpha Q})}{\trace(\rho_{\Upsilon}^n)}
\ee
and the universal ratio
\be
f_{n}^\Upsilon(\alpha) \equiv \frac{p^{\Upsilon}_n(\alpha) }{p^{\mathds 1}_n(\alpha) },
\label{univ_ratio}
\ee
where in both formulas we drop the $x$ dependence for simplicity. 
The moments entering in the definition of $f_{n}^\Upsilon(\alpha)$ above may all be expressed as correlation functions of $\Upsilon$ and ${\cal V}_\alpha$ 
on the $n$-sheeted Riemann surface. 
Compared to the correlations defining $F^{(n)}_\Upsilon(x)$ in Eq.~\eqref{FU} we only need to insert ${\cal V}_\alpha$ 
on an arbitrary sheet at the branch points of the Riemann surface. 
Using the same conventions of Eq. \eqref{FU} for the insertions of $\Upsilon$ and $\Upsilon^\dagger$  (located at $\{z_k^{\mp}\}$, i.e. the past/future infinite respectively 
of the $k$-th copy), we have
\begin{equation} \label{fU}
f_n(\alpha) = \frac{\displaystyle\left\langle \mathcal{V}_\alpha(u_1)\mathcal{V}_{-\alpha}(v_1)\prod_{k=1}^n \Upsilon(z_k^-) \Upsilon^\dagger(z_k^+) \right\rangle_{\mathcal{R}_n}    }{\displaystyle \langle \mathcal{V}_\alpha(u_1)\mathcal{V}_{-\alpha}(v_1) \rangle_{\mathcal{R}_n}\left\langle   \prod_{k=1}^n \Upsilon(z_k^-) \Upsilon^\dagger(z_k^+)   \right\rangle_{\mathcal{R}_n} }.
\end{equation}
Here $u_1$ and $v_1$ are the points where the flux is inserted (coinciding with the branch points), which are identified with the points $0$ and $\ell$ of the first replica.

\subsubsection{The charged moments for relative entropies and distances}

The charged moments necessary for the relative entropies and subsystem distances are in Eqs. \eqref{exp} and \eqref{exp2}.
They can all be written in terms of $\trace(e^{i\alpha Q}\rho_{1}\dots\rho_n)$ with properly chosen $\rho_i$.
The corresponding correlation functions are then the ones for the neutral moments reported in \cite{Rel_Paola,Trace_Paola} 
with the insertion of two charge operators ${\cal V}_\alpha$, as done for the charged moments 
for the entanglement entropies in Eq. \eqref{fU}. 

\begin{figure}[t]
  \includegraphics[width=\linewidth]{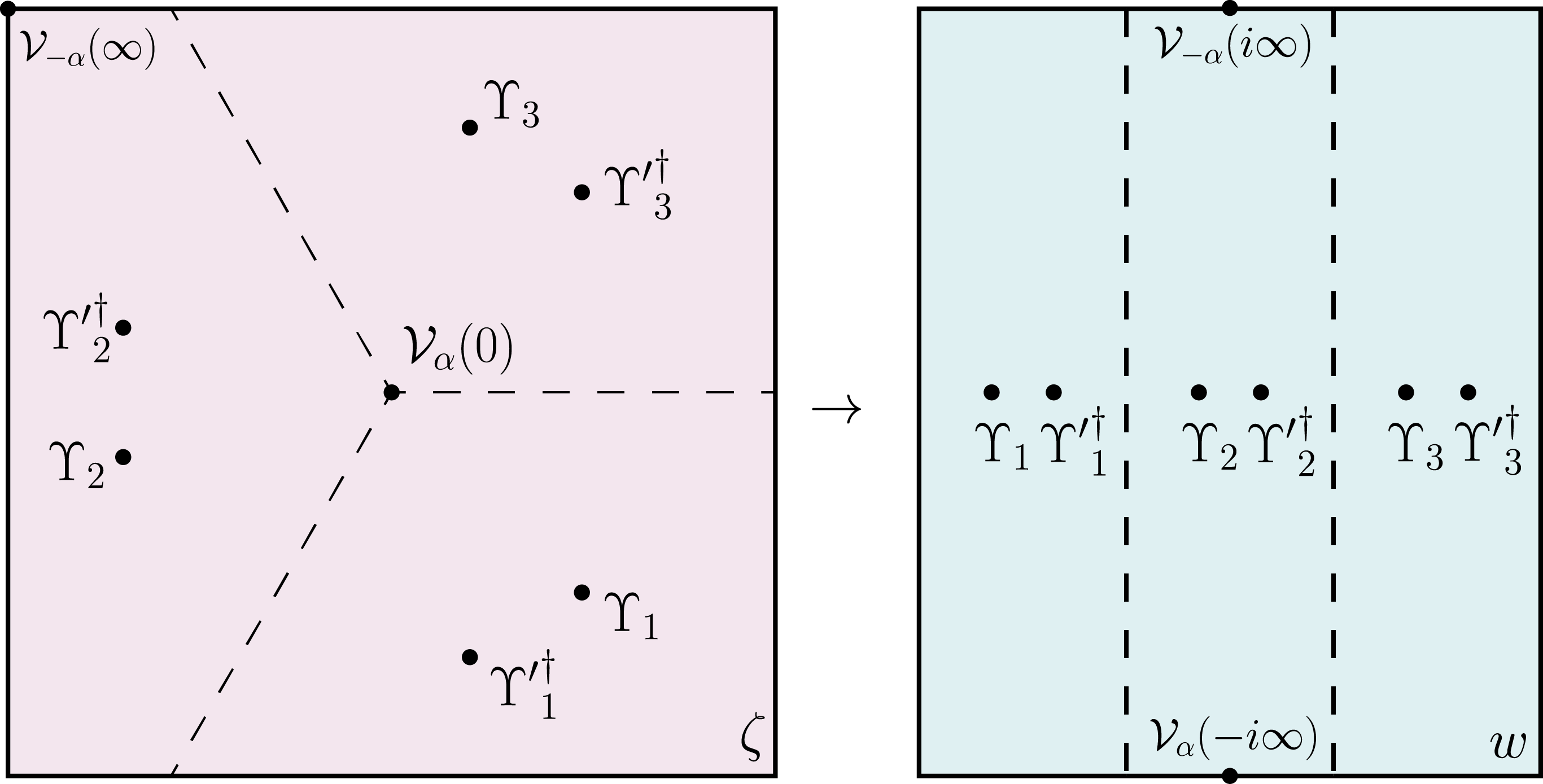}
  \caption{Points where the operators are inserted in the correlations $\left\langle {\cal V}_\alpha{\cal V}_{-\alpha}\prod_{k=1}^{n} \Upsilon_k{\Upsilon'}_k^\dagger \right\rangle$
  in Eq. \eqref{corr_ratio}. We report $n=3$ for the two geometries, planar (left) and cylindrical (right).}
\label{fig1} 
\end{figure}

Given some RDMs $\rho_j \equiv \trace_B(\ket{\Upsilon_j}\bra{{\Upsilon}_j})$, 
the charged moments of interest are conveniently parametrised as
\begin{multline}
\frac{\trace(e^{i\alpha Q}\rho_{1}\dots\rho_n)}{\trace(e^{i\alpha Q} \rho_{\mathds 1}^n)}\frac{\trace(\rho_{\mathds 1}^n)}{\trace(\rho_{1}\dots\rho_n)} =  \frac{\displaystyle\left\langle {\cal V}_\alpha(0){\cal V}_{-\alpha}(\infty)\prod_{k=1}^{n} \Upsilon_k(\zeta_k^-) \Upsilon_k^\dagger(\zeta_k^+) \right\rangle_{\mathbb{C}}    }{       \displaystyle  \langle \mathcal{V}_\alpha(0)\mathcal{V}_{-\alpha}(\infty) \rangle_{\mathbb{C}}\left\langle   \prod_{k=1}^{n} \Upsilon_k(\zeta_k^-) \Upsilon_k^\dagger(\zeta_k^+)   \right\rangle_{\mathbb{C}} } =\\
 \frac{\displaystyle\left\langle {\cal V}_\alpha(-i\infty){\cal V}_{\alpha}(i\infty)\prod_{k=1}^{n} \Upsilon_k(w_k^-) \Upsilon_k^\dagger(w_k^+) \right\rangle_{\rm cyl}    }{       \displaystyle  \langle \mathcal{V}_\alpha(-i\infty)\mathcal{V}_{\alpha}(i\infty) \rangle_{\rm cyl}\left\langle   \prod_{k=1}^{n} \Upsilon_k(w_k^-) \Upsilon_k^\dagger(w_k^+)   \right\rangle_{\rm cyl} }.
 \label{corr_ratio}
\end{multline}
The points $\zeta^{\mp}_k, w^{\mp}_k$ correspond respectively to the infinite past/future points in the $k$-th sheet of the Riemann surface ($k=1,\dots, n$). 
Their explicit expression is read off from Eq. \eqref{wkpm}, i.e.
\be
\zeta^{\mp}_k = \exp(-i\frac{2\pi (k-1)}{n} + \frac{i\pi(1\pm x)}{n}), \quad w^{\pm}_k = \frac{2\pi (k-1)}{n} + \frac{\pi(1\pm x)}{n},
\label{points}
\ee
and $x = \ell/L$. The locations of these operator insertions in the $\zeta$ and $w$ planes are reported in Fig. \ref{fig1}.

In the calculation of relative entropies and distances, we are dealing with just two (primary) fields at a time, says $\Upsilon$ and $\chi$, 
and we need to work with combinations of the form $\rho_\Upsilon^{m_1}\rho_\chi^{m_2}\rho_\Upsilon^{m_3}\dotso$.
Hence, each partition $\mathcal{S} = (m_1,\dots , m_k)$  of $n$ ($m_1+\dots +m_k=n$) is related to a product of RDMs according to the rule
\be
\mathcal{S}=(m_1,\dots , m_k) \rightarrow  \mathcal{A}_\mathcal{S} \equiv \rho_\Upsilon^{m_1}\rho_\chi^{m_2}\rho_\Upsilon^{m_3}\dots.
\ee
We define also the following quantities
\be
p^{\Upsilon, \chi}_{\mathcal{S}}(\alpha) \equiv \frac{\trace(\mathcal{A}_\mathcal{S} e^{i\alpha Q})}{\trace(\mathcal{A}_\mathcal{S})}, \qquad 
f^{\Upsilon, \chi}_{\mathcal{S}}(\alpha) \equiv \frac{p^{\Upsilon, \chi}_{\mathcal{S}}(\alpha)}{p^{\mathds 1}_{n}(\alpha)}.
\label{definition}
\ee
With a slight abuse of notation, we will refer to $p_{\mathcal{S}}(\alpha)$ as probability generating function. 
Although $p_{\mathcal{S}}(\alpha)$ is normalised  as $p_{\mathcal{S}}(\alpha=0)=1$, it is not guaranteed that $\mathcal{A}_{\mathcal{S}}$ is hermitian, 
nor that it has non-negative spectrum. However, none of these complications is a problems for our aims and we can safely define
\be
p^{\Upsilon, \chi}_{\mathcal{S}}(q) \equiv \frac{\trace(\mathcal{A}_\mathcal{S} \Pi_q)}{\trace(\mathcal{A}_\mathcal{S})} = \int_{-\pi}^{\pi} \frac{d\alpha}{2\pi}p^{\Upsilon, \chi}_{\mathcal{S}}(\alpha)e^{-i\alpha q},
\label{ppdef}
\ee
although it does not have a direct interpretation as a probability, like it happens for the entropy.
%
The function $f^{\Upsilon, \chi}_\mathcal{S}(\alpha)$ is universal and scale invariant. 

After having set up the framework for our calculation, we are already in position to make a first fundamental observation, without doing any calculation.
Indeed, since by construction, the universal functions $f^{\Upsilon, \chi}_{\mathcal{S}}(\alpha)$ are scale invariant (i.e. function only of $x=\ell/L$), they are of order one in $L$.
As a consequence, the only diverging piece in the generating function $p^{\Upsilon, \chi}_{\mathcal{S}}(\alpha)$ in Eq. \eqref{definition} comes from 
the vacuum contribution $p^{\mathds 1}_{n}(\alpha)$. 
For the latter, it is well known that the second derivative wrt $\alpha$ (i,e, the variance of the distribution) diverges as $\log L$,  while all other cumulants are finite \cite{G1,ll-93,kl-09,dk-06}. 
Hence $p^{\Upsilon, \chi}_{\mathcal{S}}(q)$ at the leading order in $L$ is always a Gaussian shaped probability 
with a variance growing like $\log L$, exactly as it happens for $p^{\Upsilon}_n(q)$ for any $\Upsilon$ \cite{Capizzi}. 
The generalised probabilities $p^{\Upsilon, \chi}_{\mathcal{S}}(q)$ and $p^{\mathds 1}_n(q)$ are 
different at order $L^0$, with an excess of variance related to the second derivative wrt $\alpha$ of $f_n^\Upsilon(\alpha)$. 
Hence, we proved that in the large $L$ limit we have
\be
\frac{p^{\Upsilon, \chi}_\mathcal{S}(q)}{p^{\mathds 1}_n(q)}\stackrel{L\to\infty}{\longrightarrow} 1.
\label{equip}
\ee
We can now rewrite the symmetry resolved relative entropies \eqref{Rel_res} in terms of the generalised probabilities as
\begin{multline}
S_n(\rho_\Upsilon \| \rho_\chi)(q) = \frac{1}{1-n}\log\frac{\trace(\rho_\Upsilon\rho_\chi^{n-1}\Pi_q)}{\trace(\rho_\Upsilon^n\Pi_q)}+ 
\log \frac{\trace(\rho_\chi^n\Pi_q)}{\trace(\rho_\Upsilon^n\Pi_q)}=\\
S_n(\rho_\Upsilon \| \rho_\chi)+\frac{1}{1-n}\log\frac{p^{\Upsilon,\chi}_{(1,n-1)}(q)}{p^\Upsilon_n(q)}+ \log \frac{p^\chi_1(q)}{p^\Upsilon_1(q)}.
\label{Svsp}
\end{multline}
Consequently, whenever \eqref{equip} holds, all the ratio of probabilities go to $1$ and
\be
S_n(\rho_\Upsilon \| \rho_{\chi})(q)\stackrel{L\to\infty}{\longrightarrow} {S_n(\rho_\Upsilon \| \rho_{\chi})}.
\label{Sreq}
\ee

Similarly, we show that 
\be
\frac{D_1(\rho_\Upsilon,\rho_\chi)(q)}{D_1(\rho_\Upsilon,\rho_\chi)} \rightarrow 1,
\label{Deq}
\ee
but in this case {the argument is slightly more involved}. First, let us write
\be
D_1(\rho_\Upsilon,\rho_\chi)(q) = \frac{1}{2}\trace\l \left|\frac{\rho_\Upsilon \Pi_q}{\trace(\rho_\Upsilon \Pi_q)} - \frac{\rho_\chi \Pi_q}{\trace(\rho_\chi \Pi_q)} \right|\r \simeq \frac{1}{2\trace(\rho_{\mathds 1} \Pi_q)}\trace\l |\rho_\Upsilon-\rho_\chi|\Pi_q\r.
\ee
Then, we express the second term through an analytical continuation over the even integers
\be
\trace\l |\rho_\Upsilon-\rho_\chi|\Pi_q\r = \underset{n_e \rightarrow 1}{\lim} \trace \l (\rho_\Upsilon-\rho_\chi)^{n_e}\Pi_q\r.
\ee
Doing so, one can expand $(\rho_\Upsilon-\rho_\chi)^{n_e}$ as a sum of products of $\rho_\Upsilon$ and $\rho_\chi$; in each term we approximate 
$\trace{{\cal A_S} \Pi_q} =(\trace{{\cal A_S}}) p^{\mathds 1}_{n_e}(q) $ so that 
$\trace\l (\rho_\Upsilon-\rho_\chi)^{n_e}\Pi_q\r \simeq \trace\l (\rho_\Upsilon-\rho_\chi)^{n_e}\r p^{\mathds 1}_{n_e}(q)$ and then perform the limit $n_e\rightarrow 1$. 
As a practical example,  we show what happens explicitly when $n_e=2$, i.e.
\begin{multline}
\trace\l (\rho_\Upsilon-\rho_\chi)^{2}\Pi_q\r = \trace\l \rho_\Upsilon^2\Pi_q\r - 2\trace(\rho_\Upsilon\rho_\chi\Pi_q) +\trace\l \rho_\chi^2\Pi_q\r =\\
 \trace\l \rho_\Upsilon^2\r \frac{\trace\l \rho_\Upsilon^2\Pi_q\r}{\trace\l \rho_\Upsilon^2\r} - 2\trace(\rho_\Upsilon\rho_\chi)\frac{\trace\l \rho_\Upsilon\rho_\chi\Pi_q\r}{\trace\l \rho_\Upsilon\rho_\chi\r} +\trace\l \rho_\chi^2\r \frac{\trace\l \rho_\chi^2\Pi_q\r }{\trace\l \rho_\chi^2\r } \simeq\\
 \trace(\rho^2_{\mathds 1}\Pi_q)\l \trace\l \rho_\Upsilon^2\r - 2\trace(\rho_\Upsilon\rho_\chi) +\trace\l \rho_\chi^2\r \r = \trace(\rho^2_{\mathds 1}\Pi_q)\trace\l (\rho_\Upsilon-\rho_\chi)^{2}\r,
\end{multline}
where in the last line we used Eq. \eqref{equip}.

Eqs. \eqref{Sreq} and \eqref{Deq} represent a first main result of this paper: exactly like the entanglement of a single state \cite{Sierra3}, 
also the subsystem measures of distinguishability (relative entropies and
distance) satisfy {\it equipartition}, i.e. do not depend on the symmetry sector $q$\footnote{This is a property of the thermodynamic limit in which $q$ is kept fixed. When $L$ is finite, equipartition is expected to hold only if $q$ is much smaller than the typical fluctuation scale of order $\sqrt{\log L}$}.

\section{Correlation functions for the compact boson}\label{CBos}

In this section, we provide some explicit expressions for the universal functions of the correlation functions necessary for relative entropies and subsystems distances, 
generically given by Eq. \eqref{corr_ratio}, specialising to the massless compact boson. 
In some cases, for generic $n$, we are only able to work out analytically  the short distance expansion via the OPE of composite twist fields.
In the following two sections, we are going to explicitly use these results to give predictions for entanglement measures. 

The CFT of the compact boson (or Luttinger liquid)  is described by the euclidean action \cite{pagialle}
\be
S[\varphi] = \frac{1}{8\pi K}\int d^2x(\partial_\mu \varphi)^2 ,
\ee
with the additional requirement that the bosonic field is compact
\be
\varphi \sim \varphi + 2\pi.
\ee
This CFT has central charge $c=1$. 
Left and right modes are decoupled,  so one can write in complex coordinates
\be
\varphi(z,\bar{z}) = \phi(z)+\bar{\phi}(\bar{z}).
\ee
This theory admits a topological $U(1)$ symmetry generated by the following vertex operator
\be
\mathcal{V}_{\alpha}(z,\bar{z}) = e^{i\frac{\alpha}{2\pi}\phi(z)+i\frac{\alpha}{2\pi}\bar{\phi}(\bar{z})}.
\ee
The primaries of this CFT and their conformal weights $(h,\bar{h})$ are respectivly 
\be
(i\partial \phi)(z) \quad (1,0), \qquad (i\bar{\partial} \bar{\phi})(\bar{z}) \quad (0,1), \qquad V_{\beta,\bar{\beta}}(z,\bar{z})\equiv e^{i\beta \phi(z)+i\bar{\beta}\bar{\phi}(\bar{z})} \quad \l \frac{K\beta^2}{2},\frac{K\bar{\beta}^2}{2}\r.
\ee
Not all the values of $(\beta,\bar{\beta})$ give rise to physical states, but the set of the allowed values is quantised (see \cite{pagialle}); 
however, this discussion is not important for our purposes. Without loss of generality,  we will deal only with the holomorphic part 
of the vertex operator ($\bar{\beta}=0$), keeping $\beta$ as a free parameter.
Moreover, in what follows we will fix $K=1$ (the results for $K\neq 1$ can be easily obtained as mentioned in Appendix \ref{AppCorr}). 
This value of $K$ is related to a free Dirac fermion via bosonization, corresponding to an XX spin chain
which we will use to numerically test the analytic predictions obtained in the following. In that case the symmetry is the internal $U(1)$ charge of a Dirac fermion i.e number of fermions minus number of antifermions. The explicit correspondence between microscopic low energy excitations of the XX chain and the primary operators of the compact boson, 
via bosonisation techniques, has been discussed in the work by Alcaraz et al. \cite{Sierra1}.

Let us briefly recall the OPE among primaries \cite{pagialle}, which can be obtained by their 3-point functions, see Appendix \ref{AppCorr}. The following fusions are present
\be
[V_\beta] \times [V_{-\beta}] \rightarrow [1]+ [i\partial \phi],  \qquad [i\partial \phi] \times [i\partial \phi] \rightarrow [1].
\ee
The only nontrivial (the others are $1$) OPE coefficient is
\be
C^{i\partial \phi}_{V_\beta V_{-\beta}} = \beta.
\ee
We will also use the OPE coefficient associated to the generation of the stress-energy tensor $T = \frac{1}{2}(i\partial \phi)^2$ 
($T = \frac{1}{2K}(i\partial \phi)^2$ for $K\neq 1$),  fixed by Virasoro algebra (see \cite{pagialle}) as
\be
\qquad C^{T}_{i\partial \phi i\partial \phi} = 2, \qquad \qquad C^{T}_{V_\beta V_{-\beta}} = \beta^2.
\ee

For the charged twist fields,  we will focus on the fusion channels\footnote{To be precise, we are considering the generation of the following operators in the orbifold theory: $1\otimes \dots \otimes 1$, $1\otimes \dots i \partial\phi \dots \otimes 1$, $1\otimes \dots T \dots \otimes 1$, where $i \partial \phi$ and $T$ are inserted in any of the $n$ replicas.}
\be
\mathcal{T}_{n,\alpha} \times \l\mathcal{T}_{n,\alpha}\r^\dagger \rightarrow 1,i\partial \phi,T.
\ee 
Although the vertex operators are generated in the OPE, their expectation value is zero for the states we consider (by neutrality condition), 
and thus they do not contribute to $f^{\Upsilon, \chi}_{\mathcal{S}}(\alpha)$.

In the forthcoming subsections, we will characterise $f^{\Upsilon, \chi}_{\mathcal{S}}(\alpha)$ for different states, 
giving the exact results when possible or the leading order, obtained by OPE expansion, for $x=\frac{\ell}{L} \rightarrow 0$ in the other cases.

\subsection{Universal function for the pair of states $\Upsilon = V_{\beta_1}$ and $\chi = V_{\beta_2}$}
\label{fn_vertex}

Let us start from the states being both vertex operators with weight $\beta_1$ and $\beta_2$, i.e. $\Upsilon = V_{\beta_1}$ and $\chi = V_{\beta_2}$.
We first consider the universal function $f^{V_{\beta_1}, V_{\beta_2}}_{\mathcal{S}}(\alpha)$ in Eq.  \eqref{univ_ratio}
for the partition $\mathcal{S} = (m_1,m_2)$ given as
\begin{multline}
f^{V_{\beta_1},V_{\beta_2}}_\mathcal{S}(\alpha) =  \frac{\displaystyle\left\langle {V}_{\alpha/2\pi}(-i\infty){V}_{-\alpha/2\pi}(i\infty)\prod_{k=1}^{m_1} V_{\beta_1}(w_k^-) V_{-\beta_1}(w_k^+) \prod_{k=m_1+1}^{m_2} V_{\beta_2}(w_k^-) V_{-\beta_2}(w_k^+)\right\rangle_{\rm cyl}    }{       \displaystyle  \langle V_{\alpha/2\pi}(-i\infty)V_{-\alpha/2\pi}(i\infty) \rangle_{\rm cyl}\left\langle   \prod_{k=1}^{m_1} V_{\beta_1}(w_k^-) V_{-\beta_1}(w_k^+) \prod_{k=m_1+1}^{m_2} V_{\beta_2}(w_k^-) V_{-\beta_2}(w_k^+)  \right\rangle_{\rm cyl} } =\\
\prod_{k=1}^{m_1} \la {V}_{\alpha/2\pi}(-i\infty) V_{\beta_1}(w_k^-)\ra_{\rm cyl}\la {V}_{-\alpha/2\pi}(i\infty) V_{\beta_1}(w_k^-)\ra_{\rm cyl} \\
 \la {V}_{\alpha/2\pi}(-i\infty) V_{-\beta_1}(w_k^+)\ra_{\rm cyl} \la{V}_{-\alpha/2\pi}(i\infty) V_{-\beta_1}(w_k^+)\ra_{\rm cyl}\\
\prod_{k=m_1+1}^{m_2} \la {V}_{\alpha/2\pi}(-i\infty) V_{\beta_2}(w_k^-)\ra_{\rm cyl}\la {V}_{-\alpha/2\pi}(i\infty) V_{\beta_2}(w_k^-)\ra_{\rm cyl}\\   \la {V}_{\alpha/2\pi}(-i\infty) V_{-\beta_2}(w_k^+)\ra_{\rm cyl} \la {V}_{-\alpha/2\pi}(i\infty) V_{-\beta_2}(w_k^+)\ra_{\rm cyl}.
\label{VV1}
\end{multline}
%
Here we used the correlation function between vertex operators on the cylinder 
\be
\la \prod_j V_{\beta_j}(w_j)\ra_{\rm cyl} = \prod_{i<j} \l \frac{L}{\pi} \sin \frac{\pi (w_i-w_j)}{L}\r^{\beta_i\beta_j}, \quad \text{if} \quad \sum_{j}\beta_j=0,
\ee
while it vanishes if $\quad \sum_{j}\beta_j\neq 0$: the requirement $\sum_{j}\beta_j=0$ is the neutrality condition. 
All the correlation functions appearing in the first line of  Eq. \eqref{VV1} satisfy the neutrality condition and so the previous formula can be safely applied. 
With a slight abuse of notation, we identify (for notational convenience) the putative two-point correlators in (the rhs of) Eq. \eqref{VV1} as
\be
\la V_{\beta_i}(w_i)V_{\beta_j}(w_j)\ra_{\rm cyl} = \l \frac{L}{\pi} \sin  \frac{\pi(w_i-w_j)}{L}\r^{\beta_i\beta_j}.
\ee

Eq. \eqref{VV1} has to be regularised due to the insertion of the vertex operators at infinity. 
One way to do so is through their insertion at $\pm i\Lambda$ and, only at the end, take the limit $\Lambda\rightarrow +\infty$. 
Doing so, with  calculations similar to those in  Ref. \cite{Capizzi}, one straightforwardly gets
\begin{multline}
\la {V}_{\alpha/2\pi}(-i\infty) V_{\beta}(w_k^-)\ra_{\rm cyl}\la {V}_{-\alpha/2\pi}(i\infty) V_{\beta}(w_k^-)\ra_{\rm cyl}\\
 \la {V}_{\alpha/2\pi}(-i\infty) V_{-\beta}(w_k^+)\ra_{\rm cyl} \la{V}_{-\alpha/2\pi}(i\infty) V_{-\beta}(w_k^+)\ra_{\rm cyl} =e^{i\frac{\alpha \beta x}{n}}.
\end{multline}
Taking the product over the different $k$'s we finally obtain the extremely simple form
$f^{V_{\beta_1},V_{\beta_2}}_\mathcal{S}(\alpha) = e^{\frac{i\alpha x}{n}[m_1\beta_1+m_2\beta_2]}$.
Clearly, because of the factorisation property, if we would have chosen a different partition ${\cal S}$, i.e. a different order of the replicas $V_{\beta_1},V_{\beta_2}$
we would have get the same result, i.e. 
\be
f^{V_{\beta_1},V_{\beta_2}}_\mathcal{S}(\alpha) = e^{\frac{i\alpha x}{n}[m_1\beta_1+m_2\beta_2]}, \qquad \forall {\cal S}.
\label{VVCFT}
\ee

This striking simple result implies that $p^{V_{\beta_1},V_{\beta_2}}_{\mathcal{S}}(q)$ is simply obtained by a shift of the average value of the charge, while the other cumulants are not affected by the insertion of the vertex operators. The same conclusion was already pointed out in \cite{Capizzi} for the symmetry-resolved Rényi entropy of these states.

The $O(x)$ term of $f^{V_{\beta_1},V_{\beta_2}}_\mathcal{S}(\alpha)$, i.e. 
\be
f^{V_{\beta_1},V_{\beta_2}}_\mathcal{S}(\alpha) \simeq 1+\frac{i\alpha x}{n}[m_1\beta_1+m_2\beta_2] + O(x^2),
\ee
should be also interpreted in terms of the OPE expansion of modified twist fields. 
To show that explicitly, let us express $f^{V_{\beta_1},V_{\beta_2}}_\mathcal{S}(\alpha)$ as a charged twist field correlation:
\be
f^{V_{\beta_1},V_{\beta_2}}_\mathcal{S}(\alpha) = \frac{\bra{V_{\beta_1},\cdots ,V_{\beta_2},\cdots}\mathcal{T}_{n,\alpha}(0)\tilde{\mathcal{T}}_{n,\alpha}(\ell)\ket{V_{\beta_1},\cdots, V_{\beta_2},\cdots}}{\bra{V_{\beta_1},\cdots ,V_{\beta_2},\cdots}\mathcal{T}_n(0)\tilde{\mathcal{T}}_n(\ell)\ket{V_{\beta_1},\cdots, V_{\beta_2},\cdots}}\frac{ \bra{0,\dots,0}\mathcal{T}_n(0)\tilde{\mathcal{T}}_n(\ell)\ket{0,\cdots,0} }{ \bra{0,\dots,0}\mathcal{T}_{n,\alpha}(0)\tilde{\mathcal{T}}_{n,\alpha}(\ell)\ket{0,\cdots,0}}.
\ee
Restricting our analysis to order $O(x)$ and focusing on the terms of the OPE with non vanishing expectation value, we can approximate
\bea
\mathcal{T}_n(0)\tilde{\mathcal{T}}_n(\ell) &\simeq& \la \mathcal{T}_n(0)\tilde{\mathcal{T}}_n(\ell) \ra_{\mathbb{C}}(1+o\l \ell \r),\\
\mathcal{T}^\alpha_n(0)\tilde{\mathcal{T}}^\alpha_n(\ell) &\simeq& 
\la \mathcal{T}_n(0)\tilde{\mathcal{T}}_n(\ell) \ra_{\mathbb{C}} \Big( 1+\frac{\ell}{n}C^{i\partial\phi}_{V_{\alpha/2\pi}V_{-\alpha/2\pi}}\sum_j (i\partial \phi)^j(0)+o(\ell) \Big).
\eea
The expectation value of $\sum_j (i\partial \phi)^j(0)$ is the sum of the contributions of each single replica, namely
\bea
\bra{0,\dots,0} \sum_j (i\partial \phi)^j(0) \ket{0,\dots,0} &=&0,\\
 \bra{V_{\beta_1},\cdots ,V_{\beta_2},\cdots} \sum_j (i\partial \phi)^j(0)  \ket{V_{\beta_1},\cdots ,V_{\beta_2},\cdots} &=& \frac{i2\pi}{L}\l m_1\beta_1+m_2\beta_2\r.
\eea
Putting all the pieces together
\begin{multline}
\frac{\bra{V_{\beta_1},\cdots ,V_{\beta_2},\cdots}\mathcal{T}_{n,\alpha}(0)\tilde{\mathcal{T}}_{n,\alpha}(\ell)\ket{V_{\beta_1},\cdots, V_{\beta_2},\cdots}}{\bra{V_{\beta_1},\cdots ,V_{\beta_2},\cdots}\mathcal{T}_n(0)\tilde{\mathcal{T}}_n(\ell)\ket{V_{\beta_1},\cdots, V_{\beta_2},\cdots}}\frac{ \bra{0,\dots,0}\mathcal{T}_n(0)\tilde{\mathcal{T}}_n(\ell)\ket{0,\cdots,0} }{ \bra{0,\dots,0}\mathcal{T}_{n,\alpha}(0)\tilde{\mathcal{T}}_{n,\alpha}(\ell)\ket{0,\cdots,0}} \simeq\\
\l 1+\frac{\ell}{n}\frac{\alpha}{2\pi} \frac{i2\pi}{L}\l m_1\beta_1+m_2\beta_2\r \r = \l 1+i\frac{\alpha x}{n}\l m_1\beta_1+m_2\beta_2\r\r.
\end{multline}

\begin{figure}[t]
	\centering
	\includegraphics[width=0.5\linewidth]{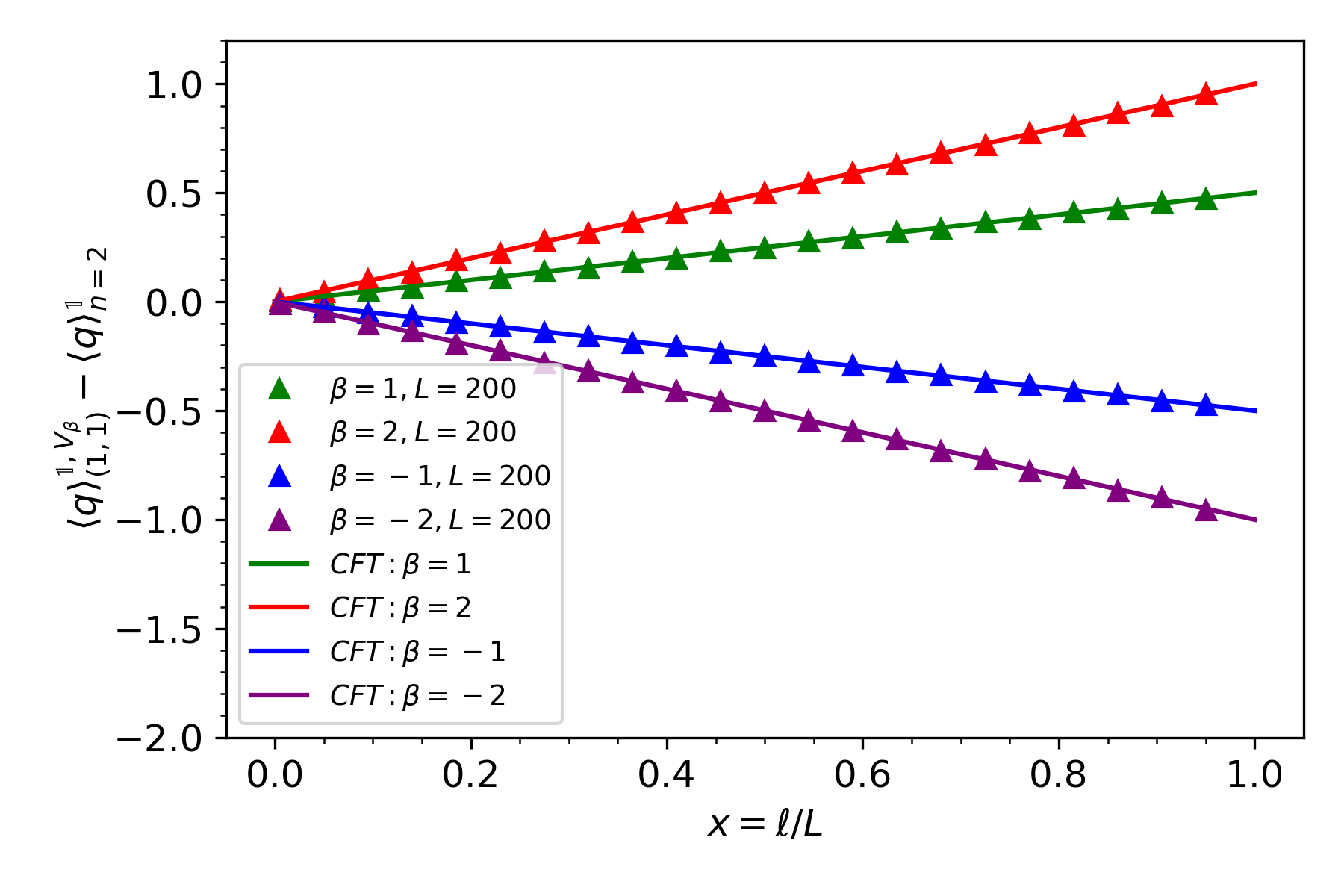}
	\caption{Excess of the average charge $\la q\ra^{{\mathds 1},V_\beta}_{(1,1)} - \la q\ra^{\mathds 1}_{2}$ 
	as a function of $x=\ell/L$ for different values of $\beta$ ($\beta = -2,-1,1,2$). 
	The universal CFT results, which are linear functions of $x$, cf. Eq. \eqref{exc1} are tested against exact numerical result for the XX chain at half-filling ($L=200$). 
	}
\label{fig2} 
\end{figure}

\begin{figure}[t]
\centering
\includegraphics[width=0.484\textwidth]{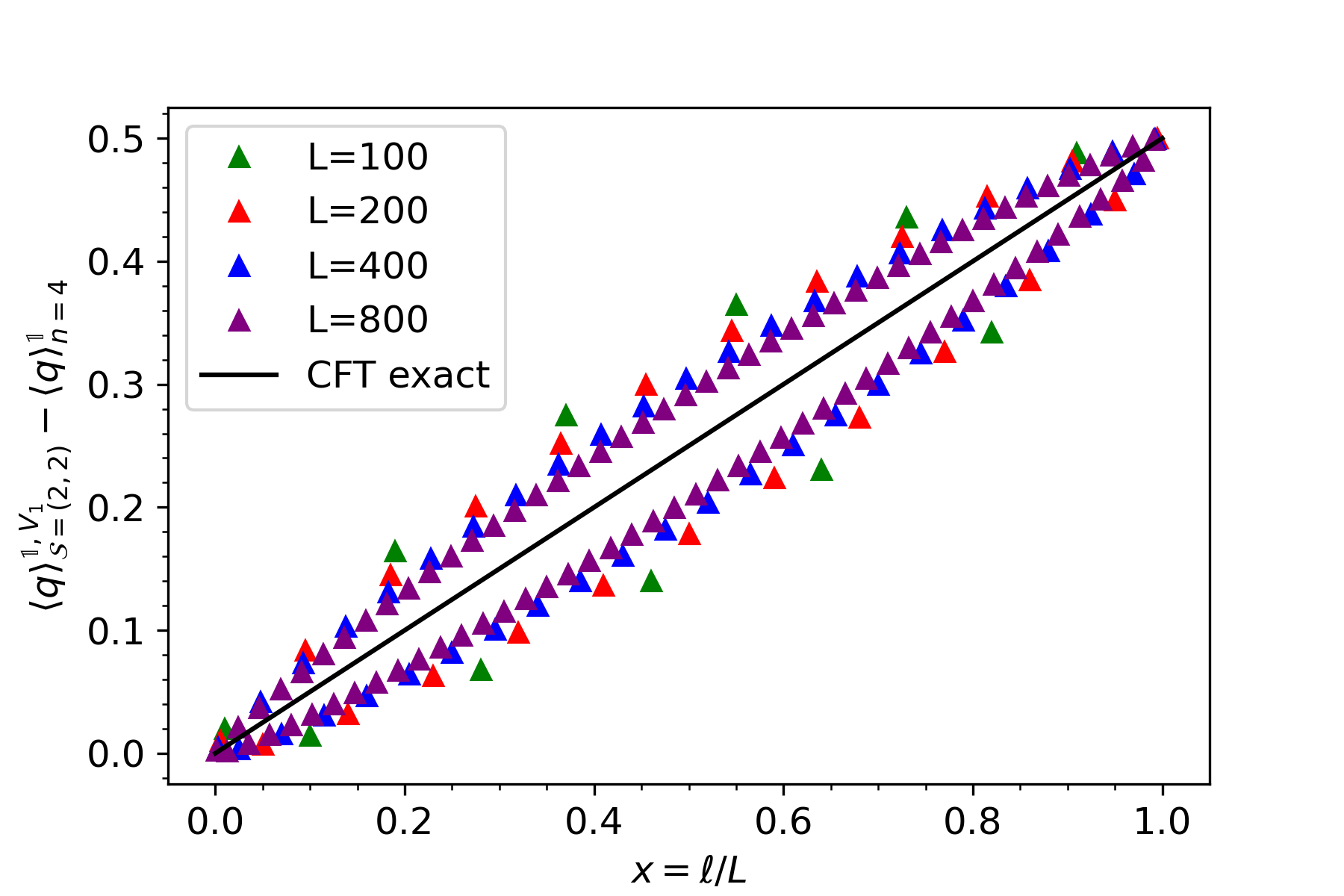}
\includegraphics[width=0.484\textwidth]{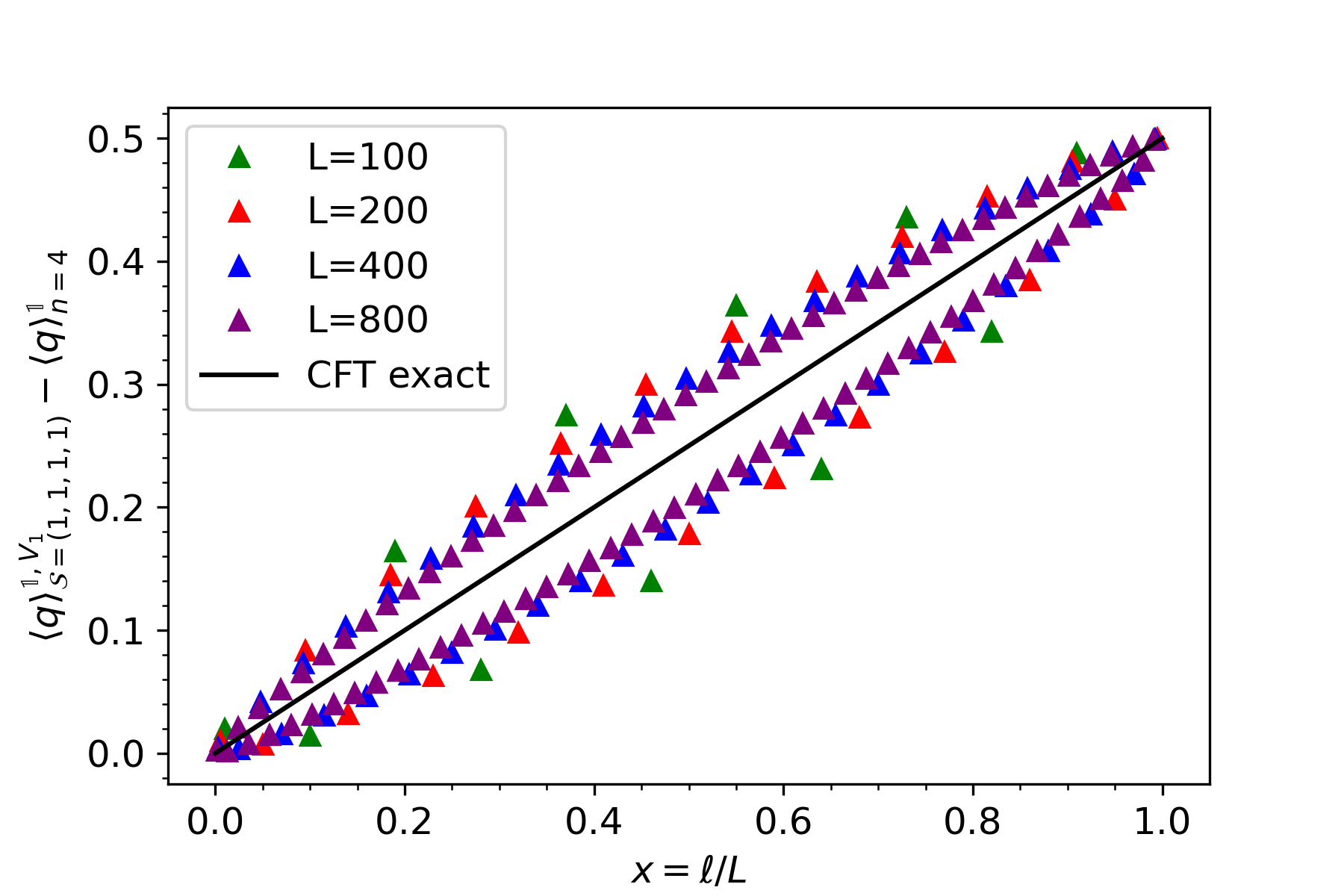}
\caption{Excess of average charge $\la q\ra^{{\mathds 1},V_1}_{\mathcal{S}} - \la q\ra^{\mathds 1}_{4}$ for the partitions $\mathcal{S} = (2,2)$ and $\mathcal{S} = (1,1,1,1)$ (left/right panel respectively) as a function of $x=\ell/L$. The CFT predictions are tested against numerical data for the XX chain for different system sizes ($L = 100,200,400,800$). The order of insertion of the operator is different in the two cases, indeed $V_1$ is present in the first and second replica for $\mathcal{S} = (2,2)$ while for $\mathcal{S} = (1,1,1,1)$ it is inserted in the second and fourth replica; nevertheless, the analytical prediction is the same which is a special feature of the vertex states.
}
\label{fig3} 
\end{figure}

In Figs. \ref{fig2} and \ref{fig3} we test the CFT prediction \eqref{VVCFT} against numerics for the XX chain (obtained with the methods of Appendix \ref{App2}). 
We focus on the excess of average charge
\be
\la q \ra^{V_{\beta_1},V_{\beta_2}}_{\mathcal{S}} - \la q \ra^{\mathds 1}_{n} \equiv \frac{1}{i}\frac{d}{d\alpha} f^{V_{\beta_1},V_{\beta_2}}_{\mathcal{S}}(\alpha)\Big|_{\alpha=0} 
= \frac{m_1 \beta_1 + m_2\beta_2}{n}x,
\label{exc1}
\ee
and plot it as a functions of $x = \ell/L$. As shown in Fig. \ref{fig2}, for $n=2$ and $\mathcal{S}=(1,1)$ the agreement with numerical data (system size $L=200$) 
is remarkable for different vertex states and no significant corrections are visible for this relatively small system size. 
In Fig. \ref{fig3} we consider instead $n=4$ considering the partitions $\mathcal{S} = (2,2)$ and $\mathcal{S}=(1,1,1,1)$.
We emphasise that now there are evident deviations of the numerics from the CFT predictions with the 
numerical data oscillating around the analytical value with an amplitude of going to zero as the system size increases. 
This behaviour is expected from the exact analysis of the symmetry-resolved R\'enyi entropies of the XX chain performed in \cite{Ric1}, 
where the deviations from CFT are more severe as the number of replicas $n$ is increased (and it is a consequence of the presence 
of well known unusual corrections to the scaling \cite{cc-10}). 

\subsection{Universal function for the pair of states $\Upsilon = i\partial \phi$ and  $\chi = 1$}
\label{Der_GS}

Here we consider the insertion of $\Upsilon= i\partial \phi$ while the other state is the ground state, i.e. $\chi = 1$.
In Ref. \cite{Capizzi}, we showed that $f^{i\partial \phi}_n(\alpha)$ can be expressed as a characteristic polynomial of a certain matrix. 
The argument of \cite{Capizzi} was based on the fact that
\be
\frac{\displaystyle\left\langle {V}_{\alpha/2\pi}(-i\infty){V}_{-\alpha/2\pi}(i\infty)\prod_{k=1}^{n} (i\partial \phi)(w_k^-) (i\partial \phi)(w_k^+) \right\rangle_{\rm cyl}    }{       \displaystyle  \langle V_{\alpha/2\pi}(-i\infty)V_{-\alpha/2\pi}(i\infty) \rangle_{\rm cyl}} 
\ee
has a certain diagrammatic expansion (see the Appendix \ref{AppCorr}) which can be recast in a clever way. 
For instance, the order $O(\alpha^0)$ is given by the contractions of the derivative operators among themselves which can be expressed as a determinant using Wick theorem. 
At order $O(\alpha^2)$ two derivative operators are contracted with ${V}_{\pm\alpha/2\pi}$, while the remaining $2(n-1)$ ones are contracted among themselves, and so on.

The same argument can be applied to $f^{i\partial \phi,{\mathds 1}}_\mathcal{S}$.  
The only difference is the explicit form of the resulting matrix for the characteristic polynomial. 
For instance, for any partition $\mathcal{S}$ where $\rho_{i\partial\phi}$ appears $m_{i\partial \phi}$ times in the product $\mathcal{A}_\mathcal{S}$, 
we construct an antisymmetric matrix $M$ of dimension $2m_{i\partial \phi} \times 2m_{i\partial \phi}$ with elements
\be
M_{ij} \equiv \begin{cases} \frac{1}{2\sin(\frac{w_i-w_j}{2})} &\quad i\neq j, \\ 0 &\quad i=j,\end{cases}
\ee
with $\{w_i\}$ being the set of points in which $i\partial \phi$ is inserted in the cylindrical geometry, cf. Eq. \eqref{points}. 
In terms of $M$, $f^{i\partial \phi,{\mathds 1}}_\mathcal{S}(\alpha)$ is expressed as follows
\be
f^{i\partial \phi,{\mathds 1}}_\mathcal{S}(\alpha) = \frac{\det \left(M \pm \frac{i\alpha}{2\pi}\right)}{\det(M)}.
\label{fder}
\ee
Eq.  \eqref{fder} provides and analytic expression of the function $f^{i\partial \phi,{\mathds 1}}_\mathcal{S}(\alpha)$  for any integer $n$ and for any partition $\mathcal{S}$.
However, its form becomes more and more cumbersome as more and more derivative operators are inserted. 
As an example, let us see what happens in the simplest case, namely $f^{i\partial \phi,{\mathds 1}}_{(1,n-1)}(\alpha)$, so that the matrix $M$ is
\be
M = \begin{pmatrix} 0 & \frac{1}{2 \sin \frac{\pi x}{n}} \\ -\frac{1}{2 \sin \frac{\pi x}{n}} & 0\end{pmatrix},
\qquad \Rightarrow\qquad
\det\l M + i \frac{\alpha}{2\pi}\r = \frac{1}{4 \sin^2 \frac{\pi x}{n}} - \l \frac{\alpha}{2\pi}\r^2.
\ee
Plugging this expression  in Eq. \eqref{fder}, we get
\be
f^{i\partial \phi,{\mathds 1}}_{(1,n-1)}(\alpha) = 1-\frac{\alpha^2}{\pi^2}\sin^2 \frac{\pi x}{n}. 
\label{VICFT}
\ee
Similarly, for $f^{i\partial \phi,{\mathds 1}}_{(2,n-2)}(\alpha)$ the matrix $M$ is
\be
M = \begin{pmatrix} 0 & \frac{1}{2\sin\l \frac{\pi x}{n} \r} & \frac{1}{2\sin\l \frac{\pi }{n} \r} & \frac{1}{2\sin\l \frac{\pi x}{n} + \frac{\pi}{n}\r} \\  -\frac{1}{2\sin\l \frac{\pi x}{n} \r}  & 0 & \frac{1}{2 \sin \l  \frac{\pi}{n} - \frac{\pi x}{n}\r } & \frac{1}{2\sin\l \frac{\pi }{n} \r }\\
-\frac{1}{2\sin\l \frac{\pi }{n} \r}  & -\frac{1}{2 \sin \l  \frac{\pi}{n} - \frac{\pi x}{n}\r} & 0 & \frac{1}{2\sin\l \frac{\pi x}{n} \r} \\
-\frac{1}{2\sin\l \frac{\pi x}{n} + \frac{\pi}{n}\r} & -\frac{1}{2\sin\l \frac{\pi}{n}\r} & -\frac{1}{2 \sin \l\frac{\pi x}{n}\r} & 0
\end{pmatrix},
\ee
so
\begin{multline}
f^{i\partial \phi,{\mathds 1}}_{(2,n-2)}(\alpha) = 1-\frac{2 \csc ^2\left(\frac{\pi  x}{n}\right)+\csc ^2\left(\frac{\pi  x}{n}+\frac{\pi }{n}\right)+\csc ^2\left(\frac{\pi }{n}-\frac{\pi  x}{n}\right)+2 \csc ^2\left(\frac{\pi }{n}\right)}{\pi ^2 \left(-\csc ^2\left(\frac{\pi  x}{n}\right)-\csc \left(\frac{\pi }{n}-\frac{\pi  x}{n}\right) \csc \left(\frac{\pi  x}{n}+\frac{\pi }{n}\right)+\csc ^2\left(\frac{\pi }{n}\right)\right)^2}\alpha^2+\\
\frac{1}{\pi ^4 \left(-\csc ^2\left(\frac{\pi  x}{n}\right)-\csc \left(\frac{\pi }{n}-\frac{\pi  x}{n}\right) \csc \left(\frac{\pi  x}{n}+\frac{\pi }{n}\right)+\csc ^2\left(\frac{\pi }{n}\right)\right)^2}\alpha^4.
\end{multline}
It is then clear that, for any specific partition ${\cal S}$, it is possible to write down  $f^{i\partial \phi,{\mathds 1}}_{\cal S}(\alpha)$, but a closed form is very likely impossible to write.
The other extreme that can be analytically handled is the case $\mathcal{S} = (n,0)$, when $f^{i\partial \phi,{\mathds 1}}_\mathcal{S} = f_n^{i\partial \phi}(\alpha)$ 
and its explicit expression, analytical continued to non integer values of $n$, is known \cite{Capizzi} and reads
\begin{multline}
f^{i\partial \phi}_n(\alpha) = \prod_{p=1}^n \left(   1-\left(  \frac{\alpha}{\pi} \right)^2 \frac{1}{(\frac{n}{\sin(\pi x)} -n-1+2p)^2}        \right) =\\
\left(\frac{\Gamma(1+n + \frac{1}{2}(\frac{n}{\sin(\pi x)} -n-1) +\frac{\alpha}{2\pi})}{\Gamma(1+ \frac{1}{2}(\frac{n}{\sin(\pi x)} -n-1)+\frac{\alpha}{2\pi})} \right) \\ 
 \left(\frac{\Gamma(1+n + \frac{1}{2}(\frac{n}{\sin(\pi x)} -n-1) -\frac{\alpha}{2\pi})}{\Gamma(1+ \frac{1}{2}(\frac{n}{\sin(\pi x)} -n-1)-\frac{\alpha}{2\pi})} \right) 
  \left(\frac{\Gamma(1+ \frac{1}{2}(\frac{n}{\sin(\pi x)} -n-1))}{\Gamma(1+n + \frac{1}{2}(\frac{n}{\sin(\pi x)} -n-1))} \right)^2.
\label{Capizzif_n}
  \end{multline}

Instead, the small $x$ behaviour of $f^{i\partial \phi,{\mathds 1}}_\mathcal{S}(\alpha)$ for a general partition $\mathcal{S}$ can be obtained analytically via OPE.   
Let us start with the partition $\mathcal{S} = (m_{i\partial\phi},n-m_{i\partial\phi})$. 
The function $f^{i\partial \phi,{\mathds 1}}_{\mathcal{S}}$ is written in terms of twist fields as
\be
f^{i\partial \phi,{\mathds 1}}_{\mathcal{S}}(\alpha) = \frac{\bra{i\partial\phi,\cdots,0,\cdots}\mathcal{T}_{n,\alpha}(0)\tilde{\mathcal{T}}_{n,\alpha}(\ell)\ket{i\partial\phi,\cdots,0,\cdots}}{\bra{i\partial\phi,\cdots,0,\cdots}\mathcal{T}_n(0)\tilde{\mathcal{T}}_n(\ell)\ket{i\partial\phi,\cdots,0,\cdots}}\frac{ \bra{0,\dots,0}\mathcal{T}_n(0)\tilde{\mathcal{T}}_n(\ell)\ket{0,\cdots,0} }{ \bra{0,\dots,0}\mathcal{T}_{n,\alpha}(0)\tilde{\mathcal{T}}_{n,\alpha}(\ell)\ket{0,\cdots,0}} ,
\ee
where $\ket{i\partial\phi,\dots , 0,\dots}$ stands for the state where $i\partial \phi$ appears $m_{i\partial\phi}$ times in the first replicas. 
The non-vanishing terms at order $O(x^2)$ come from the expectation value of the stress energy-tensor. Thus we keep the following terms in the OPE of twist fields
\be
\mathcal{T}_n(0)\tilde{\mathcal{T}}_n(\ell) = \la \mathcal{T}_n(0)\tilde{\mathcal{T}}_n(\ell) \ra \l 1 + \ell^2 \frac{2h_{\mathcal{T}_{n,\alpha}}}{n}\sum_j T^j(0) + o(\ell^2)\r,
\ee
where we used that the central charge of this model is $c=1$. Putting the pieces together, we get
\be
f^{i\partial \phi,{\mathds 1}}_{\mathcal{S}}(\alpha)  \simeq 1+\ell^2\frac{2h_{\mathcal{V}_\alpha}}{n^2}\sum_j \l \bra{i\partial\phi,\cdots,0,\cdots}T^j(0)\ket{i\partial\phi,\cdots,0,\cdots}- \bra{0,\dots,0}T^j(0)\ket{0,\dots,0}\r.
\label{ope22}
\ee
Using
\be
h_{\mathcal{V}_\alpha} = h_{V_{\alpha/2\pi}} = \frac{1}{2}\l\frac{\alpha}{2\pi}\r^2
\ee
and (see e.g. \cite{Zhang2})
\be
\sum_j \bra{i\partial\phi,\cdots,0,\cdots}T^j(0)\ket{i\partial\phi,\cdots,0,\cdots}- \bra{0,\dots,0}T^j(0)\ket{0,\dots,0} = -\frac{4\pi^2}{L^2}m_{i\partial\phi},
\ee
one finally obtains the desired result
\be
f^{i\partial \phi,{\mathds 1}}_\mathcal{S}(\alpha) \simeq 1-\frac{m_{i\partial \phi}}{n^2}x^2\alpha^2.
\label{fderOPE}
\ee
 As a consistency check, for $m_{i\partial \phi}=1$, Eq. \eqref{Capizzif_n} reduces to
\be
f_n^{i\partial \phi}(\alpha) \simeq 1-\frac{\alpha^2x^2}{n}.
\ee
This simple result has an interesting physical meaning: 
in the small $x$ regime, the replicas appear to be decoupled at order $O(x^2)$ the contributions of $i\partial \phi$ in the various replicas just sum up 
(it is a consequence of the additivity of the stress-energy tensor).
This is not at all the case at higher order in $x$.

\begin{figure}[t]
	\centering
	\includegraphics[width=0.5\linewidth]{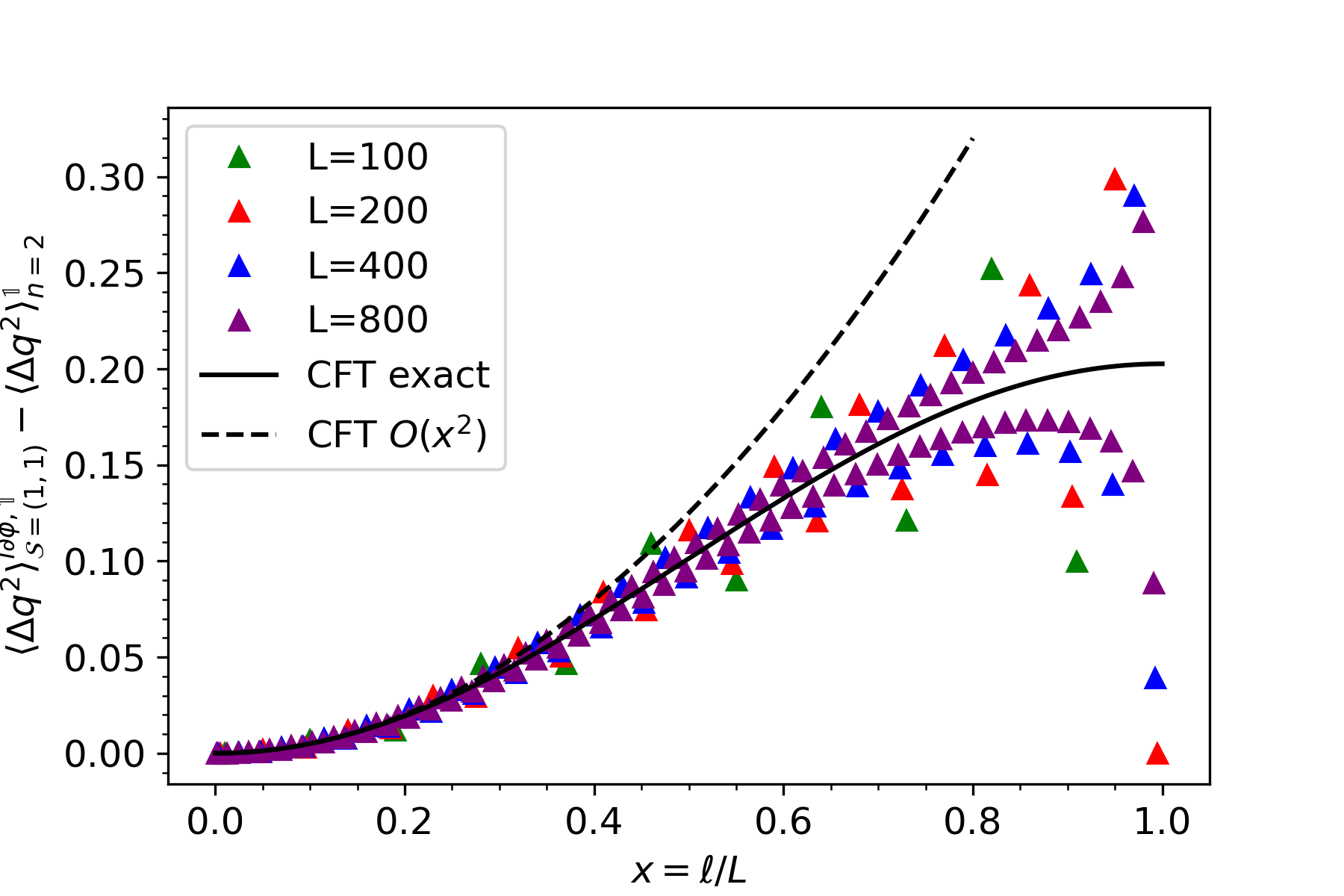}
	\caption{Excess of variance associated to the partition $\mathcal{S}=(1,1)$ of the operators $(\Upsilon,\chi)=(i\partial\phi,{\mathds 1})$. 
	The universal CFT results are compared to the XX chain at half-filling. 
	The numerical data for different sizes ($L=100,200,400,800$) are compared with both the full CFT result (solid line) and the small 
	$x$ expansion at order $O(x^2)$ (dashed line).}
\label{fig4}
\end{figure}

We now test how the CFT results in this subsection match with the numerical data for the XX chain. 
We start from $f^{i\partial\phi,1}_{(1,n-1)}(\alpha)$ in Eq. \eqref{VICFT}. In Fig. \ref{fig4} we plot the CFT result of excess of variance
\be
\la \Delta q^2\ra^{i\partial\phi,{\mathds 1}}_{(1,1)}- \la \Delta q^2\ra^{i\partial\phi,{\mathds 1}}_{2} 
= \frac{1}{(i)^2}\frac{d^2}{d\alpha^2}\log f^{i\partial\phi,{\mathds 1}}_{(1,1)}(\alpha)\Big|_{\alpha=0} = \frac{2}{\pi^2}\sin^2 \frac{\pi x}{2},
\ee
as a function of $x= \ell/L$ against the numerics for different sizes ($L=100,200,400,800$). 
The $O(x^2)$ approximation is indistinguishable from the full result  up to $x\sim0.3$.
Clearly, the numerical data approach the prediction when $x$ is kept fixed and $L$ gets larger. 
The finite-size corrections are small for $x$ close to $0$, but they explode in the opposite regime $x\rightarrow 1$.

\begin{figure}[t]
\centering
\includegraphics[width=0.484\textwidth]{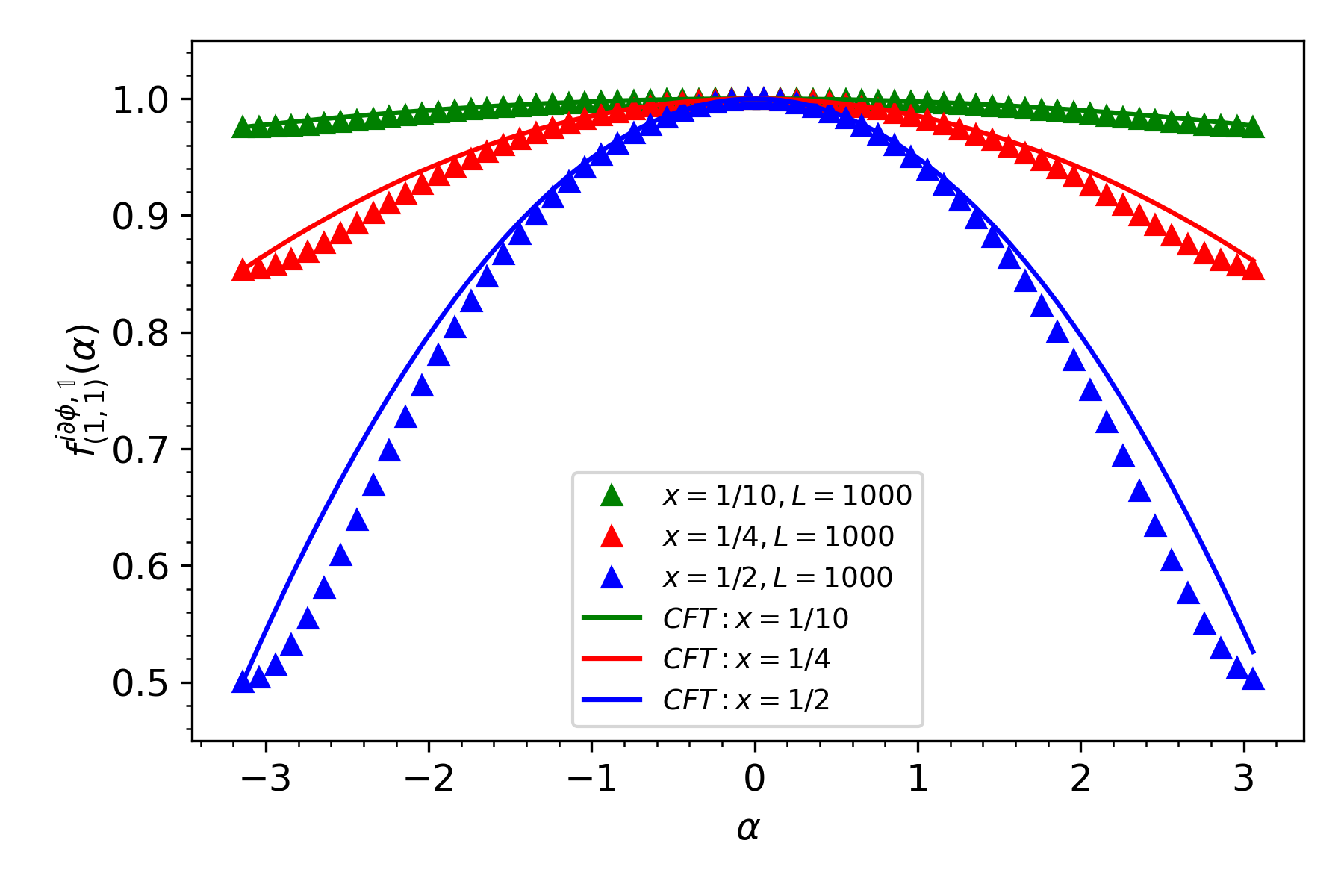}
\includegraphics[width=0.484\textwidth]{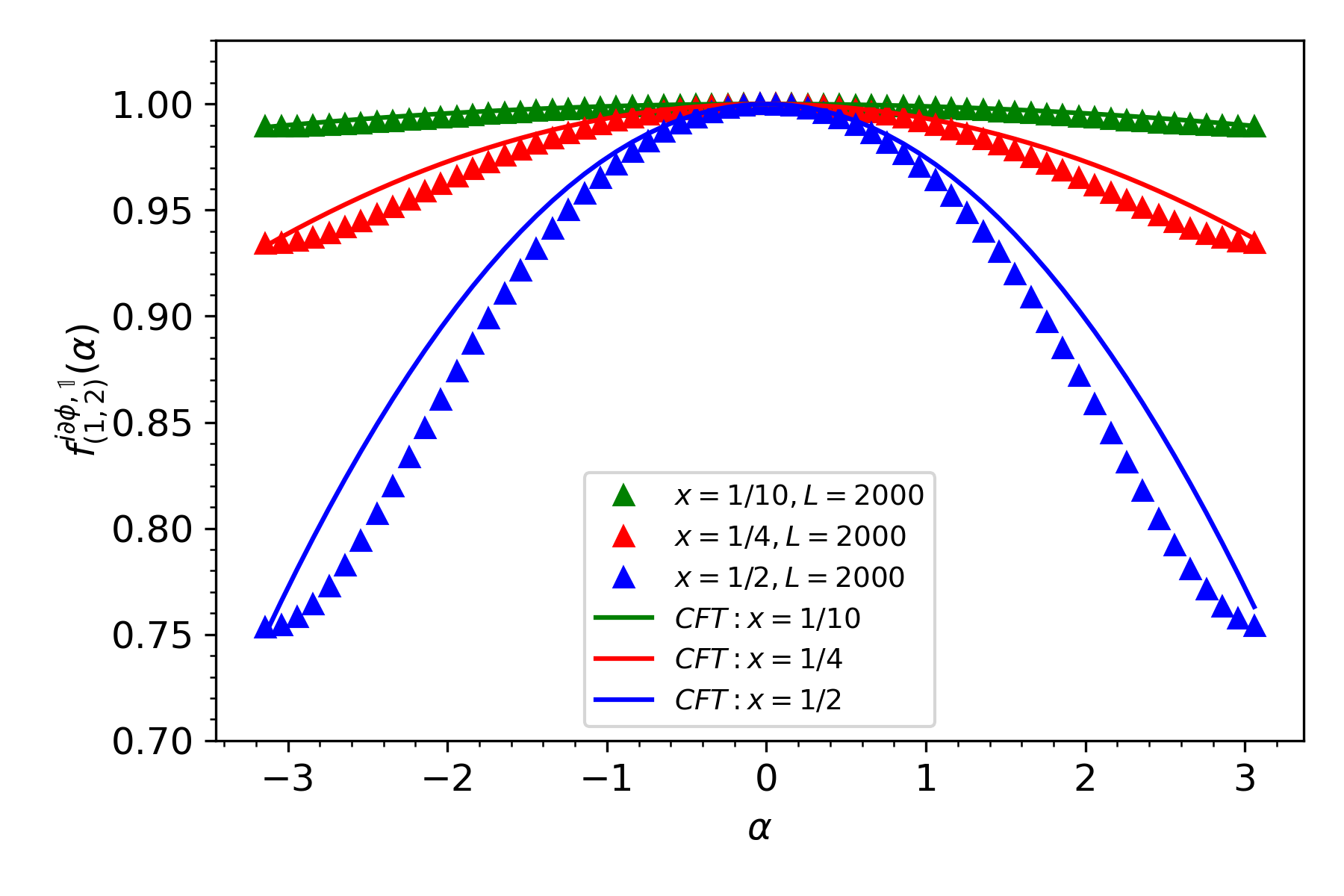}
\caption{The universal functions $f^{i\partial\phi,{\mathds 1}}_{(1,1)}(\alpha)$ and $f^{i\partial\phi,{\mathds 1}}_{(1,2)}(\alpha)$ (left/right panels respectively) as functions of $\alpha$ for different values of $x$ ($x=1/10,1/4,1/2$). 
The agreement with numerical data for the XX chain at half-filling, is good for small $\alpha$ but it worsens as $\alpha$ gets closer to $\pm \pi$, as discussed in the text.
 }
\label{fig5}
\end{figure}

In Fig. \ref{fig5} we plot $f^{i\partial\phi,{\mathds 1}}_{(1,n-1)}(\alpha)$ for $n=2,3$ (left/right panel respectively) as functions of $\alpha$ with fixed $x$ and compare it with numerical data.
We consider system sizes large enough so that the finite size-corrections of the excess of variance are negligible;
in particular as $n$ increases a larger $L$ is required to satisfy the latter requirement (as expected from the ground state results \cite{Ric1}). 
The numerical data give a function $f^{i\partial\phi, {\mathds 1}}_{(1,n-1)}(\alpha)$ which is always smooth and periodic under $\alpha \rightarrow \alpha+2\pi$. 
Although we expect a singularity at $\alpha=\pm \pi$ from the analytical predictions of $f^{i\partial\phi,{\mathds 1}}_{(1,n-1)}(\alpha)$, 
the convergence of the numerics to this singularity is slow.
This is the reason why in the neighbourhood of  $\alpha = \pm \pi$ numerics and CFT do not yet match well and much larger system sizes are required 
to generate the singularity. A full and detailed explanation of this phenomena is given in Ref. \cite{G4}.
We just notice that as $x$ gets larger the phenomenon is amplified as also clear at the level of the variance in Fig. \ref{fig4}.

\subsection{Universal function for the pair of states $\Upsilon = V_\beta $ and $\chi = i\partial \phi$}
Finally we move the most cumbersome combination of vertex and derivative operator. 
In this case, we have been able to compute an explicit analytic expression just for the partition $\mathcal{S} = (n-1,1)$ with final result 
\begin{multline}
f^{V_\beta, i\partial \phi}_{\mathcal{S}}(\alpha) = e^{i\alpha\beta x\frac{n-1}{n}}  \frac{1}{ \frac{1}{4\sin^2 \frac{x\pi}{n}}+\frac{\beta^2}{4}\left(n\cot(\pi x)-\cot(\frac{\pi x}{n})\right)^2 } \\
\left(  \frac{1}{4\sin^2 \frac{x\pi}{n}}+\frac{\beta^2}{4}\left(n\cot(\pi x)-\cot(\frac{\pi x}{n})\right)^2 -i\frac{\alpha}{2\pi}\beta \left(n\cot(\pi x)-\cot(\frac{\pi x}{n}) \right) -\left(\frac{\alpha}{2\pi}\right)^2 \right),
\label{f99}
\end{multline}
which we are going to prove in the following. 

The derivation of  Eq. \eqref{f99} is rather cumbersome and we exploit the diagrammatic interpretation reported in Appendix \ref{AppCorr}. 
We refer moreover to Ref. \cite{Rel_Paola}, where similar calculations appear in the context of relative entropy between the same states, i.e. $V_\beta$ and $i\partial\phi$. 
Let us start with the following representation of $f^{V_\beta, i\partial \phi}_{(n-1,1)}(\alpha)$
\begin{multline}
f^{V_\beta, i\partial \phi}_{(n-1,1)}(\alpha) =  \frac{\displaystyle\left\langle {V}_{\alpha/2\pi}(-i\infty){V}_{-\alpha/2\pi}(i\infty)(i\partial\phi)(w_n^-)(i\partial\phi)(w_n^+)\prod_{k=1}^{n-1} V_{\beta_1}(w_k^-) V_{-\beta_1}(w_k^+)\right\rangle_{\rm cyl}    }{       \displaystyle  \langle V_{\alpha/2\pi}(-i\infty)V_{-\alpha/2\pi}(i\infty) \rangle_{\rm cyl}\left\langle  (i\partial\phi)(w_n^-)(i\partial\phi)(w_n^+)\prod_{k=1}^{n-1} V_{\beta_1}(w_k^-) V_{-\beta_1}(w_k^+)  \right\rangle_{\rm cyl} }.
\end{multline}
We now provide a diagrammatic representation of
\be
\frac{\displaystyle\left\langle {V}_{\alpha/2\pi}(-i\infty){V}_{-\alpha/2\pi}(i\infty)(i\partial\phi)(w_n^-)(i\partial\phi)(w_n^+)\prod_{k=1}^{n-1} V_{\beta_1}(w_k^-) V_{-\beta_1}(w_k^+)\right\rangle_{\rm cyl} }{ \displaystyle  \langle V_{\alpha/2\pi}(-i\infty)V_{-\alpha/2\pi}(i\infty) \rangle_{\rm cyl}},
\ee
whose ratio with its value at $\alpha=0$ gives directly $f^{V_\beta, i\partial \phi}_{(n-1,1)}(\alpha)$.
The diagrammatic rules are the following (see again Appendix \ref{AppCorr} for all needed definition)
\begin{itemize}
\item The contractions between the vertex operators $V_{\pm \beta}$ and $V_{\pm \alpha/2\pi}$ are present for each diagram and their contribution is factorised out. 
The resulting contribution is $e^{i\alpha \beta x \frac{n-1}{n}}$, similarly to $f^{V_{\beta_1},V_{\beta_2}}_{\mathcal{S}}(\alpha)$.
\item A diagram with the contraction between $(i\partial\phi)(w_n^+)$ and $(i\partial\phi)(w_n^-)$ always appears. 
Its contribution is $\l\frac{1}{2\sin \frac{\pi x}{n}}\r^2$, which is simply the two-point correlation function in a cylindrical geometry. 
\item We have a set of diagrams where the derivative operators are contracted with $V_{\pm \beta}$. 
Summing all their contributions, we get in the end $-\frac{\beta^2}{4}\l \sum^{n-1}_{k=1} \cot \frac{\pi}{n}(k+x)\r \l \sum^{n-1}_{k=1} \cot \frac{\pi}{n}(k-x)\r$.
\item A set of diagrams where one of the two $i\partial\phi$ is contracted to $V_{\pm \alpha/2\pi}$ and the other to one of the vertex $V_{\pm \beta}$ is present. 
The sum amounts to $-i\frac{\alpha}{2\pi}\beta \sum^{n-1}_{k=1} \cot \frac{\pi}{n}(k+x)$.
\item A final set of four diagrams is the one where each of the two derivative operators is contracted with one of $V_{\pm \alpha/2\pi}$, 
and it contributes with $- \l \frac{\alpha}{2\pi}\r^2$.
\end{itemize}
Summing up all these contributions, already gives the desired correlation, but a last useful manipulation is to provide the analytical continuation of 
$\sum^{n-1}_{k=1} \cot \frac{\pi}{n}(k+x) =-\cot \frac{\pi x}{n}+ \sum^{n-1}_{k=0} \cot \frac{\pi}{n}(k+x)$. 
By looking to the periodicity of  $\sum^{n-1}_{k=0} \cot \frac{\pi}{n}(k+x)$ under $x\rightarrow x+1$ and its poles/zeros structure we can identify
\be
\sum^{n-1}_{k=0} \cot \frac{\pi}{n}(k+x) = n \cot \pi x,
\ee
and so
\be
\sum^{n-1}_{k=1} \cot \frac{\pi}{n}(k+x) = n \cot \pi x -\cot \frac{\pi x}{n}.
\ee
Taking into account all these contributions, we finally arrive to the form of $f^{V_\beta, i\partial \phi}_{(n-1,1)}(\alpha)$ reported in Eq. \eqref{f99}.

In principle one can follow the same diagrammatic rules to express $f^{V_\beta, i\partial \phi}_{\mathcal{S}}(\alpha)$ for other partitions $\mathcal{S}$. 
However, the number of partitions grows rapidly with $n$ and we are not aware of any systematic treatment, but we have to work them out in a case by case manner. 
Consequently, it is impossible to obtain a close form for general $n$ that can be used for the analytical continuations. 
For this reason, we are not going to investigate other partitions.

Conversely, with a relatively small effort we can provide the leading term at order $O(x)$ of $f^{V_\beta, i\partial \phi}_\mathcal{S}(\alpha)$ for an aribitary partition with 
$m$ insertions of the vertex $V_\beta$.
The calculation closely follows the one for $f^{V_{\beta_1},V_{\beta_2}}(\alpha)$ and so we just sketch the derivation here. 
The order $O(x)$ comes from the generation of $i\partial \phi$ in the OPE expansion of $\mathcal{T}_{n,\alpha} \tilde{\mathcal{T}}_{n,\alpha}$; 
moreover the state $\ket{i\partial \phi}$, of the unreplicated theory, is neutral and so it does not contribute to the expectation value of the charge density $i\partial\phi$, 
while we get a nontrivial contribution from $\ket{V_\beta}$.
At order $O(x^2)$, one has contributions from the stress energy tensor and a double insertion (in two different replicas) of $i\partial \phi$. 
The starting point is as usual the rewriting of $f^{V_\beta,i\partial \phi}_{\mathcal{S}}(\alpha)$ in terms of twist fields 
\be
f^{V_\beta,i\partial \phi}_{\mathcal{S}}(\alpha) = \frac{\bra{V_\beta,\cdots,i\partial\phi,\cdots}\mathcal{T}_{n,\alpha}(0)\tilde{\mathcal{T}}_{n,\alpha}(\ell)\ket{V_\beta,\cdots,i\partial\phi,\cdots}}{\bra{V_\beta,\cdots,i\partial\phi,\cdots}\mathcal{T}_n(0)\tilde{\mathcal{T}}_n(\ell)\ket{V_\beta,\cdots,i\partial\phi,\cdots}}\frac{ \bra{0,\dots,0}\mathcal{T}_n(0)\tilde{\mathcal{T}}_n(\ell)\ket{0,\cdots,0} }{ \bra{0,\dots,0}\mathcal{T}_{n,\alpha}(0)\tilde{\mathcal{T}}_{n,\alpha}(\ell)\ket{0,\cdots,0}} ,
\ee
and then consider the OPE expansion at order $O(\ell^2)$
\begin{multline}
\mathcal{T}_{n,\alpha}(0) \tilde{\mathcal{T}}_{n,\alpha}(\ell) \simeq \la \mathcal{T}_{n,\alpha}(0) \tilde{\mathcal{T}}_{n,\alpha}(\ell) \ra \\
\l 1+\ell a_1(\alpha) \sum_j (i\partial\phi)^j(0) + \ell^2 \sum_{j<j'}a_{jj'}(\alpha)(i\partial\phi)^j(0)(i\partial\phi)^{j'}(0) +\ell^2 a_2(\alpha)\sum_j T^j(0) \r,
\end{multline}
where $a_1(\alpha),a_2(\alpha),a_{jj'}(\alpha)$ are the OPE coefficients
\be
a_1(\alpha) = \frac{\alpha}{2\pi n}, \qquad a_2(\alpha)-a_2(0) = a_{jj'}(\alpha)-a_{jj'}(0) = \frac{1}{n^2}\l\frac{\alpha}{2\pi}\r^2.
\ee
Indeed, $a_1(\alpha)$ and $a_2(\alpha)$ have already been reported in Eq. \eqref{ope22}, while $a_{jj'}(\alpha)$ is simply fixed matching the previous exact computation
of $f^{V_\beta,V_{\beta'}}_{\mathcal{S}}(\alpha)$ with he OPE expansion in terms of twist fields at order $O(x^2)$.
Expanding $f^{V_\beta,i\partial \phi}_{\mathcal{S}}(\alpha)$ at the same order, we obtain
\begin{multline}
f^{V_\beta,i\partial \phi}_{\mathcal{S}}(\alpha)  \simeq 1+\ell a_1(\alpha)\sum_j \bra{V_\beta,\cdots,i\partial\phi,\cdots}(i\partial\phi)^j(0)\ket{V_\beta,\cdots,i\partial\phi,\cdots} +\\
\ell^2\sum_{j<j'}(a_{jj'}(\alpha)-a_{jj'}(0))\bra{V_\beta,\cdots,i\partial\phi,\cdots}(i\partial\phi)^j(0)(i\partial\phi)^{j'}(0)\ket{V_\beta,\cdots,i\partial\phi,\cdots} +\\
\ell^2(a_2(\alpha)-a_2(0))\sum_j \l \bra{V_\beta,\cdots,i\partial\phi,\cdots} T^j(0)\ket{V_\beta,\cdots,i\partial\phi,\cdots} - \bra{0,\cdots,0}T^j(0)\ket{0,\cdots,0} \r.
\end{multline}
Evaluating all the expectation values, we finally get the desired result
\be
f^{V_\beta,i\partial \phi}_{(m,n-m)}(\alpha) \simeq 1+i\alpha \beta x\frac{m}{n}-\frac{1}{2}\l \alpha \beta x\frac{m}{n}\r^2 - \frac{(n-m)^2}{n^2}\alpha^2 x^2.
\ee
One interesting feature is that, at order $O(x^2)$, the insertion of the vertex operators only shifts the average charge, i.e. 
the quadratic term in $\alpha$ of $\log f^{V_\beta,i\partial \phi}_{(m,n-m)}(\alpha)$ does not depend on $\beta$. 
Instead, the excess of variance, encoded in the $O(\alpha^2)$ term of $\log f^{V_\beta,i\partial \phi}_{(m,n-m)}(\alpha)$, is entirely due to the presence of the derivative operators. Clearly, this observation is no longer true at higher order in $x$, where the correlation effects between the different replicas matter, as can be seen explicitly 
from the exact result for $f^{V_\beta,i\partial\phi}_{(n-1,1)}(\alpha)$ in Eq. \eqref{f99}.

\begin{figure}[t]
\centering
\includegraphics[width=0.484\textwidth]{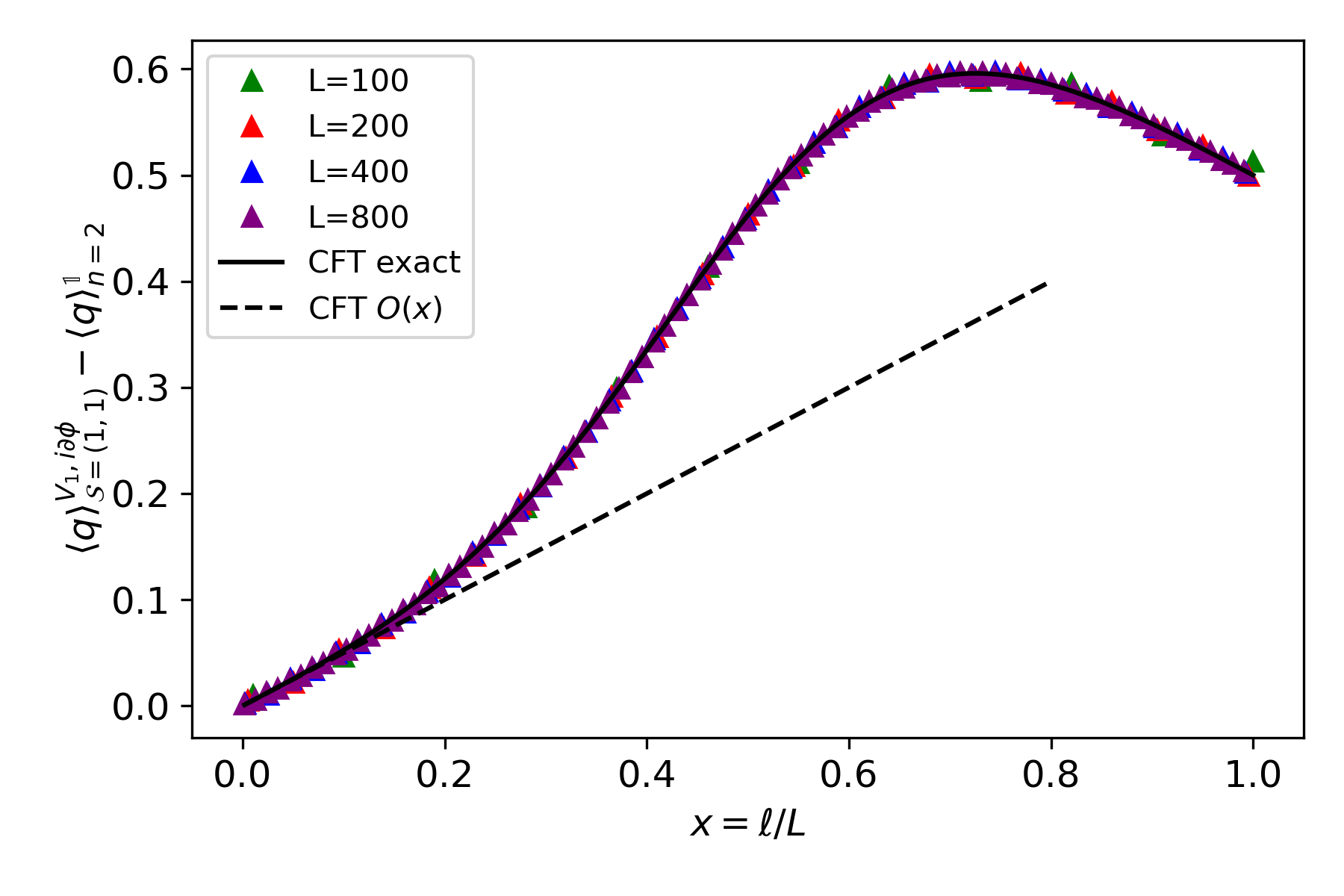}
\includegraphics[width=0.484\textwidth]{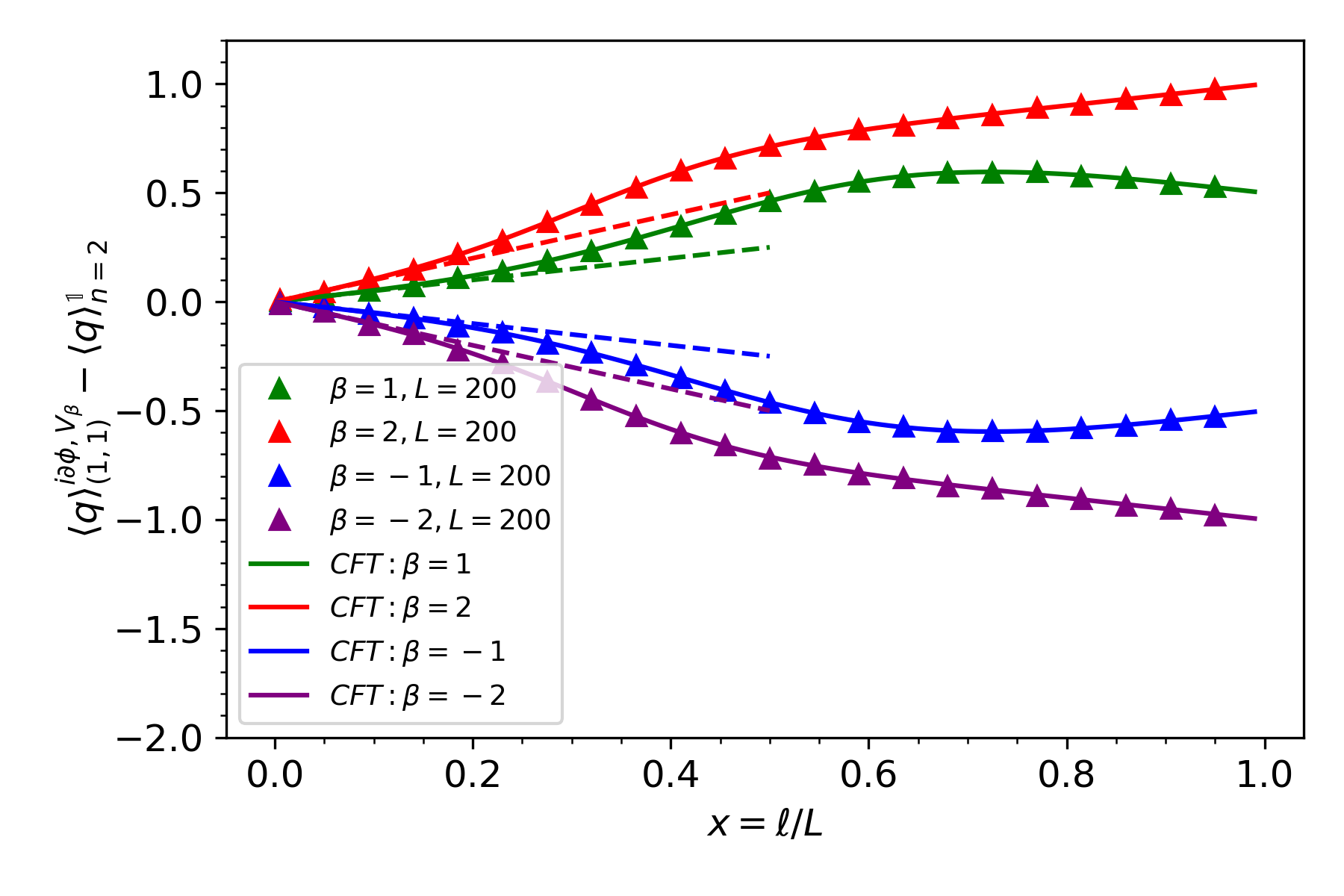}
\caption{Left: The CFT prediction of the excess of average $\la  q\ra^{V_1,i\partial\phi}_{(1,1)}-\la q\ra^{\mathds 1}_{2}$ as a function of $x$ is plotted against the numerical data  
for the XX chain at different system sizes ($L=100,200,400,800$). The numerics match very well with CFT predictions even for relatively small system sizes (such as  $L=100$).
Right: $\la q\ra^{V_\beta,i\partial\phi}_{(1,1)}-\la q\ra^{\mathds 1}_{2}$ is plotted for different $\beta$ ($\beta=-2,-1,1,2$) as a function of $x = \ell/L$. 
Numerical data refer to the low energy excitations of the half-filled XX chain for $L=200$. 
The $O(x)$ predictions from OPE calculation (dashed lines) are clearly working well only for small values of $x$, up to $x\sim0.2$.
 }
\label{fig6}
\end{figure}

We now test the CFT result for $f^{V_\beta,i\partial\phi}_{(n-1,1)}(\alpha)$ given by Eq. \eqref{f99} against the numerical data for the XX chain. 
We focus again on the excess of charge 
\begin{multline}
\la q \ra^{V_\beta,i\partial\phi}_{(n-1,1)}-\la q \ra^{\mathds 1}_{n} = \frac{1}{i} \frac{d}{d\alpha} f^{V_\beta,i\partial\phi}_{(n-1,1)}(\alpha)\Big|_{\alpha=0}=\\
\frac{\beta  (n-1) x}{n}-\frac{\beta  \left(n \cot (\pi  x)-\cot \left(\frac{\pi  x}{n}\right)\right)}{(2 \pi ) \left(\frac{1}{4} \beta ^2 \left(n \cot (\pi  x)-\cot \left(\frac{\pi  x}{n}\right)\right)^2+\frac{1}{4 \sin ^2\left(\frac{\pi  x}{n}\right)}\right)},
\label{eq113}
\end{multline}
and of variance
\begin{multline}
\la \Delta q^2 \ra^{V_\beta,i\partial\phi}_{(n-1,1)}-\la \Delta q^2 \ra^{\mathds 1}_{n} = \frac{1}{(i)^2} \frac{d^2}{d\alpha^2} \log f^{V_\beta,i\partial\phi}_{(n-1,1)}(\alpha)\Big|_{\alpha=0}=\\
 -\frac{\left(\frac{\beta  \left(n \cot (\pi  x)-\cot \left(\frac{\pi  x}{n}\right)\right)}{2 \pi }\right)^2}{ \left(\frac{1}{4} \beta ^2 \left(n \cot (\pi  x)-\cot \left(\frac{\pi  x}{n}\right)\right)^2+\frac{1}{4 \sin ^2\left(\frac{\pi  x}{n}\right)}\right)^2}+
2\frac{\left(\frac{1}{2 \pi }\right)^2}{\frac{1}{4} \beta ^2 \left(n \cot (\pi  x)-\cot \left(\frac{\pi  x}{n}\right)\right)^2+\frac{1}{4 \sin ^2\left(\frac{\pi  x}{n}\right)}}.
\label{eq114}
\end{multline}
In Fig. \ref{fig6} we plot the excess of average charge $\la q \ra^{V_\beta,i\partial\phi}_{(n-1,1)}-\la q \ra^{\mathds 1}_{n}$ for $n=2$ as a function of $x=\ell/L$ and compare it with numerical data. From the figure we see that the discrepancies of numerics from the analytical predictions are quite negligible also for a system as small as $L=200$.

\begin{figure}[t]
\centering
\includegraphics[width=0.484\textwidth]{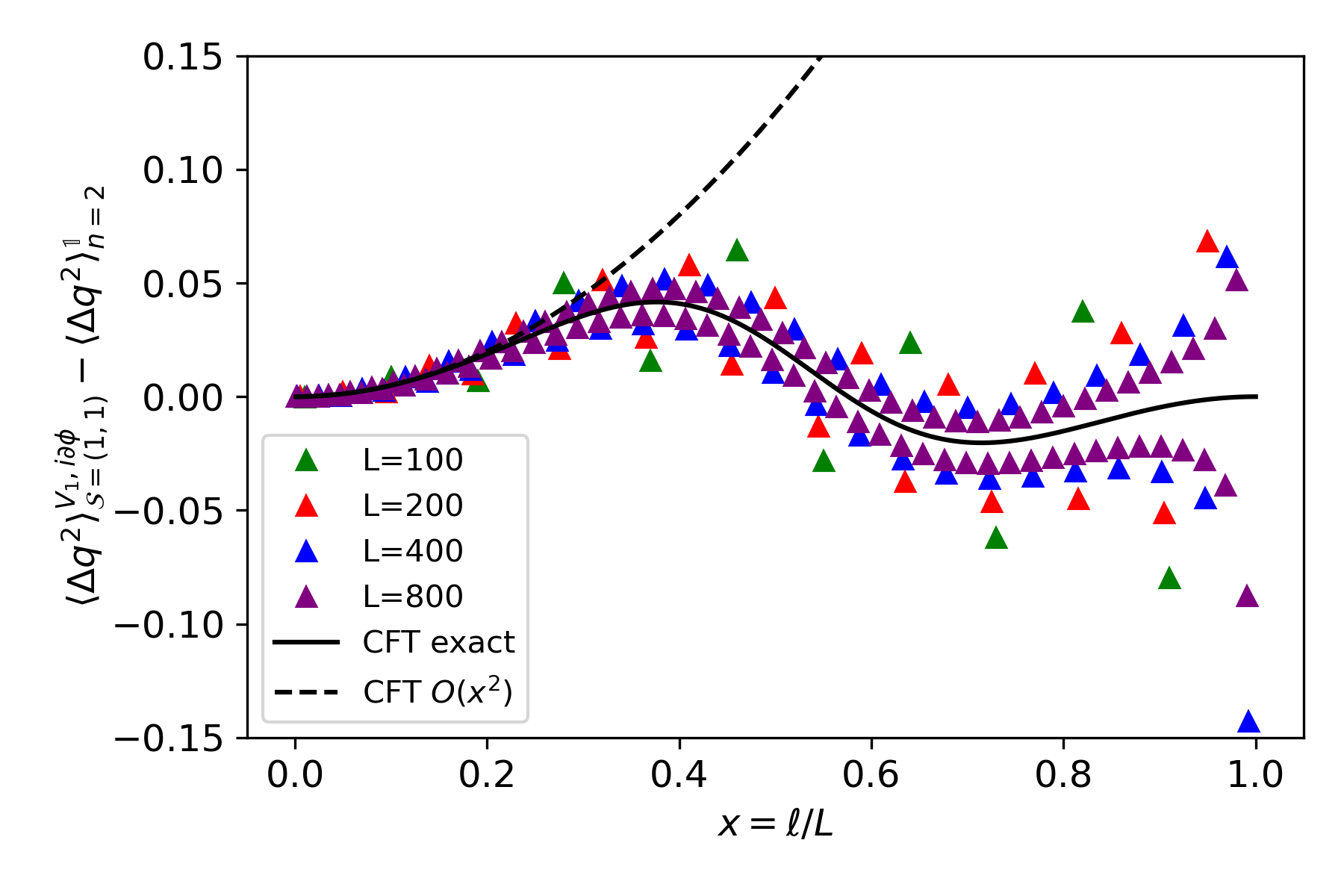}
\includegraphics[width=0.484\textwidth]{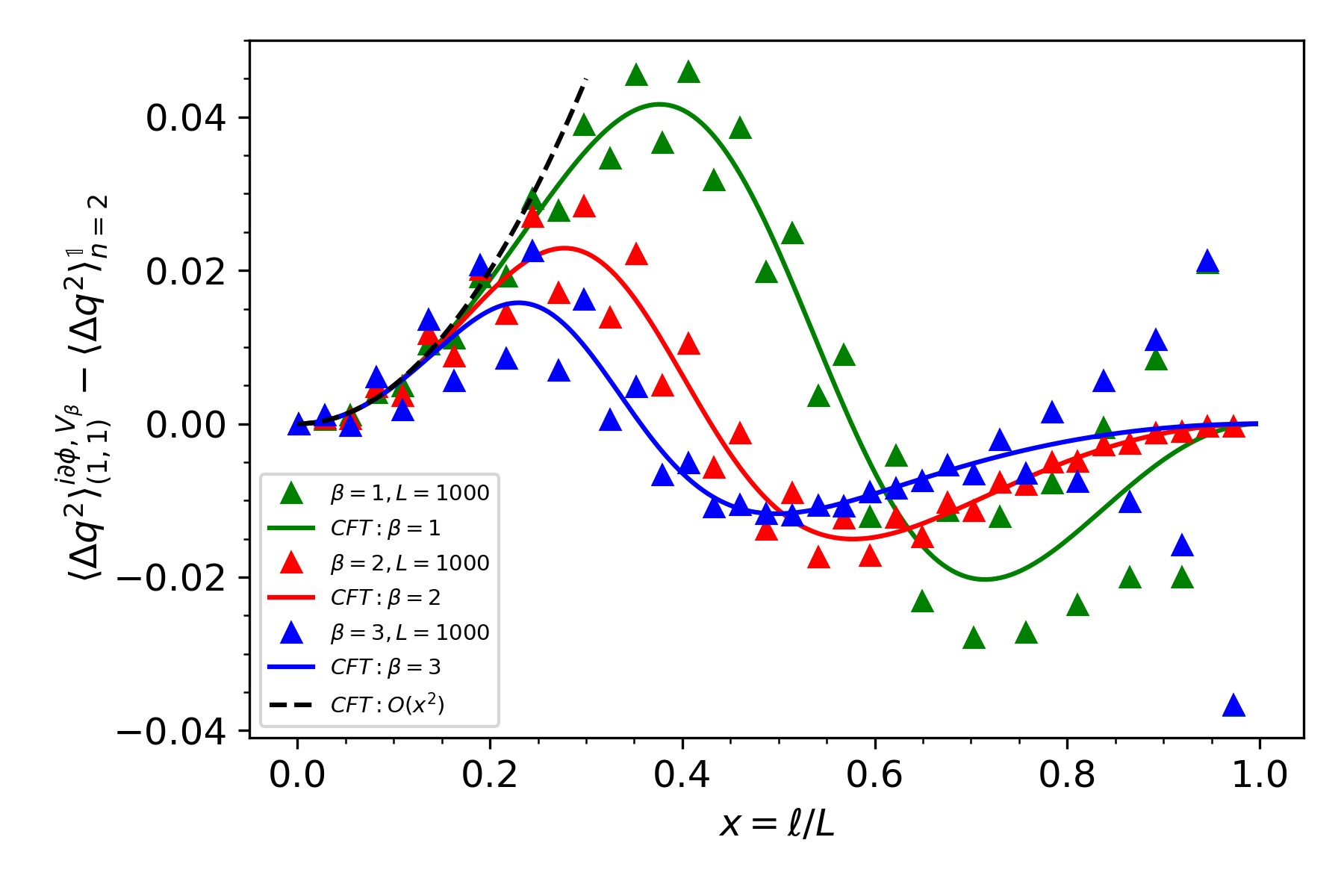}
\caption{Left: The CFT prediction of the excess of variance $\la \Delta q^2\ra^{V_1,i\partial\phi}_{(1,1)}-\la \Delta q^2\ra^{\mathds 1}_{2}$ as a function of $x$ is plotted against the 
numerical data  for the XX chain for different system sizes ($L=100,200,400,800$).
Right: $\la \Delta q^2\ra^{V_\beta,i\partial\phi}_{(1,1)}-\la \Delta q^2\ra^{\mathds 1}_{2}$ is plotted for different values of $\beta$ ($\beta=1,2,3$) as a function of $x = \ell/L$. 
Numerical data refer to the low energy excitations of the half-filled XX chain for $L=1000$. 
As known from OPE arguments, the excess of variance at order $O(x^2)$ does not depend on $\beta$, but this is no longer true for larger $x$. 
 }
\label{fig7}
\end{figure}

The numerical data for the excess of variance $\la \Delta q^2 \ra^{V_\beta,i\partial\phi}_{(n-1,1)}-\la \Delta q^2 \ra^{\mathds 1}_{n}$ are reported in Fig. \ref{fig7}. 
In the left panel we consider $n=2$, $\beta=1$ and compare different values of the system size $L$. 
In contrast to the data for the excess of charge at the same size, the deviations from the CFT predictions are now evident and they become larger as $x\rightarrow 1$. 
As usual for the variance (see also the previous subsection), the numerical data oscillate around the CFT result.
In the right panel of Fig. \ref{fig7} , we fix the system size ($L =1000$) and compare different values of $\beta$. 
A peculiar feature manifest from this plot is that for $\beta=2$ the numerical data agree quite well with the predictions close to $x= 1$, 
which is definitely not the case for $\beta = 1,3$.

\section{Symmetry-resolved relative entropy}
\label{sec5}

In this section, we use the results of the previous one to obtain explicit predictions for the symmetry-resolved relative $n$-th relative entropy, defined by Eq. \eqref{Rel_res}
specialised to low-lying CFT states $\rho=\rho_\Upsilon$ and $\sigma=\rho_\chi$, generated by two primary operators $\Upsilon $ and $\chi $ in the 
compact boson theory. 

We use Eq. \eqref{Svsp} for the relative entropy in terms of the generalised probability, that we also rewrite here 
\be
S_n(\rho_\Upsilon \| \rho_\chi)(q) = S_n(\rho_\Upsilon \| \rho_\chi)+\frac{1}{1-n}\log\frac{p^{\Upsilon,\chi}_{(1,n-1)}(q)}{p^\Upsilon_n(q)}+ \log \frac{p^\chi_1(q)}{p^\Upsilon_1(q)}
\label{ratio3}.
\ee
As we have discussed already, in the thermodynamic limit $L\rightarrow \infty$ ($x=\ell/L$ fixed), the ratio of probabilities tends to $1$ (cf. Eq. \eqref{equip}) 
leading relative entropy equipartition \eqref{Sreq}, i.e.  
${S_n(\rho_\Upsilon \| \rho_{\chi})(q)}\simeq{S_n(\rho_\Upsilon \| \rho_{\chi})}$.
Starting from Eq. \eqref{ratio3}, we can systematically characterise the corrections to this asymptotic behaviour and identify the terms breaking the equipartition.
The leading and physically most important corrections to equipartition comes from the orders $O(\alpha)$ and $O(\alpha^2)$ of $f^{\Upsilon,\chi}_{(1,n-1)}(\alpha)$.
At this order, the generalised probabilities appearing in Eq. \eqref{ratio3} are still Gaussian \footnote{This leading Gaussian behaviour is a consequence of the $\alpha$-dependence of the scaling dimension of the modified twist fields, which leads to a diverging variance in the continuum limit. The other non-universal but finite cumulants of the ground-state probability distribution come instead from the $\alpha$-dependent proportionality constant appearing in the relation $\text{tr}\l \rho^n_{\mathds 1}e^{i\alpha Q}\r \sim \la \mathcal{T}_{n,\alpha}(0)\tilde{\mathcal{T}}_{n,\alpha}(\ell)\ra$.}, but with renormalised values of 
average charge and variance compared to the vacuum, explicitly computed, also numerically, in the previous section. 

Let us now use this Gaussian behaviour to compute at the first subleading order the symmetry resolved relative entropies $S_n(\rho_\Upsilon \| \rho_\chi)(q)$.
We need the Fourier transform of $p^{\Upsilon,\chi}_{\mathcal{S}}(\alpha) = p^{\mathds 1}_{n}(\alpha)f^{\Upsilon,\chi}_{\mathcal{S}}(\alpha)$, 
for the partitions ${\cal S}=(1,n-1)$ of interest,
and $p^{\Upsilon/\chi}(\alpha) = p^{\mathds 1}_{n}(\alpha)f^{\Upsilon/\chi}(\alpha)$. 
At the leasing order, we write $p^{\mathds 1}_n(q)$ as (we adopt the convention $\langle q\rangle^{\mathds 1}_n=1$) 
\be
p_n^{\mathds 1}(\alpha) \simeq \exp(-\frac{\alpha^2}{2}\la \Delta q^2\ra_n^{\mathds 1}), \qquad p^{\mathds 1}_n(q) \simeq \frac{1}{\sqrt{2\pi \la \Delta q^2\ra_n^{\mathds 1}}}\exp(-\frac{q^2}{2 \la \Delta q^2\ra_n^{\mathds 0}}). 
\label{pGS}
\ee
We recall that
\be
\la \Delta q^2\ra_n^{\mathds 1}=\frac{1}{\pi^2n} \log \left[\frac{L}{\pi} \sin \Big(\pi \frac\ell{L}\Big)\right]+\kappa_n+ o(1)\,,
\ee
where the additive constant $\kappa_n$ is not universal (and it is known \cite{Ric1} in the XX chain used for the numerics). 
Also $p^{\Upsilon,\chi}_{\cal S}(q)$ is Gaussian and we write
\be
p^{\Upsilon,\chi}_{\cal S}(\alpha) \simeq 
\exp( i \alpha \la q \ra^{\Upsilon,\chi}_{\cal S}
-\frac{\alpha^2}{2}\la \Delta q^2\ra^{\Upsilon,\chi}_{\cal S}), 
\qquad p^{\Upsilon,\chi}_{\cal S}(q) \simeq \frac{1}{\sqrt{2\pi \la \Delta q^2\ra^{\Upsilon,\chi}_{\cal S}}}\exp(-\frac{(q- \la q\ra^{\Upsilon,\chi}_{\cal S})^2}{2 \la \Delta q^2\ra^{\Upsilon,\chi}_{\cal S}}). 
\label{pgen}
\ee
The average charge and variance of these distributions are read directly from the expansion of the universal functions $f^{\Upsilon,\chi}_{\cal S}(\alpha)$ 
\be
\log f^{\Upsilon,\chi}_{\cal S}(\alpha) = i \alpha a^{\Upsilon,\chi}_{\cal S}
-\frac{\alpha^2}{2} b^{\Upsilon,\chi}_{\cal S} + O(\alpha^4),
\label{fexp0}
\ee
and so
\be
\la q \ra^{\Upsilon,\chi}_{\cal S}=a^{\Upsilon,\chi}_{\cal S}, \qquad 
\la \Delta q^2 \ra^{\Upsilon,\chi}_{\cal S} = \la \Delta q^2 \ra_n^{\mathds 1}  +b^{\Upsilon,\chi}_{\cal S}    .
\ee
Obviously, the very same formulas remain valid if there is the insertion of a single operator in $p_n^\Upsilon(q)$ (which indeed correspond to ${\cal S}=(n,0)$. 

Plugging these Gaussian approximations for all the probabilities into Eq. \eqref{ratio3}, awe easily get a completely general result for the relative entropy with the first 
subleading order that reads
\begin{multline}
S_n(\rho_\Upsilon \| \rho_\chi)(q) = S_n(\rho_\Upsilon \| \rho_\chi) +\\
+\frac{1}{1-n}\l -\frac{(q- \la q\ra^{\Upsilon,\chi}_{(1,n-1)})^2}{2 \la \Delta q^2\ra^{\Upsilon,\chi}_{(1,n-1)}} +
\frac{(q- \la q\ra^{\Upsilon}_n)^2}{2 \la \Delta q^2\ra^{\Upsilon}_n}
\r +
\l \frac{(q- \la q\ra^{\Upsilon}_1)^2}{2 \la \Delta q^2\ra^{\Upsilon}_{1}} -
 \frac{(q- \la q\ra^{\chi}_1)^2}{2 \la \Delta q^2\ra^{\chi}_1}
\r +\dots
\label{Srel_gen}.
\end{multline}
Now we are going to use that the variances can be written as in Eq. \eqref{exc1} the sum of the diverging piece from the vacuum, plus an $O(1)$ term, obtaining   
\begin{multline}
S_n(\rho_\Upsilon \| \rho_\chi)(q) = S_n(\rho_\Upsilon \| \rho_\chi) +\\
\frac{1}{2(1-n) \la \Delta q^2\ra^{\mathds 1}_n} 
\l -{(q- a^{\Upsilon,\chi}_{(1,n-1)})^2} \l1-\frac{b^{\Upsilon,\chi}_{(1,n-1)}}{\la \Delta q^2\ra^{\mathds 1}_n}\r 
 + {(q- a^{\Upsilon}_n)^2} \l1-\frac{b^{\Upsilon}_{n}}{\la \Delta q^2\ra^{\mathds 1}_n}\r 
\r \\
+ \frac{1}{ 2\la \Delta q^2\ra^{\mathds 1}_1}
\l {(q-a^{\Upsilon}_1)^2} \l1-\frac{b^{\Upsilon}_{1}}{\la \Delta q^2\ra^{\mathds 1}_1}\r -
 {(q- a^{\chi}_1)^2}\l1-\frac{b^{\chi}_{1}}{\la \Delta q^2\ra^{\mathds 1}_1}\r
\r +\dots.
\label{Srel_gen2}
\end{multline}
This shows that quite generally equipartition is broken at order $(\log L)^{-1}$ (unless some cancellations take place).
It is also clear that in order to compute the relative entropies, all we have to do is just to extract the coefficients $a^{\Upsilon,\chi}_{\cal S}$ and $b^{\Upsilon,\chi}_{\cal S}$
from the CFT universal functions and plug into Eq. \eqref{Srel_gen2}.

In the next subsection, we will carefully analyse the symmetry resolved relative entropies for specific pairs of states case by case.
Before doing so, we briefly review some useful results for short $\ell$ expansion of the standard relative entropies \cite{Rel-CFT1} and mention the main difference with the 
charged counterpart.
For two primary states $\Upsilon,\chi$ the relative entropy $S_1(\rho_\Upsilon \| \rho_\chi)$ at leading order scales as
\be
S_1(\rho_\Upsilon \| \rho_\chi) \propto \l \la \Psi\ra_{\Upsilon} - \la \Psi \ra_{\chi} \r^2 \l \ell \r^{2\Delta_\Psi} + o\Big(\Big( \frac{\ell}{L} \Big)^{2\Delta_\Psi}\Big),
\label{S1_rel}
\ee
where $\Psi$ is the lightest quasi-primary for which $\la \Psi\ra_\Upsilon - \la \Psi \ra_{\chi}\neq 0$. 
The prefactor has been worked out explicitly in Ref. \cite{Rel-CFT1} for the case in which $\Psi$ is a primary operator or the stress-energy tensor $T$. 
A simple argument to understand this behaviour is that, for any primary operator $\mathcal{O}$ the fusion channel
\be
\mathcal{T}_n \times \tilde{\mathcal{T}}_n \rightarrow \mathcal{O}\otimes 1 \otimes \cdots \otimes 1
\ee
is never there, since the expectation value of $\mathcal{O}$ is zero. Moreover, although the fusion channel
\be
\mathcal{T}_n \times \tilde{\mathcal{T}}_n \rightarrow T \otimes 1 \otimes \cdots \otimes 1
\ee
provides a non-vanishing contribution for any integer $n>1$, when the analytical continuation $n\rightarrow 1$ is performed this fusion effectively disappears and 
$T$ behaves almost like a primary operators \cite {calab-2int2,Zhang7,2int-Paola} (we will return on this point later on). 

Conversely, in the presence of a flux, 
\be
\frac{1}{1-n} \log \frac{\trace\l \rho_\Upsilon\rho_\chi^{n-1}e^{i\alpha Q}\r}{\trace\l \rho_\Upsilon^n e^{i\alpha Q}\r}
\ee
is in general of order $\l \frac{\ell}{L}\r^{\Delta_\Psi}$, also in the limit $n\rightarrow 1$ beacuse, as we discussed in section \ref{TFields}, the fusion channel
\be
\mathcal{T}_{n,\alpha} \times \tilde{\mathcal{T}}_{n,\alpha} \rightarrow \Psi \otimes 1 \otimes \cdots  \otimes 1
\ee
has a non-vanishing contribution even for $n=1$, and for $\Psi$ being a primary a field the correspondent OPE coefficient is directly related to the 3-point function 
$\la \mathcal{V}_\alpha(0)\Psi(z)\mathcal{V}_{-\alpha}(\infty)\ra$.

\subsection{Vertex-vertex symmetry resolved relative entropies}

The universal function $f^{V_\beta,V_{\beta'}}_{\mathcal{S}}(\alpha)$ necessary for this calculation
has been computed for all partitions $\mathcal{S}$, and the final result is given by  Eq. \eqref{VVCFT}.
For the two possible relative entropy, we only need the partition ${\cal S}=(1,n-1)$ and ${{\cal S}=(n-1,1)}$.
We start from the relative entropy between a vertex state and the vacuum, for which Eq. \eqref{VVCFT} simplifies to
\be
f^{{\mathds 1},V_{\beta}}_{(1,n-1)}(\alpha) = e^{i\alpha \beta x \frac{n-1}{n}}, \qquad f^{V_{\beta},{\mathds 1}}_{(1,n-1)}(\alpha) = e^{i\alpha \beta x/n}, \quad f_n^{V_\beta}(\alpha) = e^{i\alpha \beta x}.
\label{eq127}
\ee
In this cases then, the variances are always the same as in the ground states and there are only shifts of the average charges, encoded in the factors $a^{\Upsilon,\chi}_{\cal S}$
in Eq. \eqref{fexp0} that in our cases read
$a^{{\mathds 1},V_{\beta}}_{(1,n-1)} =  \beta x \frac{n-1}{n}$,  $a^{V_{\beta},{\mathds 1}}_{(1,n-1)} =  \beta x/n$, and $a_n^{V_\beta}(\alpha) =  \beta x$.
For completeness it is also worth to report the explicit values of the generalised probabilities
\be
p^{{\mathds 1},V_{\beta}}_{(1,n-1)}(q)  \simeq \frac{1}{\sqrt{2\pi \la \Delta q^2\ra_n^{\mathds 1}}}\exp(-\frac{\l q- \frac{\beta x(n-1)}{n}\r^2}{2 \la \Delta q^2\ra_n^{\mathds 1}}),
\label{eq128}
\ee
\be
p^{V_{\beta},{\mathds 1}}_{(1,n-1)}(q)  \simeq \frac{1}{\sqrt{2\pi \la \Delta q^2\ra_n^{\mathds 1}}}\exp(-\frac{\l q- \frac{\beta x}{n}\r^2}{2 \la \Delta q^2\ra_n^{\mathds 1}}),
\label{eq129}
\ee
\be
p^{V_{\beta}}_{n}(q)  \simeq \frac{1}{\sqrt{2\pi \la \Delta q^2\ra_n^{\mathds 1}}}\exp(-\frac{\l q-\beta x\r^2}{2 \la \Delta q^2\ra_n^{\mathds 1}}).
\label{eq130}
\ee

In order to use Eq. \eqref{Srel_gen2}, we also need the symmetry-resolved $n$-th R\'enyi relative entropy. 
An explicit expression for $S_n(\rho_{\mathds 1}\| \rho_{V_\beta}) = S_n( \rho_{V_\beta} \| \rho_{\mathds 1})$ has been provided in \cite{Rel_Paola} and reads
\be
S_n(\rho_{\mathds 1}\| \rho_{V_\beta}) =  S_n( \rho_{V_\beta} \| \rho_{\mathds 1}) = \frac{\beta^2}{1-n}\log \l \frac{\sin \pi x}{n \sin \frac{\pi x}{n}}\r.
\label{eq131}
\ee
Plugging Eqs. \eqref{eq128}, \eqref{eq129} and \eqref{eq130} into Eq. \eqref{ratio3}, or \eqref{Srel_gen2}, for
$S_n(\rho_{\mathds 1}\| \rho_{V_\beta})(q)$ and $S_n( \rho_{V_\beta} \| \rho_{\mathds 1})(q)$, one obtains
\begin{multline}
S_n(\rho_{\mathds 1}\| \rho_{V_\beta})(q) = S_n(\rho_{\mathds 1}\| \rho_{V_\beta}) + \frac{1}{1-n}\log \frac{p^{{\mathds 1},V_{\beta}}_{(1,n-1)}(q) }{p_n^{\mathds 1}(q)}-\log \frac{p^{\mathds 1}_1(q)}{p^{V_\beta}_1(q)} \simeq \\
\frac{\beta^2}{1-n}\log \l \frac{\sin \pi x}{n \sin \frac{\pi x}{n}}\r -\frac{1}{\la \Delta q^2\ra_n^{\mathds 1}}\l \frac{\beta x}{n}q + \frac{1-n}{n^2}\frac{\beta^2 x^2}{2}\r +\frac{1}{\la \Delta q^2\ra_1^{\mathds 1}}\l \beta x q -\frac{\beta^2 x^2}{2}\r,
\label{eq132}
\end{multline}
and
\begin{multline}
S_n( \rho_{V_\beta} \| \rho_{\mathds 1})(q) = S_n( \rho_{V_\beta} \| \rho_{\mathds 1}) + \frac{1}{1-n}\log \frac{p^{V_{\beta},{\mathds 1}}_{(1,n-1)}(q) }{p_n^{V_\beta}(q)}-\log \frac{p^{V_\beta}_1(q)}{p^{\mathds 1}_1(q)} \simeq \\
\frac{\beta^2}{1-n}\log \l \frac{\sin \pi x}{n \sin \frac{\pi x}{n}}\r +\frac{1}{\la \Delta q^2\ra_n^{\mathds 1}}\l \frac{\beta x}{n} q - \frac{1+n}{n^2}\frac{\beta^2x^2}{2}\r +\frac{1}{\la \Delta q^2\ra_1^{\mathds 1}}\l -\beta x q + \frac{\beta^2 x^2}{2}\r.
\label{eq132b}
\end{multline}
A remarkable feature of these results is that 
\be
S_n( \rho_{V_\beta} \| \rho_{\mathds 1})(q) \neq S_n(\rho_{\mathds 1}\| \rho_{V_\beta})(q),
\ee
while $S_n( \rho_{V_\beta} \| \rho_{\mathds 1}) =S_n(\rho_{\mathds 1}\| \rho_{V_\beta})$.
Furthermore, also the small $x$ behaviour is different in the two cases. 
In fact  $S_n( \rho_{V_\beta} \| \rho_{\mathds 1})$ scales as $\sim x^2$ for small $x$, while $S_n( \rho_{V_\beta} \| \rho_{\mathds 1})(q)$ behaves as $x$. 
This is understood from OPE expansion, because the fusion
\be
\mathcal{T}_{n,\alpha} \times \tilde{\mathcal{T}}_{n,\alpha} \rightarrow i\partial \phi \otimes 1 \otimes \cdots \otimes 1
\ee
is present in general for $\alpha \neq 0$, but vanishes whenever $\alpha=0$. 
Taking the Fourier transform, this $O(x)$ contribution produces the correction seen in Eqs. \eqref{eq132} and \eqref{eq132b}.
Another important feature concerns the limit $n\to1$ when the two relative entropies become equal and simplify to 
\be
S_1( \rho_{V_\beta} \| \rho_{\mathds 1})(q) = S_1(\rho_{\mathds 1}\| \rho_{V_\beta})(q)= \beta^2(1-\pi x\cot \pi x)-\frac{\beta^2 x^2}{2\la \Delta q^2\ra_1^{\mathds 1}}.
\ee
Remarkably, at this order there is no q-dependence and so the equipartition may be eventually broken at higher order in $1/\log\ell$. 

For two general vertex states the calculation is identical. Indeed from Eq. \eqref{VVCFT} we have
\be
f^{V_{\beta_1},V_{\beta_2}}_{(1,n-1)}(\alpha) = e^{\frac{i\alpha x}{n}[\beta_1+(n-1)\beta_2]},
\ee
and hence $a^{V_{\beta_1},V_{\beta_2}}_{(1,n-1)}=x [\beta_1+(n-1)\beta_2]/n$ and $b^{V_{\beta_1},V_{\beta_2}}_{(1,n-1)}=0$.
Plugging these expression into Eq. \eqref{Srel_gen2}, we get
\begin{multline}
S_n(\rho_{V_{\beta_1}}\| \rho_{V_{\beta_2}})(q)  =S_n(\rho_{V_{\beta_1}}\| \rho_{V_{\beta_2}})
-\frac{1}{\la \Delta q^2\ra_n^{\mathds 1}}\l \frac{\beta_2 -\beta_1}{n}x q + \frac{x^2}{2n^2}\Big( \beta_1^2(1+n)-2 \beta_1\beta_2  +(1-n) \beta_2^2\Big) \r
\\ 
+\frac{1}{\la \Delta q^2\ra_1^{\mathds 1}}\l (\beta_2-\beta_1) x q -x^2 \frac{\beta_2^2- \beta_1^2}{2}\r,
\label{SnVV}
\end{multline}
that in the limit $n\to1$ simplifies as 
\be
S_1( \rho_{V_{\beta_1}} \| \rho_{V_{\beta_2}})(q) = (\beta_1-\beta_2)^2(1-\pi x\cot \pi x)-\frac{(\beta_1-\beta_2)^2 x^2}{2\la \Delta q^2\ra_1^{\mathds 1}}.
\ee
Once again, in the replica limit, and at this order the relative entropy becomes symmetric in its argument and independent of $q$.
Clearly Eq. \eqref{SnVV} for $\beta_1\to 0$ or $\beta_2\to 0$ correctly reproduce Eqs. \eqref{eq132b} and \eqref{eq132} respectively.

\subsection{Vacuum-current symmetry resolved relative entropies}

Here we move to the relative entropy between the current state $i\partial \varphi$ and a vertex state, starting from the special case of the vacuum.
Again, we start from Eq. \eqref{Srel_gen2}, in which the various coefficients $a$ and $b$  require the knowledge 
of $f_n^{i\partial \phi}(\alpha)$ and $f_{(1,n-1)}^{i\partial \phi,{\mathds 1}}(\alpha)$
at order $\alpha^2$. 
The former is obtained just by expanding Eq. \eqref{Capizzif_n} and obtaining $a^{i\partial \varphi}_n=0$ and 
\be
b^{i\partial \varphi}_n = \frac{1}{2\pi^2}\left( \psi\left( \frac{1}{2} \left( \frac{n}{\sin(\pi x)} -n-1\right)+1 \right) -\psi\left( \frac{1}{2}\left( \frac{n}{\sin(\pi x)} -n-1\right)+1+n\right)\right),
\ee
where $\psi(x)$ is the digamma function. 
Combining this quadratic form with $p_n^{\mathds 1}(\alpha)$ in Eq. \eqref{pGS}, we can perform the Fourier transform of 
$p_n^{i\partial\phi}(\alpha) =p_n^{\mathds 1}(\alpha) f_n^{i\partial\phi}(\alpha)$, obtaining
\be
p_n^{i\partial\phi}(q) \simeq \frac{1}{\sqrt{2\pi \la \Delta q^2\ra_n^{\mathds 1}}}\exp (-\frac{q^2}{2 \la \Delta q^2\ra_n^{\mathds 1}}) 
\l 1+\frac{1}{2}b^{i\partial \varphi}_n\l-\frac{1}{\la \Delta q^2 \ra_n^{\mathds 1}} + \frac{q^2}{\l\la \Delta q^2 \ra_n^{\mathds 1}\r^2}\r \r.
\label{eq136}
\ee
Similarly, expanding at the second order $f^{i\partial\phi,{\mathds 1}}_{(1,n-1)}$ we get $1^{i\partial \varphi, {\mathds 1}}_{1,n-1}=0$ and
 $b^{i\partial \varphi, {\mathds 1}}_{1,n-1}$ is obtained 
\be
b^{i\partial \varphi, {\mathds 1}}_{1,n-1}=\frac{2}{\pi^2} \sin^2 \l \frac{\pi x}{n}\r.
\ee
Taking the Fourier transform of $p_{(1,n-1)}^{i\partial \phi,{\mathds 1}}(\alpha) = p_n^{\mathds 1}(\alpha)f_{(1,n-1)}^{i\partial \phi,{\mathds 1}}(\alpha)$, in the Gaussian 
approximation, we get that $p_{(1,n-1)}^{i\partial \phi,{\mathds 1}}(q)$ is the same as Eq. \eqref{eq136} with the 
replacement $b^{i\partial \varphi}_n\to b^{i\partial \varphi, {\mathds 1}}_{1,n-1}$. 
%
Finally, the $n$-th total relative entropy $S_n(\rho_{i\partial\phi}\| \rho_{\mathds 1})$ has been computed for general $n$ in Ref. \cite{Rel_Paola}, but its form 
is rather cumbersome and we report here just its value for $n\to 1$
\be
S_1(\rho_{i\partial\phi}\| \rho_{\mathds 1}) = 2 \l \log \l 2 \sin \pi x\r +1 -\pi x \text{cot} \pi x + \psi\l \frac{1}{2} \text{csc} \pi x \r + \sin \pi x\r,
\ee
with $\psi(x)$ being again the digamma function. 
Plugging Eqs. \eqref{pGS} and \eqref{eq136} into Eq. \eqref{ratio3} for $S_n(\rho_{i\partial\phi}\| \rho_{\mathds 1}) (q)$, we get
\begin{multline}
S_n(\rho_{i\partial\phi}\| \rho_{\mathds 1}) (q) = S_n(\rho_{i\partial\phi}\| \rho_{\mathds 1}) + \frac{1}{1-n}\log \frac{p^{i\partial\phi, {\mathds 1}}_{(1,n-1)}(q)}{p^{i\partial\phi}_{n}(q)} - \log \frac{p_1^{i\partial\phi}(q)}{p_1^{\mathds 1}(q)} \simeq \\
S_n(\rho_{i\partial\phi}\| \rho_{\mathds 1}) + \frac{1}{1-n} \frac{1}{2}\l b^{i\partial \varphi, {\mathds 1}}_{1,n-1}-b^{i\partial \varphi}_n \r \l -\frac{1}{\la \Delta q^2 \ra_n^{\mathds 1}} + \frac{q^2}{\l\la \Delta q^2 \ra_n^{\mathds 1}\r^2}\r -
\frac{1}{2} b^{i\partial \varphi}_1 \l -\frac{1}{\la \Delta q^2 \ra_1^{\mathds 1}} + \frac{q^2}{\l\la \Delta q^2 \ra_1^{\mathds 1}\r^2}\r.
\label{eq140}
\end{multline} 
In particular, at small $x$, the difference $b^{i\partial \varphi, {\mathds 1}}_{1,n-1}-b^{i\partial \varphi}_n$ can be read from the OPE expansion \eqref{fderOPE}: 
\be
b^{i\partial \varphi, {\mathds 1}}_{1,n-1}-b^{i\partial \varphi}_n = 2\frac{\alpha x^2}{n^2}\l 1-n\r + o(x^2).
\ee

Quite interestingly for small $x$, even for $n\rightarrow 1$,  $S_n(\rho_{i\partial\phi}\| \rho_{\mathds 1}) (q) - S_n(\rho_{i\partial\phi}\| \rho_{\mathds 1})$ in Eq. \eqref{eq140} 
scales as $\sim x^2$. The origin of this behaviour is the presence of the fusion channel 
\be
\mathcal{T}_{n,\alpha} \times \tilde{\mathcal{T}}_{n,{\alpha}} \rightarrow T \otimes 1 \cdots \otimes 1,
\ee
 also in the limit $n\rightarrow 1$ when $\alpha \neq 0$.
 
If we consider instead the other relative entropy $S_n(\rho_{\mathds 1}\| \rho_{i\partial\phi})(q)$, the computation cannot be brought till the end because 
we do not have a close expression for $f^{{\mathds 1},i\partial\phi}_{(1,n-1)}(\alpha)$ for general $n$.
The same  is also true for the total relative entropy, for which in Ref. \cite{Rel_Paola}, it was not possible to derive the analytic 
continuation to general $n$, but only a determinant form valid for integer $n$. 
Eq. \eqref{Srel_gen2} can always be used. Also in this case there is no shift of the average charge, i.e. 
$a^{{\mathds 1},i\partial\phi}_{(1,n-1)}=a^{i\partial \varphi}_n=0$. 
We do not have a closed form for $b^{{\mathds 1},i\partial\phi}_{(1,n-1)}-b^{\mathds 1}_n$ (actually, $b^{\mathds 1}_n=0$ but we prefer to keep it), but 
we can still investigate the small $x$ behaviour. 
Indeed, from the OPE \eqref{fderOPE} we have 
\be
b^{{\mathds 1},i\partial\phi}_{(1,n-1)}-b^{\mathds 1}_n=\frac{2x^2(n-1)}{n^2} + o(x^2),
\ee
which, at this order, coincides with $b^{i\partial\phi,{\mathds 1}}_{(1,n-1)}-b^{\mathds 1}_n$, but such equality is does hold at higher order.


\subsection{Vertex-current symmetry resolved relative entropies}

For what concerns the pair of states $\ket{V_\beta},\ket{i\partial \phi}$ the symmetry resolved relative entropies are always given by Eq. \eqref{Srel_gen2}.
The scaling function $f^{i\partial \phi, V_\beta}_{(1,n-1)}(\alpha)$ is given in Eq. \eqref{f99}. 
Expanding at second order, we get $a^{i\partial\phi,V_\beta}_{(1,n-1)}$ and $b^{i\partial\phi,V_\beta}_{(1,n-1)}$
which are given by Eqs. \eqref{eq113} and \eqref{eq114} respectively. 
Plugging these expression into Eq. \eqref{Srel_gen2} one gets $S_n(\rho_{i\partial\phi} \| \rho_{V_\beta})(q) $ that however we do not report here because it is long and 
not very illuminating. 
We only mention that, because of the small $x$ behaviour of $a^{i\partial\phi,V_\beta}_{(1,n-1)}$ in Eq.  \eqref{eq113},  
$S_n(\rho_{i\partial\phi} \| \rho_{V_\beta})(q)$ has a $q$-dependence at order $O(x)$ when $x$ is small. 
This is completely analogous to what already obtained for $S_n(\rho_{\mathds 1} \| \rho_{V_\beta})(q)$, whose explicit expression is Eq. \eqref{eq132}.

\section{Symmetry-resolved distances}
\label{sec6}

In this section we investigate the symmetry-resolved distance $D'_n(\rho_\Upsilon,\rho_\chi)(q)$, defined by equation \eqref{D1np}, 
considering $\rho=\rho_\Upsilon$ and $\sigma=\rho_\chi$ as RDM of low lying-states of the compact boson CFT generated by primary fields. 
To do so, we have to characterise the following quantity
\be
\trace((\rho_\Upsilon-\rho_\chi)^{n_e}\Pi_q)
\ee
with $n_e$ being an even integer and eventually analytically continue the result to any value of $n_e$ \cite{Zhang1,Zhang2}. 
Such analytic continuation is  necessary for $\trace(|\rho_\Upsilon-\rho_\chi|^{n}\Pi_q)$ when $n$ is not an even integer, in particular for the physical relevant case with $n=1$.

In any quantum field theory, in order to deal with finite distances, it is custom to normalise $D_n$ via the moments of the RDM in the vacuum  \cite{Zhang1,Zhang2}. 
Hence, for the symmetry resolved one, we introduce
\be
\mathcal{D}'_n(\rho_\Upsilon,\rho_\chi)(q) \equiv \frac{1}{2} \frac{\trace\l \Pi_{q}|\rho_\Upsilon-\rho_\chi|^n\r}{\trace(\l\rho_{\mathds 1}\r^n)},
\label{dist2}
\ee
which differs from $D'_n(\rho_\Upsilon,\rho_\chi)(q)$ for an overall $q$-independent constant.
$\mathcal{D}'_n(\rho_\Upsilon,\rho_\chi)(q)$ satisfies the following sum-rule
\be
\sum_{q} \mathcal{D}'_n(\rho_\Upsilon,\rho_\chi)(q) = \mathcal{D}_n(\rho_\Upsilon,\rho_\chi) \equiv \frac {1}{2} \frac{\trace|\rho_\Upsilon-\rho_\chi|^n}{\trace(\l\rho^{\mathds 1}\r^n)},
\label{eq149}
\ee
analogous to Eq. \eqref{Dprime}. 
The reason we do so is that $\mathcal{D}_n(\rho_\Upsilon,\rho_\chi)$ is a cut-off independent quantity, a feature which is not shared by $D_n(\rho_\Upsilon,\rho_\chi)$ 
when $n\neq 1$, as discussed in \cite{Zhang1,Zhang2}. 
In the limit $n\rightarrow 1$, since the RDM is normalised as $\trace(\rho_{\mathds 1})=1$,  $\mathcal{D}'_n(\rho_\Upsilon,\rho_\chi)(q)$ and $D'_n(\rho_\Upsilon,\rho_\chi)(q)$
become equal, but for $n\neq 1$ this is not the case.

As discussed in Ref. \cite{Zhang2} it is extremely difficult to characterise $\trace((\rho_\Upsilon-\rho_\chi)^n)$ analytically for generic $x=\ell/L$, 
even for the primary states discussed so far.
Consequently, our main focus here is the small $x$ behaviour, which can be extracted through the OPE expansion adapting the techniques 
employed in \cite{Zhang2} for the total distances. We recall the main result of \cite{Zhang2}
\be
\frac{\trace((\rho_\Upsilon-\rho_\chi)^n)}{{\trace(\l\rho_{\mathds 1}\r^n)}} = 
\sum_{\Psi_1,\dots,\Psi_n}b_{\Psi_1,\dots,\Psi_n} \ell^{\Delta_{\Psi_1}+\dots+\Delta_{\Psi_n}}
(\la \Psi_1\ra_\Upsilon - \la \Psi_1\ra_{\chi})\dots(\la \Psi_n\ra_\Upsilon - \la \Psi_n\ra_{\chi}).
\label{dist_fluxabs}
\ee
Here, the sum is over the set of orthogonal quasiprimaries $(\Psi_1,\dots,\Psi_n)$ of the CFT and $b_{\Psi_1\dots \Psi_n}$ are the universal OPE coefficients of the fusion
\be
\mathcal{T}_n \times \tilde{\mathcal{T}}_n \rightarrow \Psi_1 \otimes \cdots \otimes \Psi_n.
\ee

Eq. \eqref{dist_fluxabs} is straightforwardly generalised to the charged moments as
\be
\frac{\trace((\rho_\Upsilon-\rho_\chi)^ne^{i\alpha Q})}{\trace(\l\rho_{\mathds 1}\r^ne^{i\alpha Q})}
= 
\sum_{\Psi_1,\dots,\Psi_n}b_{\Psi_1,\dots,\Psi_n}(\alpha) \ell^{\Delta_{\Psi_1}+\dots+\Delta_{\Psi_n}}(\la \Psi_1\ra_\Upsilon - \la \Psi_1\ra_\chi)\dots(\la \Psi_n\ra_\Upsilon - \la \Psi_n\ra_\chi),
\label{dist_flux}
\ee
with $b_{\Psi_1\dots \Psi_n}(\alpha)$ being the OPE coefficient of 
\be
\mathcal{T}_{n,\alpha} \times \tilde{\mathcal{T}}_{n,\alpha} \rightarrow \Psi_1 \otimes \cdots \otimes \Psi_n.
\ee
From Eq. \eqref{dist_flux} we have that the leading term in the expansion of $\trace((\rho_\Upsilon-\rho_\chi)^ne^{i\alpha Q})$ for small $\ell/L$ 
is due to the most relevant quasiprimary $\Psi$ (with the smallest scaling dimension $\Delta_\Psi$) satisfying
\be
\la \Psi\ra_\Upsilon - \la \Psi\ra_\chi \neq 0.
\ee
Thus, at leading order, we have
\be
\frac{\trace((\rho_\Upsilon-\rho_\chi)^ne^{i\alpha Q})}{\trace(\l\rho_{\mathds 1}\r^ne^{i\alpha Q})}=b_{\Psi\cdots\Psi}(\alpha) \ell^{n\Delta_\Psi}(\la \Psi\ra_\Upsilon - \la \Psi\ra_\chi)^n
 \propto \l \frac{\ell}{L}\r^{n\Delta_\Psi}.
\label{dist_flux1}
\ee
The proportionality constant in Eq. \eqref{dist_flux1} is universal and depends on $\alpha$ and the states under consideration. 
Although its explicit determination can be difficult (see \cite{Zhang2} for vertex operators in absence of the flux), 
Eq. \eqref{dist_flux1} still represents an important result providing in full generality the scaling in $x$ of the charged moment, 
once the quasiprimary $\Psi$ has been identified.
From the knowledge of $\trace((\rho_\Upsilon-\rho_\chi)^ne^{i\alpha Q})$ one easily gets $\trace((\rho_\Upsilon-\rho_\chi)^n\Pi_q)$ via Fourier transform, 
which is the key ingredient to  compute $\mathcal{D}_n'(\rho_\Upsilon,\rho_\chi)(q)$ from Eq. \eqref{dist2}. 

\subsection{The case $n=2$}

The CFT computation of the lhs of Eq. \eqref{dist_flux} for general $n$ is made difficult by the many charged moments that needs to be calculated after
expanding the $n$-th power. 
It is then rather instructive to look at  what happens for the case $n=2$ that can be simply handled without requiring $\ell/L$ small. 
On top of the pedagogical character, the calculation has a per-se interest because $n=2$ gives a genuine distance (admittedly not the most relevant one).    
Let us define, with a small abuse of notation,
\be
\mathcal{D}'_n(\rho_\Upsilon,\rho_\chi)(\alpha) \equiv \frac{1}{2}\frac{\trace\l e^{i\alpha Q}(\rho_\Upsilon-\rho_\chi)^n \r}{\trace\l \rho_{\mathds 1}^n \r}
\label{eq163}
\ee
as the Fourier transform of $\mathcal{D}'_n(\rho_\Upsilon,\rho_\chi)(q)$ appearing in Eq. \eqref{dist2}. 
For $n=2$, we can recast $\mathcal{D}'_2(\rho_\Upsilon,\rho_\chi)(\alpha)$ as follows
\be
\mathcal{D}'_2(\rho_\Upsilon,\rho_\chi)(\alpha) = \frac{1}{2}\frac{\trace(\rho_{\mathds 1}^2e^{i\alpha Q})}{\trace(\rho_{\mathds 1}^2)}
\l \frac{\trace\l \rho_\Upsilon^2e^{i\alpha Q}\r}{\trace\l \rho_{\mathds 1}^2e^{i\alpha Q}\r}+\frac{\trace\l \rho_\chi^2e^{i\alpha Q}\r}{\trace\l \rho_{\mathds 1}^2e^{i\alpha Q}\r}-2\frac{\trace\l \rho_\Upsilon \rho_\chi e^{i\alpha Q}\r}{\trace\l \rho_{\mathds 1}^2e^{i\alpha Q}\r}\r.
\label{d21}
\ee
All the terms inside the parenthesis of Eq. \eqref{d21} are universal. Indeed, the first term gives
\be
\frac{\trace\l \rho_\Upsilon^2e^{i\alpha Q}\r}{\trace\l \rho_{\mathds 1}^2e^{i\alpha Q}\r} = \frac{\trace(\rho_\Upsilon^2)}{\trace(\rho_{\mathds 1}^2)}\l \frac{\trace\l \rho_\Upsilon^2e^{i\alpha Q}\r}{\trace\l \rho_\Upsilon^2\r}\frac{\trace(\rho_{\mathds 1}^2)}{\trace(\rho_{\mathds 1}^2e^{i\alpha Q})}\r,
\ee
and we recognize $\frac{\trace(\rho_\Upsilon^2)}{\trace(\rho_{\mathds 1}^2)}$ as a universal ratio directly related to the excess of $2$-nd Rényi entropy (see \cite{Sierra1}) while 
$\l \frac{\trace\l \rho_\Upsilon^2e^{i\alpha Q}\r}{\trace\l \rho_\Upsilon^2\r}\frac{\trace(\rho_{\mathds 1}^2)}{\trace(\rho_{\mathds 1}^2e^{i\alpha Q})}\r$ is $f^\Upsilon_{n=2}(\alpha)$ defined by Eq. \eqref{univ_ratio} and discussed explicitly in the work \cite{Capizzi}. Similarly, the last piece appearing in Eq. \eqref{d21} can be written as
\be
\frac{\trace\l \rho_\Upsilon \rho_\chi e^{i\alpha Q}\r}{\trace\l \rho_{\mathds 1}^2e^{i\alpha Q}\r} = \frac{\trace(\rho_\Upsilon \rho_\chi)}{\trace(\rho_{\mathds 1}^2)} \l \frac{\trace(\rho_\Upsilon\rho_\chi e^{i\alpha Q})}{\trace(\rho_\Upsilon\rho_\chi)}\frac{\trace(\rho_{\mathds 1}^2)}{\trace(\rho_{\mathds 1}^2e^{i\alpha Q})} \r.
\ee
The quantity $\frac{\trace(\rho_\Upsilon \rho_\chi)}{\trace(\rho_{\mathds 1}^2)}$ has been studied in the context of $n$-distances in \cite{Zhang2}, 
while $\l \frac{\trace(\rho_\Upsilon\rho_\chi e^{i\alpha Q})}{\trace(\rho_\Upsilon\rho_\chi)}\frac{\trace(\rho_{\mathds 1}^2)}{\trace(\rho_{\mathds 1}^2e^{i\alpha Q})} \r$ is related 
to $f^{\Upsilon,\chi}_{(1,1)}(\alpha)$ defined by Eq. \eqref{definition}, which has been obtained explicitly for the low-lying state of the Luttinger liquid in section \ref{CBos}. 

\subsection{Setup for general $n$}

Qualitatively, the same procedure above for $n=2$ generalises to any (even integer) $n$, but it becomes soon untreatable because of
the large numbers of terms appearing when expanding $\trace\l(\rho_\Upsilon-\rho_\chi)^n e^{i\alpha Q}\r$.
In analogy to Eq. \eqref{d21} for $n=2$, $\mathcal{D}'_n(\rho_\Upsilon,\rho_\chi)(\alpha)$ can be parametrised as
\be
\mathcal{D}_n'(\rho_\Upsilon,\rho_\chi)(\alpha) = p_n^{\mathds 1}(\alpha)\l \mathcal{D}_n(\rho_\Upsilon,\rho_\chi)-c_n^{\Upsilon,\chi}\frac{\alpha^2}{2} + o(\alpha^2)\r,
\label{eq167}
\ee
where $c_n^{\Upsilon,\chi}$ is a certain universal constant (in. $\alpha$, but $x$ dependent) that for finite and integer $n$ can be computed on a case by case basis.
In Eq. \eqref{eq167} we assume that the first $\alpha$-dependent universal contribution appear at order $\alpha^2$. 
Actually in principle there could be a linear term in $\alpha$; however it never appear in any of the states considered and so we ignore such a term here so to have 
more compact formulas. 
We neglect the higher orders in $\alpha$ because they produce subleading terms by Fourier transform. 
The probability $p_n^{\mathds 1}(\alpha)$ is Gaussian with zero mean and variance $\la \Delta q^2\ra_n^{\mathds 1}$ diverging logarithmically with $L$, 
as given by Eq. \eqref{pGS}; 
consequently the Fourier transform of Eq.  \eqref{eq167} is easily performed as
\be
\mathcal{D}'_n(\rho_\Upsilon,\rho_\chi)(q) \simeq \frac{1}{\sqrt{2\pi \la \Delta q^2\ra_n^{\mathds 1}}}\exp(-\frac{q^2}{2 \la \Delta q^2\ra_n^{\mathds 1}})\l \mathcal{D}_n(\rho_\Upsilon,\rho_\chi) +\frac{c_n^{\Upsilon,\chi}}{2 \la \Delta q^2\ra_n^{\mathds 1}}\l -1+\frac{q^2}{\la \Delta q^2\ra_n^{\mathds 1}}\r\r.
\ee
This equation manifests the equipartition for large $L$ and shows that it is broken at leading order by the term in $c_n^{\Upsilon,\chi}$. 

In the following subsections we will deal with the low-lying excited states of the compact boson in the small $x=\ell/L$ limit, since the general case is always untreatable. 
In contrast to the symmetry-resolved relative entropy, also the small $x$ limit is non trivial.
In fact, in Section \ref{CBos} we analysed via OPE expansion $f^{\Upsilon,\chi}_{\mathcal{S}}$, defined by Eq. \eqref{definition}, only up to $O(x^2)$, 
but Eq. \eqref{dist_flux1} shows  that higher orders in $x$ are needed in general to describe $\mathcal{D}'_n(\rho,\sigma)(\alpha)$ as $n$ increases.

\subsection{Vertex-vertex distance}

The distance of the total density matrices between two vertex states $V_\beta$ and $V_{\beta'}$ was characterised for small $x$ in Ref. \cite{Zhang2}
and we closely follow this reference to work out the generalisation in the presence of a flux. 

The most relevant fusion channel in Eq. \eqref{dist_flux1} is represented by the current operator $i\partial\varphi$ \cite{Zhang2}.
Consequently, we have
\be
\mathcal{D}'_n(\rho_{V_{\beta}},\rho_{V_{\beta'}})(\alpha) \frac{\trace(\rho_{\mathds 1}^n)}{\trace(\rho_{\mathds 1}^ne^{i\alpha Q})}=
\frac{1}{2}\frac{\trace\l (\rho_{V_\beta}-\rho_{V_{\beta'}})^ne^{i\alpha Q}\r}{\trace(\rho_{\mathds 1}^ne^{i\alpha Q})} \simeq \frac{1}{2}b_{i\partial \phi,\dots,i\partial\phi}(\alpha)(2\pi\beta-2\pi\beta')^n\l \frac{\ell}{L}\r^n.
\label{eq171}
\ee
In absence of flux, $b_{i\partial \phi,\dots,i\partial\phi}(\alpha=0)$ is related to an $n$-point function of $i\partial \phi$ \cite{Zhang2} inserted in the points
\be
e^{i2\pi j/n},\quad j=0,\dots,n-1
\ee
of the complex plane $\mathbb{C}$. 
The final result, analytically continued for $n_e\rightarrow 1$, is \cite{Zhang2}, 
\be
\lim_{n_e\rightarrow 1} b_{i\partial \phi,\dots,i\partial\phi}(\alpha=0) = \frac{1}{4}F^{(n'=1/2)}\l y=1/2\r,
\label{eq173}
\ee
where
\be
F^{(n')}\l y\r = \l \frac{2}{n'}\sin \pi x\r^{2n'}\frac{\Gamma^2\l \frac{1+n'+n'\text{csc}(\pi y)}{2}\r}{\Gamma^2 \l \frac{1-n'+n'\text{csc}(\pi y)}{2}\r}.
\ee
The adaption of the derivation of Ref. \cite{Zhang2} to the presence of a flux requires the insertions of two additional vertex operators $V_{\pm \alpha/2\pi}$ at $z =0,\infty$ 
in the correlation functions of the $n$ derivatives at the roots of unity. 
The computation is identical to that for $f^{i\partial \phi}_n(\alpha)$ in Eq. \eqref{Capizzif_n}, reported in subsection \ref{Der_GS}.
The resulting correlation is a characteristic polynomial that can be analytically continued. 
In the end, the generalisation of Eq. \eqref{eq173} to the presence of a flux is obtained with the replacement $F^{(n')}(y)\rightarrow F^{(n')}(y,\alpha)$ 
\be
F^{(n')}\l y,\alpha\r \equiv \l \frac{2}{n'}\sin \pi y\r^{2n'}\frac{\Gamma\l \frac{1+n'+n'\text{csc}(\pi y)+\alpha/\pi}{2}\r \Gamma\l \frac{1+n'+n'\text{csc}(\pi y)-\alpha/\pi}{2}\r }{\Gamma\l \frac{1-n'+n'\text{csc}(\pi y)+\alpha/\pi}{2}\r \Gamma \l \frac{1-n'+n'\text{csc}(\pi y)-\alpha/\pi}{2}\r}.
\ee
This form is just Eq. \eqref{Capizzif_n} with some minor adjustments due to normalisation and number of insertions.
 Specialising now it to $n'=1/2$ and $y=1/2$, we get
\be
F^{(n'=1/2)}\l y=1/2,\alpha\r = \frac{2\alpha}{\pi \tan \l \alpha/2\r},
\ee
so that
\be
\lim_{n_e\rightarrow 1} b_{i\partial \phi,\dots,i\partial\phi}(\alpha) =  \frac{\alpha}{2\pi \tan \l \alpha/2\r}.
\ee

Let us also briefly discuss what we can say for the symmetry resolved distances of vertex states beyond the small $\ell/L$ approximation. 
In Ref. \cite{Zhang2}, the following expansion has been obtained
\be
\frac{\trace\l (\rho_{V_\beta}-\rho_{V_{\beta'}})^n\r}{\trace(\rho_{\mathds 1}^n)} = \sum^{n}_{k=0}(-1)^k\sum_{0\leq j_1<\dots < j_k \leq n-1}h_n(\{j_1,\dots,j_k\})^{(\beta-\beta')^2},
\label{eq178}
\ee
where the function $h(\mathcal{S}=\{j_1,\dots,j_k\})$ is
\be
h_n(\mathcal{S}) = \l \frac{\sin{\frac{\pi\ell}{L}}}{n\sin{\frac{\pi\ell}{n L}}} \r^{|\mathcal{S}|}
         \prod^{j_1 < j_2}_{j_1,j_2 \in \mathcal{S}}
         \frac{\sin^2 \frac{\pi (j_1 - j_2)}{n}}{\sin\frac{\pi (j_1 - j_2 + \ell/L)}{n}\sin\frac{\pi (j_1 - j_2 - \ell/L)}{n}}.
\ee
If the flux is inserted, using our determination of $f^{V_\beta,V_{\beta'}}_{\mathcal{S}}(\alpha)$ in Eq. \eqref{VVCFT}, we have
\be
\frac{\trace\l (\rho_{V_\beta}-\rho_{V_{\beta'}})^ne^{i\alpha Q}\r}{\trace(\rho_{\mathds 1}^ne^{i\alpha Q})} = \sum^{n}_{k=0}(-1)^ke^{i\alpha \frac{\ell}{L}(\beta k + (n-k)\beta')}\sum_{0\leq j_1<\dots < j_k \leq n-1}h_n(\{j_1,\dots,j_k\})^{(\beta-\beta')^2}.
\label{eq172}
\ee
For $\alpha=0$, it reduces to the distance among vertex states in Eq. \eqref{eq178}. 
The exponential term $e^{i\alpha \frac{\ell}{L}(\beta k + (n-k)\beta')}$ is the additional weight due to the presence of $k$ $V_\beta$ and of $n-k$ $V_{\beta'}$ 
in the partition $\mathcal{S}$, as it follows from the function $f^{V_\beta,V_{\beta'}}_{\mathcal{S}}(\alpha)$ in Eq. \eqref{VVCFT}.
Eq. \eqref{eq172} can be worked out for any finite (not too large) even $n$, but the analytic continuation is still too difficult.

\subsection{Current-vertex distances}

The distance between  the states $\ket{V_\beta}$ and $\ket{i\partial\phi}$ can be analysed in the small $\ell/L$ limit with the same 
methods of the previous subsection. 
Whenever $\beta \neq 0$, the lightest quasiprimary $\Psi$ in Eq. \eqref{dist_flux1} is identified with $i\partial\phi$ that has 
a non trivial expectation value for the vertex states. Hence, similarly to Eq. \eqref{eq171}, we finally get 
\be
\mathcal{D}'_n(\rho_{V_{\beta}},\rho_{i\partial\phi})(\alpha)\frac{\trace(\rho_{\mathds 1}^n)}{\trace(\rho_{\mathds 1}^ne^{i\alpha Q})}=
\frac{1}{2}\frac{\trace\l (\rho_{V_\beta}-\rho_{i\partial\phi})^ne^{i\alpha Q}\r}{\trace(\rho_{\mathds 1}^ne^{i\alpha Q})} \simeq \frac{1}{2}b_{i\partial \phi,\dots,i\partial\phi}(\alpha)(2\pi\beta)^n\l \frac{\ell}{L}\r^n.
\label{eq180}
\ee
When $\beta =0$, i.e. $\rho_{V_\beta}$ becomes the vacuum RDM $\rho_{\mathds 1}$, the expectation value of $\psi= i\partial\phi$ vanishes and 
the most relevant operator in Eq. \eqref{dist_flux1} is $\Psi=T$, the stress-energy tensor. We can thus write
\be
\mathcal{D}'_n(\rho_{\mathds 1},\rho_{i\partial\phi})(\alpha) \frac{\trace(\rho_{\mathds 1}^n)}{\trace(\rho_{\mathds 1}^ne^{i\alpha Q})} =\frac12
\frac{\trace\l (\rho_{\mathds 1}-\rho_{i\partial\phi})^ne^{i\alpha Q}\r}{\trace(\rho_{\mathds 1}^ne^{i\alpha Q})} \simeq 
\frac12b_{T\dots T}(\alpha)(\la T\ra_{i\partial\phi} -\la T\ra_{\mathds 1})^n \ell^{2n} \propto \l \frac{\ell}{L} \r^{2n}.
\ee 
The OPE coefficient $b_{T\dots T}(\alpha)$ is related to an $n$-point function of the stress energy tensor with the additional insertion of two vertex operators 
$V_{\pm \alpha/2\pi}$(see \cite{Zhang2} in the absence of flux). This correlation function can be in principle calculated from the Ward identities, as explained \cite{pagialle}; 
however it is difficult and no predictions for generic $n$ are available and so the analytic continuation is still untreatable. 

\section{Conclusions}
\label{sec7}

In this manuscript we developed a systematic replica technique for the  calculation of symmetry resolved relative entropies and subsystem distances. 
In principle, our approach can be applied to a generic one dimensional quantum system and in particular to 2D quantum field theories. 
We applied this method to the analytic computation of symmetry resolved relative entropies and distances between the RDMs of one interval embedded in 
various low-lying energy eigenstates of 2D CFT, with particular focus on the free massless compact boson.
We provided analytic expressions for the charged moments corresponding to the resolution of both relative entropies and distances 
for general integer $n$. 
For the relative entropies, these formulas are manageable and the analytic continuation to $n=1$ can be worked out in most of the cases.  
Conversely, for the distances the corresponding charged moments become soon untreatable as $n$ increases. 
As a consequence, we have been able only to perform the analytic continuation for small intervals via OPE of composite twist fields. 
This problem does not come unexpectedly since it was already encountered for the total subsystem distances \cite{Zhang2}.   
We recall that, if needed, one might use known techniques for numerical analytic continuations (see, e.g., Refs. \cite{ahjk-14,dct-15}) to obtain the 
relative entropies and the trace distances from the analytically presented results at finite integer  $n$. 
Our CFT results have been compared with exact numerical computations for the XX spin-chain, with a focus on the universal functions that 
provide a more accurate test of the theory.

Our replica framework can be applied to many different physical situations.
There are two specific cases where investigating the symmetry resolution of these measures of distinguishably could be very useful and insightful.
These are 
(i) the convergence of the RDM to a thermodynamic one (either thermal or generalised Gibbs depending on the number of conserved charges 
of the model) after a quantum quench;
(ii) the effectiveness of approximating the RDM of a microscopic model with a lattice Bisognano-Wichmann form. 
In both these cases, the distances between the corresponding RDMs tends to zero in the appropriate limit, but it is very natural to wonder whether the 
same is true in all symmetry sectors.  

\textit{Note added -} 
After the completion of the calculations in this paper, but before its submission, the manuscript \cite{c-21} appeared on the ArXiv that has partial overlap 
with the result presented here, in particular with those of Section \ref{sec5}. However, the first version of \cite{c-21} presents an error that has been corrected here.

\section*{Acknowledgments}
We are grateful to Sara Murciano for useful discussions. All authors acknowledge support from ERC under Consolidator grant number 771536 (NEMO).
\begin{appendix}

\section{Correlation functions among primaries in the Luttinger liquid}\label{AppCorr}

In this appendix, following Ref. \cite{Capizzi}, we give a graphical representation for the correlation function 
\be
\langle V_{\alpha_1}(\zeta_1) \dots V_{\alpha_k}(\zeta_k) (i\partial \phi)(z_1)\dots (i\partial \phi)(z_n)\rangle_{\mathbb{C}}
\label{corr}
\ee
evaluated in the ground state of a planar geometry.

The starting point is the correlation function between vertex operators
\be
\langle V_{\alpha_1}(\zeta_1) \dots V_{\alpha_k}(\zeta_k)\rangle_{\mathbb{C}} = \prod_{i<j}(\zeta_i-\zeta_j)^{K\alpha_i \alpha_j},
\label{Corr_vertex}
\ee
valid for $\sum_{j}\alpha_j =0$ otherwise it vanishes. 
Hereafter we suppose that the neutrality condition $\sum_{j}\alpha_j =0$ is always satisfied. 
The derivative operator $i\partial \phi$ can be represented as follows
\be
(i\partial \phi)(z) = \underset{\epsilon \rightarrow 0}{\lim}\frac{1}{\epsilon}(\partial V_\epsilon)(z),
\label{der}
\ee
allowing us to write \eqref{corr} as a number of derivatives of \eqref{Corr_vertex}. 
The full expression is quite involved (see \cite{Capizzi} for some details), so we introduced some diagrammatic rules to deal with it.

\begin{figure}[t]
  \includegraphics[width=\linewidth]{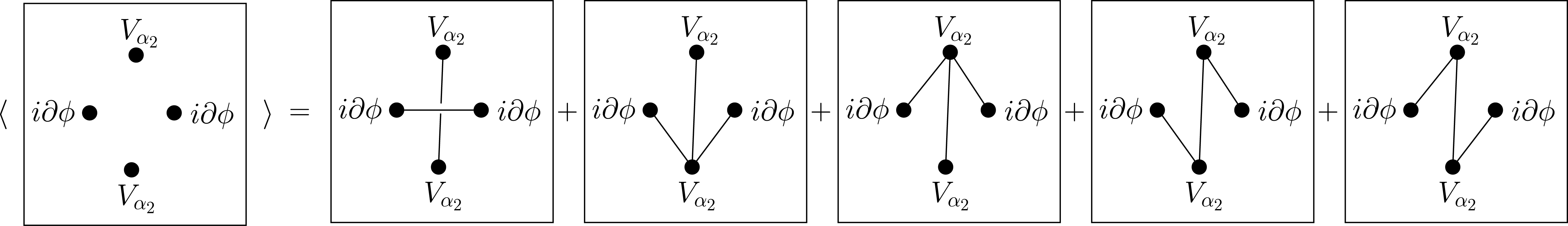}
  \caption{This is a graphical representation for $\langle V_{\alpha_1}(\zeta_1)V_{\alpha_2}(\zeta_2) (i\partial \phi)(z_1)(i\partial \phi)(z_2)\rangle$.}
  \label{Fig:con}
\end{figure}

\textbf{Diagramatic rules (for the planar geometry $\mathbb{C}$)}:
\begin{itemize}
\item The full correlation function is made by different terms containing different contractions.
\item The contraction between $(i\partial \phi)(z)$ and $V_\alpha(\zeta)$ gives a factor $\frac{K\alpha}{\zeta-z}$.
\item The contraction between $V_{\alpha_i}(z_i)$ and $V_{\alpha_j}(\zeta_j)$ gives a factor $(\zeta_i-\zeta_j)^{K\alpha_i\alpha_j}$.
\item The contraction between $(i\partial \phi)(z_i)$ and $(i\partial \phi)(z_j)$ gives a factor $\frac{K}{(z_i-z_j)^2}$.
\item Every $(i\partial \phi)(z_j)$ is contracted to just another operator.
\item Every $V_{\alpha}(\zeta_j)$ is contracted to any other operator, keeping the previous contraint.
\end{itemize}
Fig. \ref{Fig:con} reports all possible contractions for $\langle V_{\alpha_1}(\zeta_1)V_{\alpha_2}(\zeta_2) (i\partial \phi)(z_1)(i\partial \phi)(z_2)\rangle$.

In the case of a cylindrical geometry of circumference $L$, it is enough to make the following replacement in Eq.\eqref{Corr_vertex} 
\be
\zeta_i-\zeta_j \rightarrow \frac{L}{\pi}\sin \l \frac{\pi(\zeta_i-\zeta_j)}{L}\r.
\ee
Also the results for $K \neq 1$ can be obtained starting from $K=1$ and replacing
\be
\alpha_j \rightarrow \sqrt{K}\alpha_j \qquad i\partial\phi \rightarrow \sqrt{K}i\partial \phi,
\ee
a simple fact that follows directly from Eq. \eqref{Corr_vertex}.

\section{Numerical methods for the XX chain}
\label{App2}

We consider the tight-binding 1D chain of free fermions described by the hamiltonian
\be
H = - \sum_j \left[c_j^\dagger c_{j+1} + c_{j+1}^\dagger c_j -2h \Big( c_j^\dagger c_j -\frac{1}{2} \Big) \right],
\label{XX}
\ee
where $c_j^\dagger,c_j$ are the creation/annihilation operators of spinless fermions at the site $j$. 
One can study either the Ramond(R) or the Neveu-Schwarz sectors, which correspond to periodic or antiperiodic boundary conditions respectively.
By Jordan-Wigner transformation this fermion model is mapped to the XX spin-chain.

The correlation matrix of a state $\rho$ is
\be
C_{ij} = \trace(\rho c_i^\dagger c_j).
\label{Corr}
\ee
The subsystem correlation matrix $C_A$ is the restriction of $C$ to a subsystem $A$; it has dimension $\ell\times\ell$ with $\ell$ the number of sites in $A$.
A quadratic hamiltonian like \eqref{XX}  admits a basis of gaussian eigenstates, whose RDM is also gaussian, i.e.
\be
\rho_A \equiv \trace_B(\rho)\propto \exp(-\sum_{i,j}\epsilon_{ij} c_i^\dagger c_j),
\label{gstate}
\ee
for a given $\ell \times \ell$ matrix $\epsilon$. 
By Wick theorem,  $\epsilon$ and $C_A$ are related as \cite{Pesch1,Pesch2}
\be
C_A = \frac{1}{e^{\epsilon}+1}.
\ee
The proportionality constant in \eqref{gstate} ensures that $\trace(\rho_A)=1$ and it is given by $\det(C_A)$.

Given two gaussian states $\rho_1, \rho_2$ also their product $\rho = \rho_1 \rho_2$ is gaussian
The correlation matrix for the product $\rho$ is \cite{bb-69,Fagotti}
\be
C = C_1 \times C_2 \equiv C_2 \frac{1}{1-C_1-C_2+2C_1C_2}C_1.
\ee

The generating function 
\be
p(\alpha) \equiv \trace(\rho e^{i\alpha Q}).
\ee
for the statistics of the number of fermions $Q \equiv \sum_j c_j^\dagger c_j$ is written in terms of $C$ as \cite{G1,ll-93,kl-09,dk-06}
\be
p(\alpha) = \det(Ce^{i\alpha}+ (1-C)).
\ee
Consequently, the average number of particles $\langle Q\rangle$ and its variance $\langle \Delta Q^2 \rangle$ are
\be
\langle Q\rangle \equiv \frac{1}{i}\frac{d}{d\alpha}\log p(\alpha)\Big|_{\alpha=0} = \trace(C),
\ee
\be
\langle \Delta Q^2 \rangle \equiv \frac{1}{(i)^2}\frac{d^2}{d\alpha^2}\log p(\alpha)\Big|_{\alpha=0} = \trace(C)-\trace(C^2).
\ee

\end{appendix}

\end{document}